\documentclass[sigconf,screen]{acmart}
\AtBeginDocument{%
  }

\acmConference[Conference acronym 'XX]{Make sure to enter the correct
  conference title from your rights confirmation email}{June 03--05,
  2018}{Woodstock, NY}

\renewcommand\footnotetextcopyrightpermission[1]{}
\usepackage{xcolor}
\usepackage{enumitem}
\usepackage{amsmath}
\usepackage{adjustbox}
\usepackage{xspace}
\usepackage{subcaption}
\usepackage{listings}
\usepackage{tcolorbox}
\usepackage{xcolor}
\usepackage{booktabs}
\usepackage{multirow}
\usepackage{pifont} 
\usepackage{wrapfig}

\newcommand{\cmark}{\ding{51}}
\newcommand{\xmark}{\ding{55}}

\newcommand{\ncsu}[1]{{#1\textsuperscript{\includegraphics[scale=0.25]{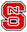}}}}
\newcommand{\uoa}[1]{{#1\textsuperscript{\includegraphics[scale=0.25]{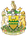}}}}

\newcommand{\ie}{\textit{i.e.,}\xspace}

\definecolor{darkgreen}{rgb}{0.0, 0.5, 0.0}
\definecolor{boxcolor}{RGB}{242,242,242}
\definecolor{bordercolor}{RGB}{0,0,0}

\newcommand{\toolname}{\textsc{Aegis}\xspace}
\newcommand{\seperate}{{\ \ \ \ \ \ \ \ }}

\usepackage{fontawesome5}

\usepackage{booktabs}
\usepackage{longtable} 
\usepackage{caption}   

\usepackage{tabularx}
\usepackage{adjustbox}
\usepackage{multicol}

\usepackage{tikz}

\usepackage{soulutf8}
\usepackage{geometry}
\usepackage{setspace}

\tcbuselibrary{theorems,skins,breakable,listings}

\definecolor{myyellow}{rgb}{1,1,0.8}
\sethlcolor{myyellow}
\definecolor{dkgreen}{rgb}{0,0.6,0}
\definecolor{mauve}{rgb}{0.58,0,0.82}
\definecolor{framegray}{gray}{0.8}

\newtcbtheorem[auto counter, number within=section]{myCodeFrame}{Listing}{
  enhanced, verbatim,
  colback=white, colframe=black, boxrule=0.5pt, arc=0.3mm,
  left=1mm, right=1mm, top=0.5mm, bottom=0.5mm,
  fonttitle=\bfseries,
  fontupper=\small\ttfamily\linespread{0.85}\selectfont,
  before upper=\vspace{-2pt},   
  after  upper=\vspace{-2pt},   
  before skip=6pt,              
  after  skip=6pt
}{lst}

\definecolor{egGreen}{HTML}{196B24}
\definecolor{egOrange}{HTML}{C04F15}

\definecolor{tmpFig2}{rgb}{0, 0, 0}

\usepackage{xurl}
\newtcolorbox{agentbox}[2][]{
    enhanced, breakable, 
    colback=#2!5, colframe=#2!80!black, 
    title=\textbf{#1}, 
    fonttitle=\bfseries,
    boxrule=1pt, arc=3pt,
    left=5pt, right=5pt, top=5pt, bottom=5pt
}

\usepackage{seqsplit}

\begin{document}

\title{\toolname: From Clues to Verdicts --- Graph-Guided Deep Vulnerability Reasoning via Dialectics and Meta-Auditing}

\author{%
  \ncsu{Sen~Fang}
  \seperate
  \ncsu{Weiyuan~Ding}
  \seperate
  \ncsu{Zhezhen~Cao}
  \seperate
  \uoa{Zhou~Yang}
  \seperate
  \ncsu{Bowen~Xu}
  \\[\bigskipamount]
  \ncsu{NC State University}
  \uoa{University of Alberta}
  \\[\bigskipamount]
  \ncsu{\{sfang9, wding8, zcao24, bxu22\}@ncsu.edu}
  \uoa{zy25@ualberta.ca}
}
\renewcommand{\shortauthors}{Fang et al.}

\thispagestyle{empty}

\begin{abstract}
Large Language Models (LLMs) are increasingly adopted for vulnerability detection, yet their reasoning remains fundamentally unsound. 
We identify a root cause shared by both major mitigation paradigms (agent-based debate and retrieval augmentation): 
reasoning in an \textit{ungrounded deliberative space} that lacks a bounded, hypothesis-specific evidence base. 
Without such grounding, agents fabricate cross-function dependencies, and retrieval heuristics supply generic knowledge decoupled from the repository's data-flow topology. Consequently, the resulting conclusions are driven by rhetorical persuasiveness rather than verifiable facts.
To ground this deliberation, we present \toolname, a novel multi-agent framework that shifts detection from ungrounded speculation to forensic verification over a closed factual substrate. 
Guided by a ``From Clue to Verdict'' philosophy, \toolname first identifies suspicious code anomalies (clues), then dynamically reconstructs per-variable dependency chains for each clue via on-demand slicing over a repository-level Code Property Graph. 
Within this closed evidence boundary, a Verifier Agent constructs competing dialectical arguments for and against exploitability, 
while an independent Audit Agent scrutinizes every claim against the trace, exercising veto power to prevent hallucinated verdicts.
Evaluation on the rigorous PrimeVul dataset demonstrates that \toolname establishes a new state-of-the-art, achieving 122 Pair-wise Correct Predictions. To our knowledge, this is the first approach to surpass 100 on this benchmark. 
It reduces the false positive rate by up to 54.40\% compared to leading baselines, at an average cost of \$0.09 per sample without any task-specific training.
\end{abstract}

\maketitle

\section{Introduction}
\label{sec:intro}
Software vulnerabilities remain a pervasive threat to critical infrastructure, with reported CVEs reaching record highs in 2024~\cite{cve23, cve24, opencve}.
However, traditional Static Application Security Testing (SAST) tools struggle to keep pace due to their reliance on rigid pattern matching and inability to interpret semantic nuances of code execution~\cite{johnson2013don, habib2018many}. 
Large Language Models (LLMs) have emerged as an increasingly prominent alternative~\cite{llm4vd, GPTLens, VulTrial, VulInstruct, llmxcpg, llm4vdre, revd, vulpo, VulSim, Vuldetectbench, llm4vuln, vultrlm, mavul}, showing promising results in understanding code semantics.
However, applying LLMs to complex, real-world software systems exposes a fundamental problem: their reasoning is \textit{ungrounded}.
Serious vulnerabilities generally manifest as local code anomalies whose exploitability depends on deep, cross-file dependency chains, such as whether a tainted variable is sanitized in a caller, or whether a buffer length is bounded by an external function.
Yet LLMs are typically forced to reason over isolated code snippets without access to a bounded, hypothesis-specific evidence base that traces these dependencies.
Without such grounding, models degenerate into contextual hallucinations, confidently inferring security properties from plausible-looking but incomplete evidence.

Two lines of work attempt to address this limitation, but neither escapes the fundamental problem of ungrounded reasoning since they differ in \textit{how} evidence is missing, not in \textit{whether} it is.
Agent-based methods~\cite{GPTLens, VulTrial, mavul} adopt multi-agent collaboration to improve reasoning reliability.
For instance, VulTrial~\cite{VulTrial} simulates adversarial debates (e.g., between a prosecutor and a defender) to refine reasoning.
While conceptually appealing, these methods often operate on \textit{isolated function slices} without access to the repository-level context.
They lack the repository-wide evidence required to reach sound conclusions, such as whether a tainted variable was sanitized in a caller function. 
Consequently, the ``consensus'' among agents is driven by \textit{contextual hallucinations}~\cite{ji2023survey, steenhoek2024comprehensive}: agents fabricate plausible but unsupported claims about cross-function behavior, rather than reasoning over actual program semantics.
Empirical evidence confirms this problem: studies on VulTrial reveal that increasing debate rounds counterintuitively degrades performance by over 15\%~\cite{VulTrial}, as agents progressively retreat from their initial assessments through mutually reinforcing concessions rather than introducing new evidence.
Retrieval-augmented approaches~\cite{llmxcpg, Vul-rag, VulInstruct} take a different strategy: they retrieve external security knowledge (e.g., CWE descriptions, historical patch patterns) and related code fragments from the repository, and concatenate them into the model's input context to provide additional information for reasoning. 
However, their retrieval follows predefined heuristics regardless of the vulnerability under analysis, such as always extracting the direct caller or retrieving context up to a fixed call depth. 
Vulnerabilities whose evidence falls within this scope may be detected, but propagation chains that require deeper or cross-file tracing are missed entirely.

These limitations are most clearly exposed under rigorous evaluation. 
On the PrimeVul benchmark, even the strongest baselines from each paradigm—fine-tuned models, retrieval-augmented approaches, and agent-based methods—suffer from false positive rates ranging from 47.1\% to 80.7\%, indicating that their predictions are driven more by superficial code patterns than by genuine semantic understanding. 
The deficiency becomes even more pronounced under the \textit{pair-wise evaluation} setting, which demands distinguishing a vulnerability from its semantically similar patch: 
even the strongest baselines fail to surpass 100 pair-wise correct predictions out of 435 test pairs~\cite{VulTrial, VulInstruct}, confirming that without a grounded evidence base, existing methods cannot comprehend the subtle semantic shifts that separate an exploit from its fix.

What is needed, therefore, is an approach that first \textit{localizes} suspicious code anomalies within a target function, then \textit{dynamically reconstructs} the repository-level context required to assess their exploitability, and finally \textit{verifies} the resulting reasoning against the assembled evidence instead of deliberating over incomplete and unverifiable context.
To this end, we present \toolname, a multi-agent framework grounded in the investigative philosophy of \textit{``From Clue to Verdict.''}
Our key insight is that vulnerabilities are not static patterns but dynamic trajectories: they manifest as local code anomalies (\textit{clues}) whose exploitability depends on their propagation through the repository's execution flow (\textit{verdict}).
To operationalize this, \toolname transforms detection from ungrounded deliberation into a transparent forensic process:
\begin{itemize}[wide=0pt, nosep]
    \item \textbf{Dynamic Evidence Construction:} 
    Our \textit{Clue-Discovery Agent} first pinpoints suspicious anomaly anchors (clues) in the target function. 
    Each identified clue then triggers the \textit{Context-Augmentation Agent}, which leverages a repository-level \textbf{Code Property Graph (CPG)}~\cite{joern-cpg} to perform on-demand slicing—dynamically reconstructing the precise dependency chain (e.g., taint propagation paths) relevant to that specific clue, rather than relying on statically predefined context boundaries.
    The resulting evidence trace forms a \textit{closed factual substrate}: a bounded, per-variable record of data provenance that defines a verifiable boundary for all downstream reasoning.

    \item \textbf{Grounded Verification via Dialectics and Meta-Auditing: } 
    Equipped with this closed evidence substrate, the \textit{Verifier Agent} replaces speculative debate with \textit{evidence-based dialectics}, constructing competing arguments (proof vs. refutation) strictly grounded in the retrieved data flow. 
    Afterward, an \textit{Audit Agent} adjudicates these arguments, filtering out unsupported reasoning leaps to render a verdict based solely on verifiable exploitability.
\end{itemize}

To validate the efficacy of \toolname, we conducted a comprehensive empirical evaluation on the PrimeVul dataset, currently the most rigorous benchmark for automated vulnerability detection. 
We benchmarked \toolname against multiple state-of-the-art (SOTA) approaches, covering agent-based methods (e.g., \textit{VulTrial}~\cite{VulTrial}), retrieval-augmented systems (e.g., \textit{VulInstruct}~\cite{VulInstruct}), and fine-tuned models (e.g., \textit{ReVD}~\cite{revd}).
Our evaluation goes beyond traditional performance metrics (Precision, Recall, F1, Accuracy) to include the stringent \textbf{Pair-wise Correct Prediction} and \textbf{False Positive Rate (FPR)}, thereby assessing not just the detection capability but also discriminative precision and practical reliability of our system.
Experimental results demonstrate that \toolname establishes a new state-of-the-art: 
on 435 test pairs, it achieves \textbf{122 Pair-wise Correct Predictions}—the first approach to surpass 100—compared to 96 for the strongest baseline. Moreover, \toolname reduces FPR by 21.96\%, 54.40\%, and 37.01\% compared to \textit{VulTrial} (trained), \textit{VulInstruct}, and \textit{ReVD} respectively, demonstrating superior capability in distinguishing true vulnerabilities from false alarms—all at an average cost of \$0.09 per sample without any task-specific training.

In summary, this paper makes the following contributions:
\begin{itemize}[wide=0pt, nosep]
    \item \textbf{New Framework Paradigm:} 
    We propose \toolname, a novel multi-agent framework that shifts vulnerability detection from ungrounded deliberation to grounded forensic verification through a \textit{``From Clue to Verdict''} workflow, which explicitly decouples vulnerability localization from reasoning verification over a closed factual substrate, effectively mitigating the contextual hallucinations prevalent in prior approaches.
    
    \item \textbf{Dynamic Graph-Guided Context Augmentation:} 
    We design the first clue-anchored, demand-driven context augmentation mechanism for LLM-based vulnerability detection, where agents dynamically reconstruct cross-function and cross-file dependency chains over Code Property Graphs (CPG) guided by each identified clue, rather than relying on statically predefined context boundaries or learned query generation.
    
    \item \textbf{State-of-the-Art Performance:} 
    \toolname establishes a new SOTA on the PrimeVul dataset, achieving \textbf{122 Pair-wise Correct Predictions}, which is the first approach to surpass 100 on this benchmark's 435 test pairs. 
    It further reduces the False Positive Rate by up to \textbf{54.40\%} compared to leading baselines, without any task-specific training and at an average cost of \$0.09 per sample. 
    We further provide a fine-grained cost breakdown across pipeline stages, offering the first systematic analysis of per-agent computational overhead in LLM-based vulnerability detection.

    \item \textbf{Data Availability.}
    To support reproducibility, the comprehensive system prompts and dialectical templates used for the Clue-Discovery, Context-Augmentation, Verification, and Audit agents are provided in Appendix~\ref{appendix:prompts}. The source code of the AEGIS framework, including the Joern-based CPG extraction scripts, is available in an Github repository at: \url{https://github.com/secureai4code/Aegis}. Furthermore, the detailed experimental results (including ablation study) on the PrimeVul dataset have been provided in a separate Google Drive at: \url{https://drive.google.com/drive/folders/13AIff2GXRu8dv9QT28RoCGa6xv4JuDMk?usp=share_link}.

\end{itemize}

\section{\toolname}

The central design principle of \toolname is the explicit separation of \textit{what is suspicious} from \textit{whether it is exploitable}. 
A local code anomaly, such as an unchecked \texttt{memcpy}, is merely a \textit{clue}; only by tracing its data provenance through the repository's execution flow can one reach a \textit{verdict}. 
\toolname operationalizes this ``From Clue to Verdict'' philosophy through four specialized agents organized in a pipeline. 
The first two agents collaborate to construct a \textit{closed factual substrate}, providing the grounded evidence base that prior approaches lack, while the latter two agents progressively verify the soundness of reasoning within this evidence boundary.

\begin{figure*}[t]
  \centering
  \includegraphics[width=\linewidth]{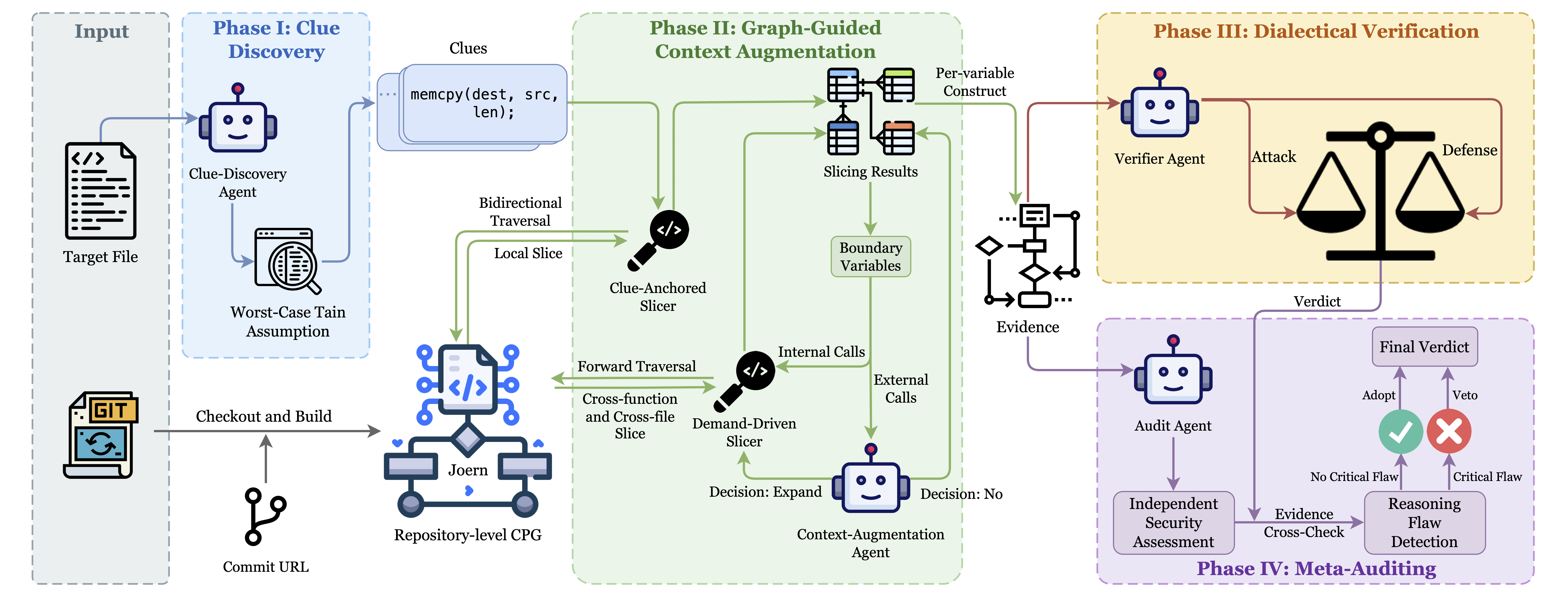}
  \caption{Overview of \toolname.}
  \label{fig:overview}
  \vspace{-5mm}
\end{figure*}

\begin{figure*}[t]
  \centering
  \includegraphics[width=\linewidth]{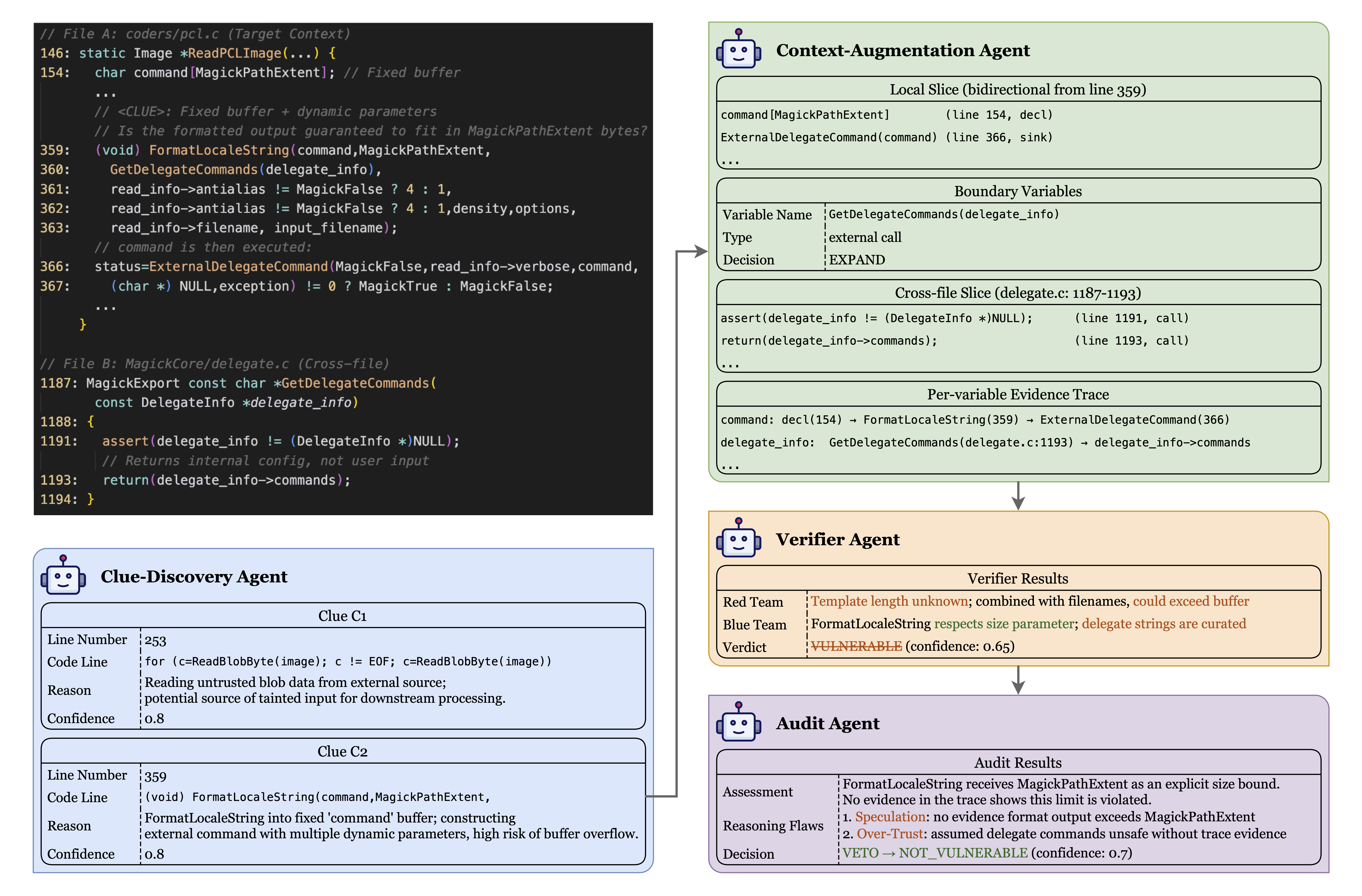}
  \caption{A running example of \toolname. \textcolor{egGreen}{Green} text indicates reasoning grounded in trace evidence; \textcolor{egOrange}{orange} text indicates claims that exceed the evidence boundary, identified and vetoed by the Audit Agent. An additional end-to-end pipeline execution example is provided in Appendix~\ref{app:case_study}.}
  \label{fig:running_example}
  \vspace{-5mm}
\end{figure*}

The overall workflow is illustrated in Figure~\ref{fig:overview}. 
Taking a target function $f_{target}$ and its encompassing repository $\mathcal{R}$ as input, \toolname produces a binary verdict $V \in \{Safe, Vulnerable\}$ alongside a structured evidence trace $\mathcal{T}$ that records every reasoning step and its supporting code fragment. 
We walk through the pipeline using the running example in Figure~\ref{fig:running_example}, \textcolor{tmpFig2}{a suspected buffer overflow involving a fixed-size buffer and external formatting functions, whose actual safety can only be confirmed by reconstructing cross-file context.}
Concretely, the pipeline operates as follows:

\begin{itemize}[wide=0pt, nosep]
    \item \textbf{Phase~I: Clue Discovery (\S\ref{subsec:discovery}).} 
    The \textit{Clue-Discovery Agent} scans $f_{target}$ and reports suspicious code locations as clues.

    \item \textbf{Phase~II: Graph-Guided Context Augmentation 
    (\S\ref{subsec:augmentation}).} 
    For each clue, the \textit{Context-Augmentation Agent} queries a repository-level Code Property Graph to dynamically reconstruct the relevant cross-function dependency chain.

    \item \textbf{Phase~III: Dialectical Verification 
    (\S\ref{subsec:verification}).} 
    The \textit{Verifier Agent} constructs dialectical arguments for and against exploitability, both grounded in the retrieved trace. It then makes a preliminary verdict from these two arguments.

    \item \textbf{Phase~IV: Meta-Auditing (\S\ref{subsec:auditing}).} 
    The \textit{Audit Agent} independently reviews each rationale output by the Verifier, detecting unsupported claims and exercising veto power when necessary to produce the final verdict for the user.
\end{itemize}

\subsection{Phase I: Clue Discovery}
\label{subsec:discovery}

Phase~I forms the foundation of the entire pipeline: if the initial clues miss the data-flow paths relevant to a true vulnerability, no amount of context augmentation or sophisticated reasoning in subsequent phases can recover it.
The \textit{Clue-Discovery Agent} is therefore deliberately designed for high recall, treating every plausible anomaly as worth investigating.

To achieve this, the agent analyzes $f_{target}$ in isolation under a \textbf{Worst-Case Taint Assumption}. 
Since the agent has no visibility into how $f_{target}$ is invoked, it treats every non-local data source, such as function parameters, global variables, and return values from external calls, as originating from an untrusted source. 
Under this assumption, the agent traces how each tainted source propagates through the function's data flow, flagging every path where tainted data reaches a security-sensitive sink (e.g., memory operations) without passing through visible sanitization logic. 
This aggressive assumption intentionally over-approximates the true attack surface; the resulting false positives are deferred to Phase~II--IV for resolution.
Each flagged location is emitted as a \textit{clue} tuple $c_i = \langle L_i, S_i, R_i, C_i \rangle$, where $L_i$ is the line number, $S_i$ the code statement, $R_i$ a natural-language explanation of the suspicion, and $C_i \in [0.1, 1.0]$ a confidence score reflecting the strength of the local evidence.

\textcolor{tmpFig2}{In Figure \ref{fig:running_example}, the agent flags two suspicious operations within the target function. Clue C1 (C=0.8) points to a \texttt{ReadBlobByte} loop at line 253, flagged as a potential taint origin since it reads untrusted data from an external source. 
Clue C2 (C=0.8) flags the \texttt{FormatLocaleString} operation at line 359, as writing multiple dynamic parameters into a fixed \texttt{command} buffer carries a high risk of buffer overflow. 
At this stage, the agent cannot determine whether these localized anomalies are safely constrained by the broader program context—such as downstream sanitization for the untrusted input in Clue C1, or whether the dynamic parameters in Clue C2 are safely bounded by the callee's size constraints and the provenance of its inputs.
This local ambiguity is precisely what triggers Phase~II: the clues are identified, but the evidence base required to assess their exploitability remains unconstructed.}

\subsection{Phase II: Graph-Guided Context Augmentation}
\label{subsec:augmentation}

Phase~I identifies suspicious anomalies within a single function, but determining their exploitability often requires context that spans function and file boundaries. 
An unchecked \texttt{memcpy} is benign if its length parameter was validated by a caller; conversely, a seemingly safe helper may become dangerous when its return value flows unchecked into a security-sensitive sink. 
The role of Phase~II is to reconstruct precisely this cross-boundary context for each clue. To this end, the \textit{Context-Augmentation Agent} extracts context directly from the repository by traversing its Code Property Graph (CPG), reconstructing only the dependency chain relevant to each clue.
The resulting evidence trace forms the \textit{closed factual substrate} that constrains all downstream reasoning to verifiable facts.

\subsubsection{Clue-Anchored Local Slicing.}
Upon receiving a clue $c_i = \langle L_i, S_i, R_i, C_i \rangle$, the agent begins with a strictly intra-procedural analysis within the target function's CPG. 
Starting from the variables involved in the suspicious statement $S_i$, it performs \textit{bidirectional} Program Dependence Graph (PDG) slicing: a backward traversal along data-dependency edges traces each variable to its origin (e.g., a function parameter), while a forward traversal tracks how each variable propagates toward downstream sinks. 
Control-dependency edges are followed in parallel to capture the conditional guards along each path. 
To ensure the slice covers complete expressions that span multiple lines, such as chained method calls or multi-line macro expansions, the agent additionally expands the slice via AST parent traversal until it reaches a statement-level boundary. 
We refer to the resulting subgraph as the \textit{local slice}, which serves two purposes: 
it provides the immediate data-flow context surrounding the clue, and it identifies \textit{boundary variables}---variables whose origins or destinations lie outside the current function scope (e.g., function parameters, return values of external calls). 
These boundary variables are the triggers for cross-function expansion.

\subsubsection{Iterative, Demand-Driven Expansion.}
Rather than blindly expanding all reachable callers or callees, which leads to path explosion in real-world repositories, \toolname employs a selective expansion strategy with two key mechanisms.

First, the agent distinguishes \textit{internal calls} (functions defined within the same file) from \textit{external calls} (functions defined in other files). 
Internal calls are expanded directly by forward-slicing within the current CPG, requiring no additional graph construction. 
External calls, by contrast, trigger an \textit{LLM-guided decision}: 
the agent presents the accumulated evidence trace and the candidate function to the LLM, which judges whether inspecting that function is necessary to confirm or refute the vulnerability hypothesis. 
This two-tier strategy avoids unnecessary graph construction for locally resolvable calls while ensuring that security-critical cross-file dependencies are not missed. 
Second, when an external expansion is approved, the agent performs \textit{on-demand graph stitching}. 
It parses the external file, constructs a temporary CPG for the callee (or caller), and synthesizes virtual edges that connect the two previously isolated graphs:

\begin{itemize}[wide=0pt, nosep]
    \item \textbf{Argument$\to$Parameter edges:} 
    For each argument at the call site, a virtual \textsc{Reaching\_Def} edge is created to the corresponding parameter in the callee, enabling forward taint propagation into the called function.
    \item \textbf{Return$\to$Call-site edges:} 
    For each \texttt{return} statement in the callee, a virtual \textsc{Reaching\_Def} edge is created back to the call-site node, capturing how the callee's return value flows into the caller's subsequent computation.
\end{itemize}

With the graphs stitched, slicing resumes in the newly expanded context. Crucially, this expansion is \textit{iterative}: 
the newly sliced context may itself reveal additional boundary variables, which are enqueued for further expansion. 
The process continues until either all boundary variables are resolved or a configurable expansion budget is exhausted.

Note that the slicing direction adapts to the expansion role: the target function is analyzed with bidirectional slicing (to trace both origins and propagation), while expanded external functions are analyzed with forward slicing only (to track how the passed arguments are used or transformed within the callee). 
This asymmetry avoids retrieving irrelevant internal logic of external functions that does not pertain to the clue under investigation.

\subsubsection{Evidence Trace Construction.}
The accumulated slicing results across all expanded files are aggregated into a structured \textit{Evidence Trace} ($\mathcal{T}$). 
Rather than concatenating raw code, the trace is organized per-variable: for each tracked variable, $\mathcal{T}$ records its backward chain (from the suspicious sink to its ultimate source) and its forward chain (from the source through transformations to downstream uses), preserving the file boundaries crossed at each step. 
This per-variable organization enables the downstream Verifier Agent (Phase~III) to reason over precise, self-contained data-flow narratives rather than navigating a monolithic code dump.

\textcolor{tmpFig2}{In the running example (Figure~\ref{fig:running_example}), the agent begins at the 
\\\texttt{FormatLocaleString} call in \texttt{coders/pcl.c} (line 359). Bidirectional slicing traces the \texttt{command} buffer backward to its declaration (line 154) and forward to its sink at \texttt{ExternalDelegateCommand} (line 366). During this local slicing, the agent identifies the external call \texttt{GetDelegateCommands} as a boundary variable. The LLM confirms that expanding this call is necessary. After stitching the graphs, cross-file slicing into \texttt{delegate.c} (lines 1187-1193) reveals how the \texttt{delegate\_info->commands} are accessed, completing the data-flow narrative required to assess the clue's exploitability.}

\subsection{Phase III: Dialectical Verification}
\label{subsec:verification}

Phase~II delivers a structured evidence trace grounded in cross-file dependencies.  However, possessing the right evidence does not guarantee sound reasoning: 
LLMs frequently exhibit \textit{confirmation bias}~\cite{echterhoff2024cognitive, jiang2024peek, guo2024bias}, anchoring on the initial suspicion from Phase~I while overlooking mitigations elsewhere in the trace.

To counteract this bias, \toolname employs a \textbf{Dialectical Verification} protocol in which a single Verifier Agent is structurally forced to argue both sides of a case before reaching a conclusion.  
We deliberately choose single-agent adversarial reasoning over multi-agent debate~\cite{VulTrial}: 
when debate agents lack repository-level evidence, successive rounds degenerate into mutually reinforcing concessions rather than introducing new facts~\cite{VulTrial}.  
Our design avoids this failure mode because the evidence trace from Phase~II provides a \emph{closed factual substrate}: 
if a security check does not appear in the trace, the agent must assume it does not exist on the analyzed path. 
This eliminates the information vacuum that destabilizes unconstrained debate and directly prevents the \textit{contextual hallucinations} identified in Section~\ref{sec:intro}.

The Verifier executes a four-step protocol over each clue--trace pair.
\textbf{(1)~Factual Comprehension.} Before any adversarial reasoning, the agent establishes a neutral foundation: the operation at the suspicious line, the provenance of each variable (trusted or untrusted), and the \emph{exact} mitigation that would make the code safe. 
This shared factual anchor prevents Red and Blue arguments from drifting onto different interpretations of the same code.
\textbf{(2)~Dialectical Attack (Red).}  
The agent constructs an \textit{Exploitability Chain} from the attacker's perspective, citing a concrete propagation path from untrusted source to vulnerable sink. Every claim must cite specific line numbers within the trace; appeals to code outside the trace are prohibited.
\textbf{(3)~Dialectical Defense (Blue).}  
The agent is then forced to switch perspectives and argue for safety by identifying mitigating factors (bounds checks, type constraints, sanitization) visible in the trace. The same evidentiary rule applies: every claim must be grounded in specific line numbers from the trace.
\textbf{(4)~Evidence-Weighted Adjudication.}  
The agent drops both personas and compares the competing arguments under a single principle: \emph{concrete trace evidence outweighs speculation}. The resulting verdict (\textsc{Vulnerable} or \textsc{Safe}) is accompanied by a calibrated confidence score and a citation of the
most decisive evidence, providing an auditable trail for Phase~IV.

\textcolor{tmpFig2}{Returning to the running example in Figure \ref{fig:running_example}, the Verifier Agent synthesizes the trace into competing arguments. The Red Team hypothesizes an exploitability chain, arguing that the unknown template length combined with filenames could exceed the buffer. Conversely, the Blue Team points to mitigating factors, noting that \texttt{FormatLocaleString} respects its size parameter and delegate strings are curated. Ultimately, the Verifier succumbs to confirmation bias, prioritizing the Red Team's argument to produce a preliminary verdict of VULNERABLE with a confidence of 0.65.}

\subsection{Phase IV: Meta-Auditing}
\label{subsec:auditing}

While the Verifier Agent enforces structural balance, it remains a single agent's judgment, still susceptible to reasoning drifts such as hallucinated mitigations or over-trust in library semantics.
Phase~IV introduces a Meta-Auditing mechanism to catch precisely these failure modes.
The Audit Agent provides an independent second review of the Verifier's reasoning.  Rather than merely voting on the previous outcome, it performs a structured \textit{Reasoning Quality Audit} that scrutinizes the logical soundness of every claim before rendering the final verdict~$V_{\mathit{final}}$.

The agent executes a four-step protocol.
\textbf{(1)~Independent Comprehension.}  
The agent analyzes the raw code and evidence trace~$\mathcal{T}$ to form its own security assessment, deliberately ignoring the Verifier's conclusion so as to avoid cascading bias.
\textbf{(2)~Evidence Cross-Check.}  
It then parses every citation in the Verifier's argument and validates that the referenced line (a)~exists in~$\mathcal{T}$ and (b)~semantically supports the claim.
For instance, if the Verifier asserts ``\textit{variable $x$ is sanitized at line~42},'' the Audit Agent verifies both the presence and the semantics of line~42.
\textbf{(3)~Reasoning Flaw Detection.}  
The agent scans both the Red and Blue Team arguments for categories of reasoning failures that are known to afflict LLM-based analysis. These categories operationalize the \textit{contextual hallucination} problem identified in Section~\ref{sec:intro} into actionable audit checks. Specifically, the agent flags \textbf{Phantom Mitigation} (citing a security check not on the analyzed execution path), \textbf{Speculation} (grounding a verdict on assumed behavior of unseen code), \textbf{Anchoring} (echoing the Phase~I suspicion without independent verification against the full trace), and \textbf{Over-Trust} (treating external libraries or APIs as inherently safe without trace-level evidence).
The Audit Agent renders one of three judgments: \emph{Agree}, \emph{Disagree}, or \emph{Defer}. 
To overturn the Verifier's verdict, the agent must identify at least one specific, material reasoning flaw from the taxonomy in Step~3 that directly undermines the conclusion, which means vague concerns or stylistic critiques are insufficient grounds for overturning. 
If such a flaw is identified, the Audit Agent substitutes its own independent judgment ($V_{\textit{final}} \leftarrow V_{\textit{audit}}$). 
If the agent has concerns but cannot pinpoint a concrete flaw that materially changes the outcome, it defers, preserving the Verifier's original verdict ($V_{\textit{final}} \leftarrow V_{\textit{verify}}$). 
This symmetric standard applies regardless of the direction of the overturn, ensuring that the audit mechanism is equally rigorous in challenging both false positives and false negatives.

\textcolor{tmpFig2}{Figure \ref{fig:running_example} demonstrates this Meta-Auditing process in action. The Audit Agent scrutinizes the Verifier's VULNERABLE verdict and identifies two critical reasoning flaws. First, it detects \textbf{Speculation}: the Verifier assumed the output could exceed the buffer, but the trace provides no evidence that the explicit \texttt{MagickPathExtent} size bound passed to \texttt{FormatLocaleString} is ever violated. Second, it flags \textbf{Over-Trust}: the Verifier assumed the delegate commands were unsafe without concrete trace evidence. Because these flaws materially undermine the Verifier's conclusion, the Audit Agent exercises its veto power, overturning the preliminary assessment to render a final verdict of NOT VULNERABLE.}

\section{Experimental Setup}
\label{sec:experimental_setup}

In this section, we detail the research questions, datasets, baselines, and implementation used to empirically evaluate \toolname.

\subsection{Research Questions}
To comprehensively assess the effectiveness, efficiency, and robustness of \toolname, we investigate the following three research questions:

\begin{itemize}[leftmargin=*]
    \item \textbf{RQ1 (Effectiveness):} 
        How does \toolname compare to the state-of-the-art approaches in vulnerability detection performance?
        
    \item \textbf{RQ2 (Ablation Study):}
        How does each stage of the \toolname pipeline contribute to overall detection capability? Specifically, (a)~how effective is the evidence construction pipeline (Phase~I--II) at localizing vulnerable code, (b)~how does the number of investigated clues affect the accuracy--cost trade-off, and (c)~how do the dialectical verification structure and meta-auditing mechanism each affect reasoning quality and detection performance?

    \item \textbf{RQ3 (Reasoning Quality):} To what extent does the
        Meta-Auditing mechanism improve reasoning soundness, and what categories of reasoning flaws does it detect and correct?
\end{itemize}

\subsection{Datasets}

We evaluate on \textbf{PrimeVul}~\cite{llm4vd}, the most rigorous benchmark for function-level vulnerability detection~\cite{VulTrial, VulInstruct, revd, llmxcpg}, comprising 6,968 vulnerable and 228,800 benign functions across 140~CWE types with chronological splits that prevent data leakage.  
Its test set contains 435 vulnerable--patched pairs for stringent pair-wise evaluation~\cite{VulTrial, VulInstruct}, which share at least 80\% of the string to ensure a highly challenging setting that requires models to distinguish subtle vulnerability semantics rather than relying on superficial textual differences.  
Each pair includes the corresponding commit URL, allowing us to clone the repository at the exact commit and construct the Code Property Graph required by \toolname.

\subsection{Baseline}
We compare \toolname against three categories of baselines, covering the major paradigms for LLM-based vulnerability detection.
\begin{itemize}[leftmargin=*]

    \item \textbf{Fine-tuned Models:} 
    We include three LLMs evaluated under the PrimeVul benchmark~\cite{llm4vd}: \textbf{CodeBERT}~\cite{codebert}, \textbf{CodeT5}~\cite{codet5}, and \textbf{UniXCoder}~\cite{unixcoder}, each fine-tuned on the PrimeVul training set for binary vulnerability classification.  
    We additionally include \textbf{ReVD}~\cite{revd}, a recent fine-tuning approach that employs curriculum preference optimization with synthesized reasoning data.

    \item \textbf{Retrieval-Augmented Approaches:} 
    \textbf{VulInstruct}~\cite{VulInstruct} retrieves reusable security specifications from historical patches and CVEs to guide LLM-based detection.  
    \textbf{LLMxCPG}~\cite{llmxcpg} uses a fine-tuned LLM to generate CPG queries that extract vulnerability-relevant code slices, which a second fine-tuned LLM then classifies. 

    \item \textbf{Agent-based Approaches:} 
    \textbf{VulTrial}~\cite{VulTrial} and \textbf{GPTLens}~\cite{GPTLens} both simulate a mock-court debate among multiple LLM agents. We also include a \textbf{Chain-of-Thought (CoT)} prompting baseline~\cite{llm4vd} to isolate the effect of structured reasoning from multi-agent orchestration.
    
\end{itemize}

\subsection{Implementation Details}

We implement \toolname in Python, using \texttt{Joern}~\cite{joern-cpg} for Code Property Graph construction and parsing.  
All four agents use \textbf{DeepSeek-V3.1}~\cite{deepseekv31} as the backbone LLM, accessed via the TensorBlock API~\footnote{\url{https://www.tensorblock.co/}}. The first three agents (Clue Discovery, Context Augmentation, and Verification) use the default sampling temperature to preserve reasoning diversity, while the Audit Agent uses a temperature of $0$ to produce maximally deterministic judgments, reflecting its role as the final adjudicator whose verdict must be as decisive and consistent as possible.
Two pipeline-level hyperparameters govern the trade-off between detection thoroughness and computational cost. 
First, the Clue-Discovery Agent (Phase~I) may identify multiple suspicious anomalies per function, ranked by confidence score; we forward the top-$k$ highest-confidence clues to subsequent phases. The choice of $k$ directly affects both detection performance and computational overhead; we provide a systematic sensitivity analysis in Section~\ref{sec:rq2}. 
Second, the Context-Augmentation Agent (Phase~II) operates under two resource limits: a per-variable slicing depth limit of 10 (i.e., traversal terminates after following 10 consecutive dependency edges) and a global cap of 50 cross-function expansions per sample, preventing unbounded traversal along deep call chains and path explosion in large repositories.
We select DeepSeek-V3.1 as it offers the lowest effective cost for our workload profile: at \$0.56 per million input tokens and \$1.68 per million output tokens, it is the most economical option on TensorBlock given that our pipeline is dominated by input token consumption (long evidence traces fed to each agent).  
All experiments are conducted on a Linux server with an AMD EPYC CPU and 256\,GB RAM; since we rely entirely on API-based inference, no local GPU is required.

\subsection{Evaluation Metrics}
\label{sec:metrics}

\textit{Standard Detection Metrics.}
We report \textbf{Precision}, \textbf{Recall}, \textbf{F1-score}, and \textbf{Accuracy} following prior work~\cite{llm4vd, VulTrial, VulInstruct, llmxcpg, revd}.  
We additionally report the \textbf{False Positive Rate} (FPR) to expose a blind spot in standard metrics:
on PrimeVul's balanced test set (435 vulnerable, 435 patched), a trivial strategy that labels every function as vulnerable could achieve Recall\,=\,1.0, Precision\,=\,0.5, and
F1\,=\,0.67, yet its FPR is 100\%, rendering it entirely useless in practice.
FPR directly quantifies this failure mode by measuring the proportion of benign functions incorrectly flagged as vulnerable.

\noindent\textit{Pair-wise Discriminative Metrics.}
Standard metrics cannot reveal whether a model genuinely understands vulnerability semantics or merely exploits superficial patterns.  
To assess this, we adopt pair-wise evaluation~\cite{llm4vd,
revd}.  
Let $\mathcal{D}_{\mathit{pair}} = \{(x_v, x_p)_i\}_{i=1}^{N}$ be the set of test pairs, where $x_v$ is a vulnerable function and $x_p$ its patched version, and let $M(x) \in \{0,1\}$ denote the model's prediction ($1$ = vulnerable).  
We report:
\begin{itemize}[wide=0pt, nosep]
  \item \textbf{P-C} (Pair-wise Correct Prediction): the fraction of
    pairs where the model correctly identifies both the vulnerability
    and its fix, defined as
    $\text{P-C} = \frac{1}{N}\sum_{i=1}^{N}
        \mathbb{I}\bigl(M(x_v^{(i)})=1 \;\land\; M(x_p^{(i)})=0\bigr)$.
  \item \textbf{VP-S} (Vulnerability Prediction Score)~\cite{revd}:
    P-C penalized by the rate of reversed predictions (P-R), where the
    model flags the patch as vulnerable while missing the original flaw,
    defined as $\text{VP-S} = \text{P-C} - \text{P-R}$.
\end{itemize}

\subsection{Methodology}
\label{sec:methodology}

\paragraph{RQ1 (Effectiveness).}
We run \toolname and all baselines on the full 435-pair PrimeVul
test set and compare across all metrics defined above.  
For baselines that report results on PrimeVul in their original papers, we directly cite the published numbers; 
for those that do not, we reproduce results using their released code under identical settings.  
We additionally report the average token consumption and wall-clock time per sample of \toolname to characterize computational overhead.

\paragraph{RQ2 (Ablation).}
We evaluate the contribution of each pipeline stage along three complementary dimensions, reflecting the two-stage design of \toolname: \emph{evidence construction} (Phase~I--II) and \emph{reasoning verification} (Phase~III--IV).

\smallskip\noindent\emph{(a) Clue Localization Quality.}
Since each PrimeVul pair includes the fixing commit, we extract the modified lines from the commit diff as ground-truth vulnerable locations.
We measure Phase~I's precision and recall over these locations to validate its high-recall design philosophy.
We then assess whether Phase~II's graph-guided context augmentation improves localization accuracy, quantifying the marginal contribution of CPG-based expansion.

\smallskip\noindent\emph{(b) Clue Sensitivity Analysis.}
Phase~I ranks each identified clue by a confidence score $C_i \in [0.1, 1.0]$.
We vary the number of top-$k$ clues (ranked by confidence) that are forwarded to subsequent phases and measure the resulting detection performance (P-C, F1) alongside computational cost (token consumption, wall-clock time).
This analysis reveals the accuracy--cost trade-off in practice: how many clues must be investigated to achieve near-optimal detection, and at what marginal cost.

\smallskip\noindent\emph{(c) Reasoning Component Ablation.}
We evaluate two variants that isolate the contribution of each reasoning mechanism while keeping the evidence construction pipeline (Phase~I--II) intact:
\begin{itemize}[wide=0pt, nosep]
  \item \textbf{\textit{w/o} Dialectical Structure}:
    the Verifier classifies directly over the evidence trace without the structured Red/Blue adversarial protocol, isolating the contribution of \emph{dialectical reasoning};
  \item \textbf{\textit{w/o} Meta-Auditing}:
    the Verifier's verdict is taken as final without Phase~IV review, isolating the contribution of \emph{independent reasoning audit}.
\end{itemize}
All variants use DeepSeek-V3.1 and are evaluated on the identical 435-pair test set.

\paragraph{RQ3 (Reasoning Quality).}
We analyze the Meta-Auditing mechanism along two dimensions.
\emph{Quantitatively}, 
we measure the Audit Agent's veto rate (how often it overrides the Verifier) and the correctness improvement attributable to vetoes.  
We further categorize detected reasoning flaws by type (Phantom Mitigation, Speculation, Anchoring, Over-Trust) and report their frequency distribution.
\emph{Qualitatively}, 
we present representative case studies where the Audit Agent corrected the Verifier's errors, illustrating the concrete reasoning patterns that each flaw category captures.

\section{Results}
\begin{table*}[t]
\centering
\caption{Vulnerability detection performance on PrimeVul.}
\label{tab:main_results}
\resizebox{\textwidth}{!}{
\begin{tabular}{llc | ccccc | cc}
\toprule
& & & \multicolumn{5}{c|}{\textbf{Standard (\%)}} & \multicolumn{2}{c}{\textbf{Pairwise}} \\
\textbf{Method} & \textbf{Model} & \textbf{Trained} & Acc.$\uparrow$ & Prec.$\uparrow$ & Rec.$\uparrow$ & F1-Score$\uparrow$ & FPR$\downarrow$ & P-C$\uparrow$ & VP-S$\uparrow$ \\
\midrule
\multicolumn{10}{l}{\textit{Fine-tuned Models}} \\
CodeBERT & - & \cmark & 49.77 & 48.84 & 9.66 & 16.12 & 10.11 & 5 & -2 \\
CodeT5 & - & \cmark & 49.43 & 47.90 & 13.10 & 20.58 & 14.25 & 0 & -5 \\
UniXcoder & - & \cmark & 50.23 & 51.72 & 6.90 & 12.17 & 6.44 & 6 & 0 \\
ReVD & Qwen2.5-Coder-7B-Instruct & \cmark & 57.82 & 55.90 & 74.02 & 63.70 & 58.39 & 78 & 68 \\
\midrule
\multicolumn{10}{l}{\textit{Retrieval-Augmented Approaches}} \\
LLMxCPG$^\dagger$ & Qwen2.5-Coder-32B-Instruct & \cmark & 51.15 & 51.74 & 34.25 & 41.22 & 31.95 & 28 & 10 \\
VulInstruct & DeepSeek-V3 & \xmark & 52.51 & 51.51 & 85.68 & 64.34 & 80.67 & 50 & 21 \\
\midrule
\multicolumn{10}{l}{\textit{Agent-based Approaches}} \\
Ding et al.(CoT) & GPT-3.5 & \xmark & 49.77 & 48.57 & 7.82 & 13.47 & 8.28 & 18 & -2 \\
Ding et al.(CoT) & GPT-4o & \xmark & 51.26 & 53.55 & 19.08 & 28.14 & 16.55 & 40 & 11 \\
GPTLens & GPT-3.5 & \xmark & 49.54 & 49.76 & 93.79 & 65.02 & 94.71 & 20 & -4 \\
GPTLens & GPT-4o & \xmark & 51.84 & 51.44 & 65.52 & 57.63 & 61.84 & 44 & 16 \\
VulTrial & GPT-3.5 & \xmark & 50.69 & 51.43 & 24.83 & 33.49 & 23.45 & 68 & 6 \\
VulTrial & GPT-4o & \xmark & 53.65 & 53.17 & 59.77 & 56.28 & 52.64 & 81 & 31 \\
VulTrial & GPT-4o & \cmark & 55.17 & 54.95 & 57.47 & 56.18 & 47.13 & 96 & 45 \\
\midrule
\textbf{AEGIS (k=2)} & \textbf{DeepSeek-V3.1} & \xmark & 56.78 & 57.78 & 50.34 & 53.82 & 36.78 & 122 & 59 \\
\bottomrule
\multicolumn{10}{l}{\small $^\dagger$Reproduced using released code; all other baseline results are cited from their original papers.} \\
\end{tabular}
}
\end{table*}

\subsection{RQ1: Effectiveness}
\label{sec:rq1}

Table~\ref{tab:main_results} summarizes the detection performance of \toolname and all baselines on the PrimeVul test set.
We organize baselines into three categories: fine-tuned models, retrieval-augmented approaches, and agent-based approaches, and report two groups of metrics.
\emph{Standard detection metrics} include Accuracy, Precision, Recall, F1-Score, and False Positive Rate (FPR), which evaluate binary classification quality over the 870 individual functions (435 vulnerable + 435 patched).
\emph{Pair-wise discriminative metrics} include Pair-wise Correct Prediction (P-C) and Vulnerability Prediction Score (VP-S), which operate at the \emph{pair} level: P-C counts the number of pairs where the model correctly identifies both the vulnerable function and its patched counterpart, while VP-S further penalizes reversed predictions where the model flags the patch as vulnerable but misses the original flaw (see Section~\ref{sec:metrics} for formal definitions).
The ``Trained'' column indicates whether the method requires task-specific fine-tuning on the PrimeVul training set; \toolname operates in a fully \emph{training-free} setting.

\paragraph{Overall Performance.}
As discussed in Section~\ref{sec:metrics}, we ground our analysis primarily in P-C and FPR, treating standard metrics as supplementary reference points.
As shown in Table~\ref{tab:main_results}, \toolname achieves \textbf{122 Pair-wise Correct Predictions} out of 435 test pairs, establishing a new state-of-the-art on the PrimeVul benchmark.
This represents a 27.1\% improvement over the strongest baseline, VulTrial\textsubscript{trained} (P-C\,=\,96), and a 50.6\% improvement over VulTrial\textsubscript{untrained} (P-C\,=\,81).
To our knowledge, \toolname is the first approach to surpass the 100-pair threshold on this benchmark, demonstrating that its forensic pipeline can reliably distinguish vulnerable functions from their semantically near-identical patches.
Furthermore, \toolname achieves the highest VP-S (59) among all \emph{training-free} methods, indicating strong discriminative precision with minimal reversed predictions.

\paragraph{Comparison with Fine-tuned Models.}
Fine-tuned encoder models (CodeBERT, CodeT5, UniXcoder) exhibit near-zero pair-wise discriminative ability (P-C $\leq$ 6), confirming that static representation learning fails to capture the subtle semantic shifts between vulnerabilities and their patches in PrimeVul's challenging setting.
ReVD, which employs curriculum preference optimization on Qwen2.5-Coder-7B-Instruct~\cite{qwencoder}, achieves a substantially higher P-C of 78 and the highest VP-S (68) among all methods.
However, this comes at the cost of a 58.39\% false positive rate, which means that nearly \emph{six out of ten} benign functions are incorrectly flagged as vulnerable.
ReVD's high VP-S is partially attributable to its aggressive prediction tendency: with Recall\,=\,74.02\% and FPR\,=\,58.39\%, the model rarely predicts a function as safe, which mechanically suppresses reversed predictions (P-R) and inflates VP-S.
In contrast, \toolname surpasses ReVD's P-C by 44 pairs (a 56.4\% relative improvement) while reducing FPR by 21.61 percentage points (from 58.39\% to 36.78\%), demonstrating that evidence-grounded reasoning achieves stronger discriminative capability without relying on biased prediction distributions.

\paragraph{Comparison with Retrieval-Augmented Approaches.}
LLMxCPG shares \toolname's intuition of leveraging Code Property Graphs for context-aware detection, yet achieves only P-C\,=\,28 with FPR\,=\,31.95\%.
This gap stems from a fundamental design difference in \emph{how} the CPG is queried.
LLMxCPG relies on a fine-tuned LLM to generate CPG queries for extracting vulnerability-relevant code slices, which are then classified by a second fine-tuned LLM.
This query-generation approach is inherently limited in generalizability: PrimeVul spans 140 CWE types, each exhibiting distinct vulnerability patterns and propagation semantics, making it impractical to train a single query-generation model that produces effective CPG queries across all vulnerability categories.
In contrast, \toolname sidesteps this bottleneck entirely through \emph{clue-driven} slicing: rather than learning to generate type-specific queries, the Context-Augmentation Agent uses the suspicious code locations identified in Phase~I as natural anchors for on-demand graph traversal, dynamically adapting its slicing strategy to the specific data-flow topology of each clue regardless of CWE type.
This design renders \toolname fundamentally CWE-agnostic in its context augmentation, contributing to its substantially stronger pair-wise performance.
VulInstruct achieves high recall (85.68\%) by augmenting LLMs with retrieved security specifications, but its FPR of 80.67\% renders it impractical, as roughly four out of five safe functions are misclassified.
Compared to VulInstruct, \toolname reduces FPR by 43.89 percentage points (from 80.67\% to 36.78\%) while achieving more than double the pair-wise correct predictions (122 vs.\ 50), confirming that repository-specific structural context is far more effective than generic security knowledge for grounding vulnerability reasoning.

\paragraph{Comparison with Agent-based Approaches.}
Agent-based methods represent the most directly comparable category, as they also employ LLM-driven multi-step reasoning without task-specific fine-tuning.
We first examine the simplest prompting strategy: Ding et al.'s Chain-of-Thought (CoT) baseline, proposed by the PrimeVul authors themselves to assess whether basic structured prompting suffices for vulnerability detection.
The results are sobering: even with GPT-4o, CoT achieves only P-C\,=\,40 with FPR\,=\,16.55\%, and with GPT-3.5, P-C drops to 18 with near-trivial Recall (7.82\%).
This confirms that LLMs equipped with simple prompting strategies, even powerful ones like GPT-4o, fundamentally lack the contextual grounding necessary for reliable vulnerability reasoning, motivating the need for more sophisticated agent architectures.

GPTLens attempts to address this through multi-agent debate, but exemplifies the fragility of unconstrained deliberation without factual grounding: on GPT-3.5, it achieves Recall\,=\,93.79\% but with a catastrophic FPR of 94.71\%, behaving almost identically to the trivial ``all-vulnerable'' baseline discussed above.
Upgrading to GPT-4o improves FPR to 61.84\%, yet P-C remains at 44, indicating that stronger model capacity alone cannot compensate for the lack of evidence-based reasoning.

VulTrial, the strongest agent-based baseline, introduces a structured mock-court protocol that substantially improves over both CoT and GPTLens.
Its performance progression from GPT-3.5 (P-C\,=\,68) to GPT-4o\textsubscript{untrained} (P-C\,=\,81) to GPT-4o\textsubscript{trained} (P-C\,=\,96) demonstrates the combined benefit of model capacity and task-specific fine-tuning.
Notably, \toolname surpasses even VulTrial's best variant (trained on task-specific data with GPT-4o) by 26 pairs, \emph{without any training data} and using DeepSeek-V3.1, whose API cost is significantly lower than GPT-4o.
This result validates our central hypothesis: when agents are equipped with precise, graph-guided evidence traces, structured reasoning over verified facts is more effective than unconstrained debate over incomplete context, regardless of the underlying model's raw capability.

\subsection{RQ2: Ablation Study}
\label{sec:rq2}

\paragraph{Overview.}
RQ1 establishes that \toolname achieves state-of-the-art pair-wise discriminative performance.
We now open the black box to understand \emph{why}: how does each stage of the pipeline contribute to this result?
Following the two-stage design of \toolname, \ie evidence construction (Phase~I--II) and reasoning verification (Phase~III--IV), we organize this analysis into three complementary dimensions:
(a)~whether the clue discovery and context augmentation pipeline accurately localizes vulnerable code,
(b)~how many clues need to be investigated to achieve near-optimal detection, and
(c)~how the dialectical verification structure and meta-auditing mechanism each affect detection performance.

\subsubsection{(a) Clue Localization Quality.}
\label{sec:localization}

The entire \toolname pipeline rests on a critical assumption: Phase~I must successfully identify the true vulnerable locations as clues; otherwise, no amount of context augmentation or sophisticated reasoning in subsequent phases can recover.
To validate this, we extract the modified lines from each PrimeVul pair's fixing commit as ground-truth vulnerable locations and measure the localization recall (the fraction of ground-truth vulnerable locations that are covered by the top-$k$ clues) for Phase~I alone and for the combined Phase~I\,+\,II pipeline.

\begin{figure}
  \centering
  \includegraphics[width=0.49\textwidth]{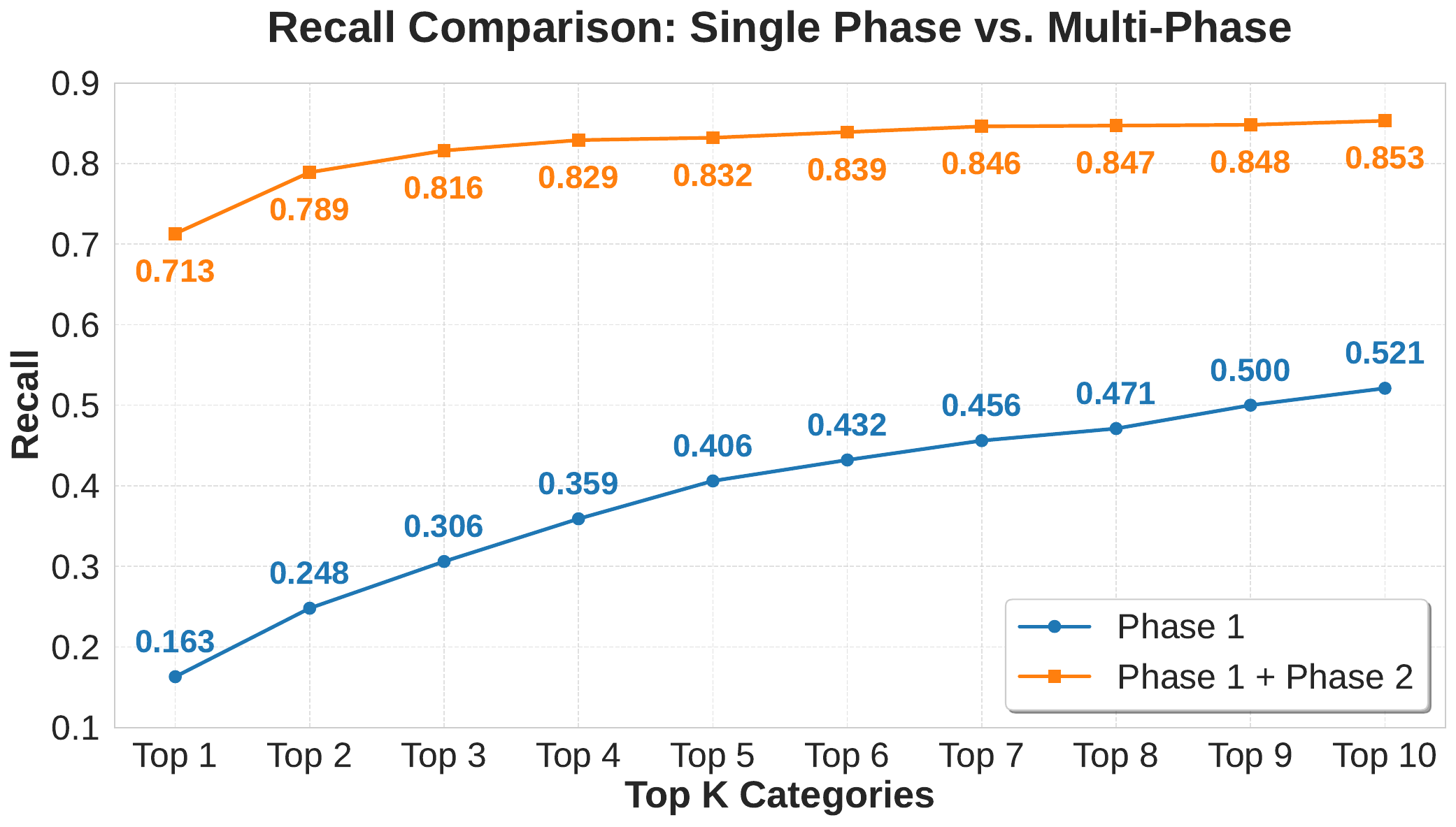}
  \caption{Localization recall of Phase~I alone vs.\ Phase~I\,+\,II across top-$k$ clues. CPG-guided context augmentation yields a substantial and consistent recall boost, particularly at low $k$.}
  \label{fig:localization_recall}
\end{figure}

Figure~\ref{fig:localization_recall} reveals two key findings.
First, Phase~I alone exhibits a gradual recall curve, reaching only 16.3\% at $k$\,=\,1 and 52.1\% at $k$\,=\,10.
This is expected: Phase~I operates under the Worst-Case Taint Assumption and analyzes the target function in isolation, so its ability to pinpoint the exact vulnerable location is limited by the absence of cross-function context.
Many true vulnerability sites involve operations that appear benign in isolation (e.g., a buffer copy whose length \emph{is} checked, but only in a caller function), causing Phase~I to assign them lower confidence ranks.

Second, and more importantly, Phase~I\,+\,II dramatically improves localization recall across all $k$ values.
At $k$\,=\,1, recall jumps from 16.3\% to 71.3\%, a 55.0 percentage point increase, and at $k$\,=\,3, it reaches 81.6\% compared to Phase~I's 30.6\%.
The gap is most pronounced at low $k$ values and narrows as $k$ increases (at $k$\,=\,10: 85.3\% vs.\ 52.1\%), exhibiting a characteristic diminishing-return pattern for the augmented pipeline.
This demonstrates that CPG-guided context augmentation does not merely add contextual volume; it fundamentally expands the coverage of each clue by tracing its data provenance through the repository. 
Even when Phase~I flags an intermediate node on the source-to-sink path rather than the exact vulnerability site, Phase~II's dependency chain reconstruction naturally encompasses the true fix location within the same data-flow trajectory.


\subsubsection{(b) Clue Sensitivity Analysis.}
\label{sec:clue:sens:ana}
Phase~I's high-recall design inevitably produces multiple candidate clues per function.
This raises a practical question: must every clue be investigated through the full pipeline, or can near-optimal detection be achieved with a subset?
To answer this, we rank Phase~I's clues by their confidence scores and vary the number of top-$k$ clues forwarded to subsequent phases, evaluating detection performance and computational cost across $k \in [1, 10]$.

\begin{figure*}[t]
  \centering
  \includegraphics[width=\linewidth]{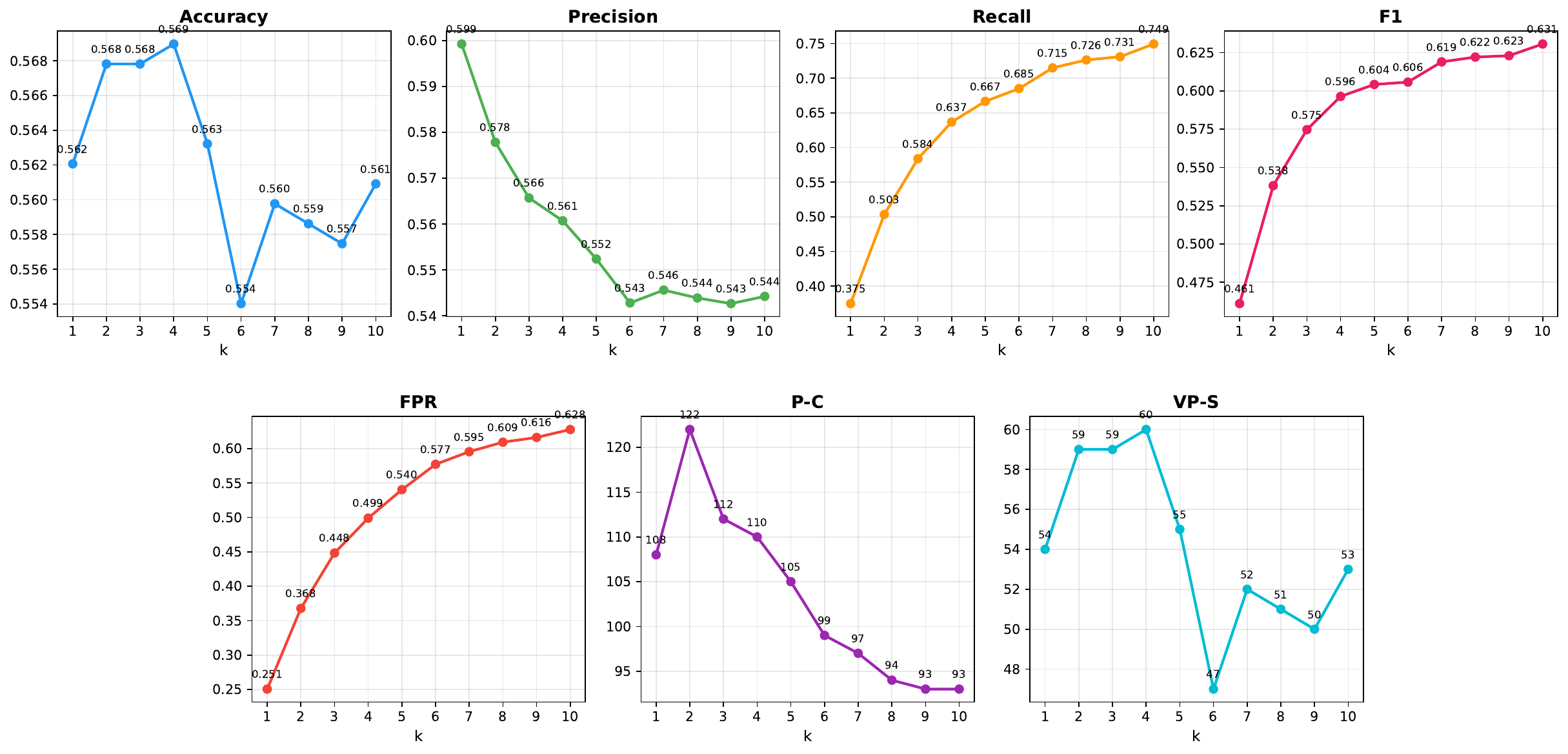}
  \caption{Detection performance metrics as a function of top-$k$ clues forwarded from Phase~I. P-C peaks at $k$\,=\,2 (122) and declines monotonically beyond $k$\,=\,3, while Recall and FPR increase steadily with $k$.}
  \label{fig:topk_metric}
\end{figure*}

\begin{figure*}[t]
  \centering
  \includegraphics[width=\linewidth]{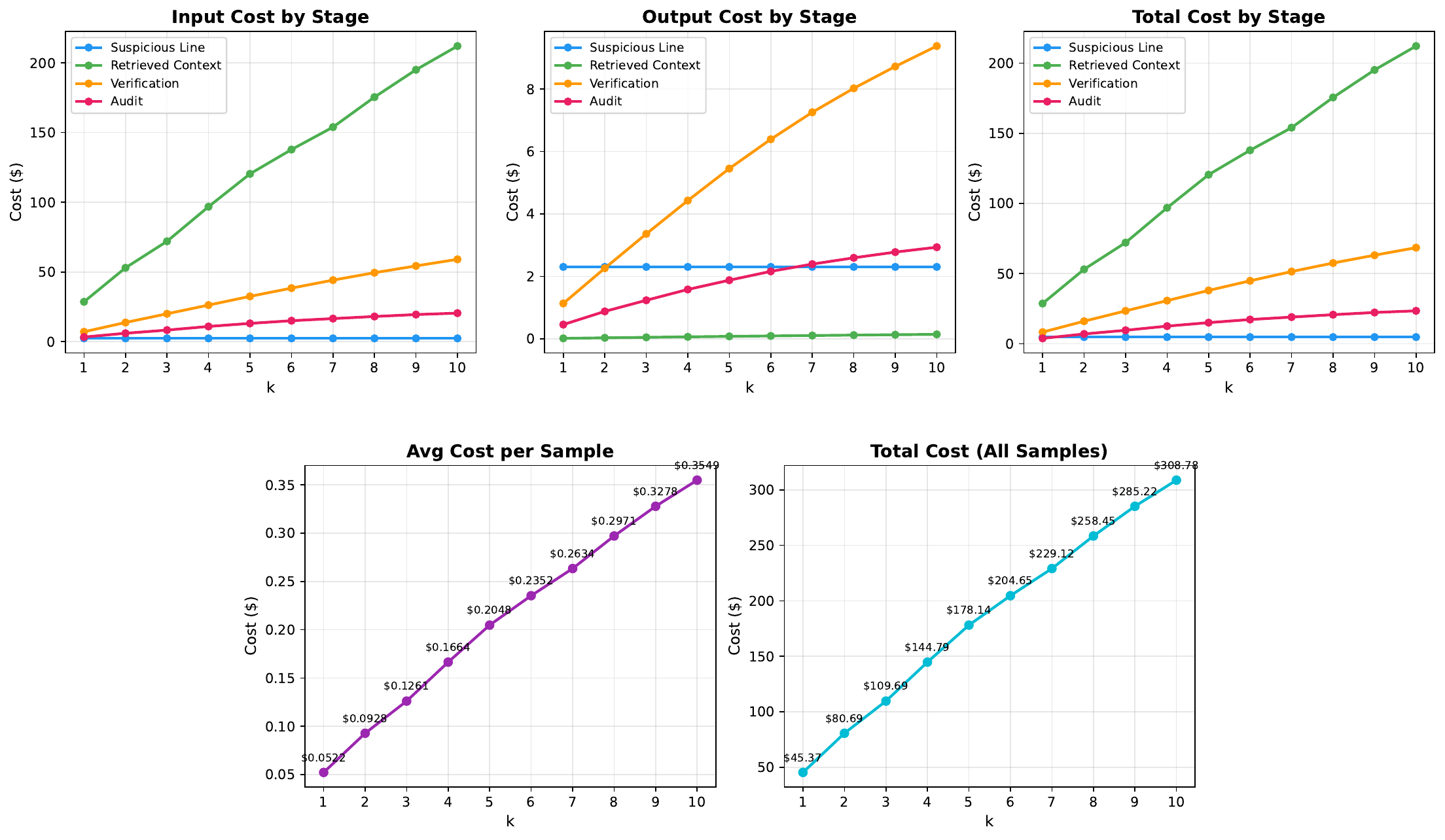}
  \caption{Computational cost analysis as a function of top-$k$ clues. Average per-sample cost grows approximately linearly from \$0.05 ($k$\,=\,1) to \$0.35 ($k$\,=\,10). The Retrieved Context stage dominates input cost, reflecting the CPG traversal overhead for each additional clue.}
  \label{fig:topk_cost}
\end{figure*}

\paragraph{Performance trends.}
Figure~\ref{fig:topk_metric} reveals a striking and non-monotonic relationship between the number of analyzed clues and pair-wise discriminative performance.
P-C peaks at $k$\,=\,2 with \textbf{122} correct pairs, 10 pairs higher than $k$\,=\,3 (112) and 29 pairs higher than $k$\,=\,10 (93), demonstrating that analyzing \emph{more} clues does not necessarily improve detection; rather, it can actively degrade it.
VP-S exhibits a similar pattern, peaking at $k$\,=\,4 (60) before declining.

This counter-intuitive finding has a clear explanation: as $k$ increases, lower-confidence clues introduce noise into the evidence trace, causing the Verifier to encounter more ambiguous or irrelevant suspicious patterns.
These additional clues dilute the signal from the genuinely critical anomalies, making it harder for the reasoning agents to maintain focused, evidence-grounded arguments.
The effect manifests most clearly in the FPR trajectory: FPR rises monotonically from 25.1\% ($k$\,=\,1) to 62.8\% ($k$\,=\,10), indicating that each additional low-confidence clue increases the likelihood of false alarms.

Conversely, standard metrics that reward detection coverage tell a different story: Recall increases steadily from 37.5\% ($k$\,=\,1) to 74.9\% ($k$\,=\,10), and F1 improves from 46.0\% to 63.1\%.
This divergence between standard metrics and pair-wise metrics reinforces our earlier observation (Section~\ref{sec:rq1}): on balanced benchmarks like PrimeVul, rising Recall and F1 can mask deteriorating discriminative precision, as the model increasingly trades specificity for sensitivity.

\paragraph{Cost analysis.}
Figure~\ref{fig:topk_cost} shows that the average per-sample cost grows approximately linearly with $k$, from \$0.052 at $k$\,=\,1 to \$0.355 at $k$\,=\,10, a 6.8$\times$ increase.
The cost breakdown by pipeline stage reveals that the Retrieved Context stage (Phase~II) dominates the total expenditure, accounting for the majority of input token consumption.
This is expected: each additional clue triggers a separate CPG traversal and graph-stitching operation, generating a proportionally larger evidence trace that must be processed by the downstream Verification and Audit agents.
Notably, the Suspicious Line stage (Phase~I) incurs negligible marginal cost as $k$ increases, since it analyzes the target function only once regardless of how many clues are subsequently forwarded.

\paragraph{Sweet spot.}
The combined analysis identifies $k$\,=\,2 and $k$\,=\,3 as the two most competitive configurations, each exhibiting distinct strengths.
At $k$\,=\,2, \toolname achieves the highest P-C (\textbf{122}) with FPR\,=\,36.8\% and higher Precision (57.8\%), at an average cost of only \$0.093 per sample.
However, its Recall (50.3\%) and F1 (53.8\%) are noticeably lower, reflecting a more conservative detection profile that prioritizes discriminative precision over coverage.
At $k$\,=\,3, P-C decreases to 112 but Recall improves substantially (58.4\%) alongside a higher F1 (57.5\%), while maintaining identical Accuracy (56.8\%) at a cost of \$0.126 per sample.
This configuration offers a more balanced trade-off across all metrics, at the expense of 10 fewer correct pairs and a higher FPR (44.8\% vs.\ 36.8\%).

The effectiveness of these low-$k$ configurations is underpinned by the localization quality demonstrated in Section~\ref{sec:localization}: Phase~I\,+\,II achieves 78.9\% and 81.6\% localization recall at $k$\,=\,2 and $k$\,=\,3 respectively, meaning the downstream reasoning agents already receive well-targeted evidence covering the vast majority of true vulnerability sites.
Increasing $k$ beyond this point yields diminishing localization gains (from 81.6\% at $k$\,=\,3 to 85.3\% at $k$\,=\,10) while introducing low-confidence clues that degrade reasoning quality, which directly explains the P-C decline observed in Figure~\ref{fig:topk_metric}.

The choice between $k$\,=\,2 and $k$\,=\,3 reflects a fundamental tension in vulnerability detection deployment.
$k$\,=\,2 is preferable in scenarios where \emph{alert fatigue} is the primary concern (e.g., integration into CI/CD pipelines where developers must triage every flagged function), as its lower FPR and higher P-C minimize wasted effort on false alarms.
$k$\,=\,3 is more suitable when \emph{comprehensive coverage} is prioritized (e.g., pre-release security audits where missing a true vulnerability carries higher risk than investigating false positives).
Beyond $k$\,=\,3, additional clues yield diminishing returns in every discriminative metric while incurring substantial cost increases, making higher $k$ values inadvisable for practical deployment.

\subsubsection{(c) Reasoning Component Ablation.}
\label{sec:reasoning:ablation}
Having validated the quality of evidence construction, we now examine whether the reasoning verification stages (Phase~III--IV) contribute independently to detection performance.
We evaluate two ablation variants, both using identical evidence traces produced by Phase~I--II:
\begin{itemize}[wide=0pt, nosep]
  \item \textbf{\textit{w/o} Dialectical Structure}: the Verifier classifies directly over the evidence trace in a single pass without the structured Red/Blue adversarial protocol, isolating the contribution of dialectical reasoning;
  \item \textbf{\textit{w/o} Meta-Auditing}: the Verifier's verdict is taken as final without Phase~IV's independent review, isolating the contribution of the Audit Agent.
\end{itemize}

\begin{figure*}[t]
  \centering
  \includegraphics[width=\linewidth]{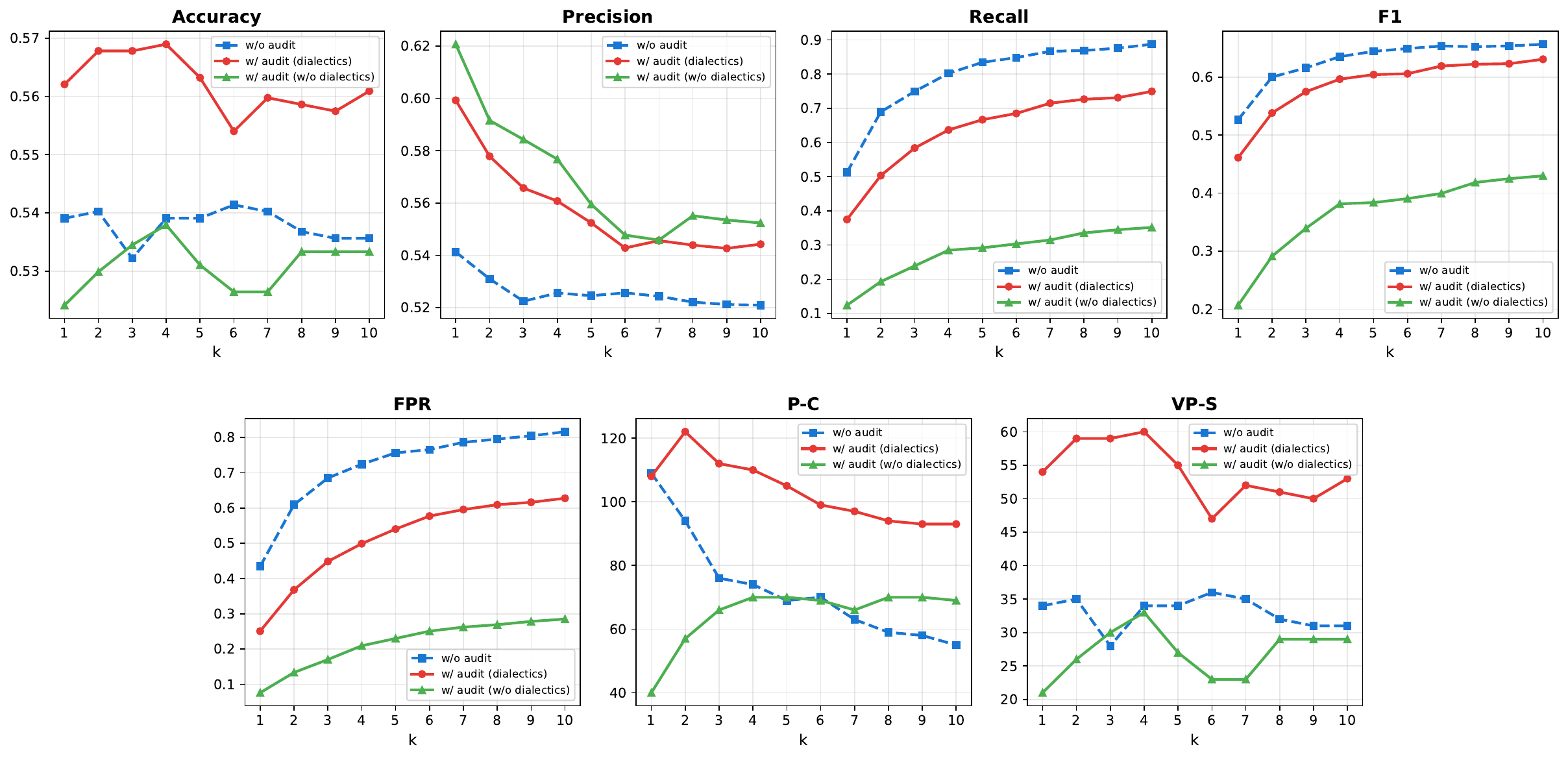}
  \caption{Three-way ablation of reasoning components across all top-$k$ configurations. Red solid: full \toolname (dialectics + audit); blue dashed: \textit{w/o} audit (dialectics only); green solid: \textit{w/o} dialectics (audit only). The full pipeline consistently dominates in P-C and VP-S.}
  \label{fig:ablation_3methods}
  \vspace{-3mm}
\end{figure*}

\paragraph{Effect of removing dialectical structure.}
As shown in Figure~\ref{fig:ablation_3methods}, removing the dialectical protocol while retaining Meta-Auditing produces the most severe performance degradation among all ablation variants.
At $k$\,=\,2, P-C drops from 122 (full pipeline) to just 57, a loss of 65 pairs (53.3\% relative decrease), and VP-S falls from 59 to 26.
The underlying cause is a dramatic collapse in Recall: without the structured Red/Blue protocol, the Verifier produces unstructured, single-pass reasoning over the evidence trace, which the Audit Agent then scrutinizes.
However, unstructured reasoning is far more likely to contain logical gaps and unsupported claims, causing the Audit Agent to exercise its veto power at a much higher rate.
At $k$\,=\,2, Recall drops to just 19.3\% (compared to 50.3\% with dialectics), while FPR is suppressed to 13.3\%.

This reveals a critical synergy: the dialectical structure not only improves the Verifier's standalone accuracy, but also produces \emph{higher-quality reasoning traces} that can survive the Audit Agent's scrutiny.
Without it, the audit mechanism becomes overly aggressive, rejecting the majority of ``vulnerable'' verdicts because the supporting arguments lack the structured evidence chains that the Red/Blue protocol forces the Verifier to construct.
Consequently, this system becomes precise but nearly inert: at $k$\,=\,1, Recall falls to a mere 12.4\% with P-C\,=\,40, rendering the system impractical for any real-world deployment.
Interestingly, the \textit{w/o} dialectics variant maintains relatively high Precision (58.4\% at $k$\,=\,3), confirming that the Audit Agent's judgments remain sound: it correctly identifies low-quality reasoning, but the absence of dialectical structure starves it of enough well-supported verdicts to achieve meaningful coverage.

\paragraph{Effect of removing Meta-Auditing.}
From Figure~\ref{fig:ablation_3methods}, we could also observe that removing the Audit Agent while retaining dialectical structure produces a qualitatively different failure mode.
At $k$\,=\,2, P-C drops from 122 to 94 (a loss of 28 pairs), but unlike the dialectics ablation, Recall \emph{increases} substantially, from 50.3\% to 69.0\%, while FPR surges from 36.8\% to 60.9\%.
This pattern reveals that without the Audit Agent's independent review, the Verifier's dialectical reasoning suffers from confirmation bias: having constructed an exploitability argument during the Red Team phase, it tends to anchor on this argument even when the Blue Team counter-argument presents mitigating evidence.
The net effect is that removing audit produces more true positives but even more false positives, resulting in a substantial P-C loss.
VP-S further underscores the difference: without audit, VP-S drops to 35, compared to 59 with the full pipeline, confirming that Meta-Auditing substantially improves discriminative precision beyond raw detection coverage.
The relatively stable P-R across both variants (63 vs.\ 59) confirms that Meta-Auditing primarily suppresses false ``vulnerable'' verdicts on patched functions rather than altering behavior on truly vulnerable functions.

\paragraph{Robustness across $k$.}
The three-way comparison reveals a consistent hierarchy across all $k$ values: the full pipeline (dialectics + audit) dominates in P-C and VP-S, followed by \textit{w/o} audit, with \textit{w/o} dialectics trailing substantially.
Notably, the gap between the full pipeline and \textit{w/o} audit \emph{widens} at higher $k$: at $k$\,=\,10, the full pipeline achieves P-C\,=\,93 versus 55 without audit, a gap of 38 pairs compared to 28 at $k$\,=\,2.
This confirms that Meta-Auditing becomes more valuable as noise from low-confidence clues increases. 
The \textit{w/o} dialectics variant exhibits a different trajectory: its P-C peaks at $k$\,=\,4--5 (up to 70) and plateaus, never approaching the full pipeline's performance at any $k$.
This ceiling effect demonstrates that without dialectical structure, even increasing evidence volume cannot compensate for the fundamental deficiency in reasoning quality.

\subsubsection{Summary.}
The ablation results confirm that \toolname's performance is not attributable to any single component but emerges from the synergy across its pipeline.
Phase~I's high-recall clue discovery ensures broad vulnerability coverage; Phase~II's graph-guided augmentation provides precise, diagnostically relevant context; Phase~III's dialectical structure ensures \emph{reasoning quality} by producing well-structured, evidence-grounded arguments; and Phase~IV's meta-auditing ensures \emph{reasoning reliability} by catching residual confirmation bias and unsupported logical leaps.
Neither reasoning mechanism alone suffices: dialectics without audit leads to high FPR, while audit without dialectics leads to critically low Recall.
The top-$k$ sensitivity analysis further demonstrates that this pipeline is practically configurable, enabling users to trade off detection thoroughness against computational cost based on their deployment constraints.

\subsection{RQ3: Reasoning Quality}

RQ2 demonstrates that Meta-Auditing significantly improves detection performance.
RQ3 opens up the Audit Agent's decision-making process to understand \emph{how} it achieves this improvement: how often does it intervene, how accurate are its interventions, and what categories of reasoning flaws does it detect?

\subsubsection{Veto Rate and Correctness.}
Table~\ref{tab:veto_analysis} summarizes the Audit Agent's veto behavior across all $k$ configurations.
At the default configuration ($k$\,=\,2), the Audit Agent overrides \textbf{186} of the Verifier's 870 function-level verdicts (21.4\% veto rate), nearly all of which reverse a ``vulnerable'' verdict to ``safe.''
Of these 186 vetoes, \textbf{105} (56.5\%) are correct, overturning false positives on patched functions, while \textbf{81} (43.5\%) are incorrect, overturning true positives on genuinely vulnerable functions.
The resulting ratio of 1.30 correct reversals per incorrect one confirms that the Audit Agent's independent judgment is substantially more often right than wrong.

\begin{table}[t]
  \centering
  \small
  \caption{Audit Agent veto analysis across top-$k$ configurations. \textit{Correct} and \textit{Incorrect} denote the number of verdicts correctly and incorrectly overturned, respectively. \textit{Rate} is the fraction of correct vetoes. \textit{$\Delta$P-C} is the net change in Pair-wise Correct Predictions attributable to auditing.}
  \label{tab:veto_analysis}
  \begin{tabular}{c rrr r r}
    \toprule
    $k$ & Total & Correct & Incorrect & Rate (\%) & $\Delta$P-C \\
    \midrule
    1  & 140 &  80 & 60 & 57.1 & $-$1 \\
    \textbf{2}  & \textbf{186} & \textbf{105} & \textbf{81} & \textbf{56.5} & \textbf{+28} \\
    3  & 175 & 103 & 72 & 58.9 & +36 \\
    4  & 170 &  98 & 72 & 57.6 & +36 \\
    5  & 167 &  94 & 73 & 56.3 & +36 \\
    6  & 153 &  82 & 71 & 53.6 & +29 \\
    7  & 149 &  83 & 66 & 55.7 & +34 \\
    8  & 143 &  81 & 62 & 56.6 & +35 \\
    9  & 145 &  82 & 63 & 56.6 & +35 \\
    10 & 142 &  82 & 60 & 57.7 & +38 \\
    \bottomrule
  \end{tabular}
  \vspace{-3mm}
\end{table}

Two patterns emerge from Table~\ref{tab:veto_analysis} that illuminate the Audit Agent's operating characteristics.

\paragraph{Stable veto accuracy.}
The correct veto rate remains remarkably stable across all $k$ values, ranging from 53.6\% ($k$\,=\,6) to 58.9\% ($k$\,=\,3) and consistently exceeding the 50\% random baseline.
This stability indicates that the Audit Agent's reasoning quality does not degrade as evidence complexity increases with higher $k$. Even when processing noisier evidence traces containing low-confidence clues, the Audit Agent maintains its ability to distinguish sound from unsound reasoning.

\paragraph{The $k$\,=\,1 anomaly.}
At $k$\,=\,1, Meta-Auditing is the only configuration where $\Delta$P-C is negative ($-$1): despite a 57.1\% correct veto rate, the audit slightly \emph{harms} pair-wise performance.
This occurs because the evidence base at $k$\,=\,1 is minimal (Phase~I\,+\,II achieves only 71.3\% localization recall with a single clue, as discussed in Section~\ref{sec:localization}), leaving the Audit Agent with insufficient context to make well-calibrated judgments.
Many of its correct FP reversals at $k$\,=\,1 do not translate into P-C gains because the corresponding vulnerable function in the pair was also incorrectly classified.
From $k$\,=\,2 onward, $\Delta$P-C is consistently positive (+28 to +38), confirming that the audit mechanism requires a minimum evidence threshold to produce net benefits.

\subsubsection{Analysis of Incorrect Vetoes.}
The 81 incorrect reversals at $k$\,=\,2, where the Audit Agent overturned a correctly-issued ``vulnerable'' verdict, represent the inherent cost of the auditing mechanism. Analysis by vulnerability type reveals that CWE-787 (Out-of-bounds Write), CWE-476 (NULL Pointer Dereference), and CWE-125 (Out-of-bounds Read) account for the highest number of incorrect vetoes. 
These three categories all rank within the 2025 CWE Top 25 Most Dangerous Software 
Weaknesses~\cite{cwe25}, underscoring that the Audit Agent's primary 
limitation manifests precisely on the most prevalent and consequential vulnerability 
types.
Specifically, they share a common characteristic: their exploitability hinges on \emph{absence-of-protection} conditions, such as whether a buffer size is correctly bounded or a pointer is guarded against NULL on a particular path. Verifying such conditions requires a \emph{negative proof} that no sufficient mitigation exists along the analyzed path. However, the evidence trace records what \emph{is} present in the code and cannot explicitly encode what is \emph{missing}. When the Verifier correctly infers exploitability from the absence of a check, the Audit Agent tends to classify this inference as Speculation, creating a systematic bias toward over-correction for these vulnerability categories. A representative case study is provided in Appendix~\ref{sec:case_study_incorrect_veto}.

This conservative behavior is by design: the Audit Agent operates at temperature\,=\,0, prioritizing consistency and rigor over recall. As Table~\ref{tab:veto_analysis} shows, incorrect vetoes decrease from 81 ($k$\,=\,2) to 60 ($k$\,=\,10) as richer evidence traces provide more complete justifications that survive audit scrutiny. Note that this effect, improved per-veto accuracy through richer evidence, is distinct from the noise introduced by low-confidence clues at higher $k$ (Section~\ref{sec:clue:sens:ana}), which degrades overall detection; the two phenomena operate at different pipeline stages and are not contradictory. This trend suggests that incorrect vetoes are primarily driven by evidence incompleteness rather than fundamental flaws in the audit mechanism itself.

\subsubsection{Reasoning Flaw Distribution.}
Since a single veto may cite multiple reasoning flaws, we count every flaw instance reported by the Audit Agent. Table~\ref{tab:flaw_distribution} presents the full distribution across all $k$ configurations. At the default setting ($k{=}2$), 186 vetoes yield 621 individual flaw instances (3.4 per veto on average). Beyond the four predefined categories from Section~\ref{subsec:auditing}, the Audit Agent organically discovers additional failure modes. The most prevalent is \emph{Pattern-Matching Flaw} (18.84\%), where the Verifier flags code as vulnerable based on superficial syntactic patterns without verifying exploitability through the trace. \emph{Semantic Misunderstanding Flaw} (16.43\%) captures cases where the Verifier misinterprets the semantics of a code construct, while \emph{Absence-as-Evidence Flaw} (11.76\%) arises when the absence of an explicit check in the trace is conflated with the absence of any protection. Together with the predefined \emph{Over-Trust} (16.43\%) and \emph{Speculation} (14.01\%), the top five categories account for 77.5\% of all detected flaws. Notably, outright hallucination (\emph{Evidence Fabrication}) accounts for only 1.77\%, indicating that when provided with a closed evidence substrate, the LLM rarely fabricates evidence; the dominant failure mode is \emph{misinterpretation} of existing evidence rather than invention of non-existent evidence~\cite{ji2023survey}. This finding directly validates the core argument of Section~\ref{sec:intro}: grounding agents in a verifiable evidence base effectively suppresses the contextual hallucinations that plague prior approaches.

Table~\ref{tab:flaw_distribution} further reveals that the distribution is remarkably stable across $k$: the proportional contribution of each category remains within $\pm$2 percentage points from $k{=}1$ to $k{=}10$. The only notable shift is a gradual increase in \emph{Speculation} (from 13.71\% to 16.46\%) accompanied by a mild decrease in \emph{Over-Trust} (from 17.45\% to 14.44\%), consistent with the noise hypothesis from Section~\ref{sec:clue:sens:ana}: lower-confidence clues at higher $k$ produce more ambiguous traces that invite speculative reasoning. This stability suggests that the detected reasoning flaws reflect intrinsic limitations of the LLM's reasoning process rather than artifacts of evidence quality, implying that targeted improvements to the reasoning protocol, such as explicit training against pattern-matching shortcuts, could yield further gains orthogonal to evidence construction improvements.

\begin{table*}[t]
\centering
\caption{Reasoning flaw distribution (\%) across top-$k$ configurations. Each cell reports the percentage of total flaw instances attributed to that category at the given $k$. Categories above the mid-rule are the four predefined types from Section~\ref{subsec:auditing}; those below are emergent types discovered by the Audit Agent. Rare types ($<$10 instances) are grouped as ``Other.''}
\label{tab:flaw_distribution}
\small
\setlength{\tabcolsep}{4pt}
\begin{tabular}{l*{10}{r}}
\toprule
\textbf{Flaw Category} & $k{=}1$ & $k{=}2$ & $k{=}3$ & $k{=}4$ & $k{=}5$ & $k{=}6$ & $k{=}7$ & $k{=}8$ & $k{=}9$ & $k{=}10$ \\
\midrule
Over-Trust          & 17.45 & 16.43 & 15.45 & 15.30 & 15.31 & 15.25 & 14.73 & 14.67 & 14.49 & 14.44 \\
Speculation         & 13.71 & 14.01 & 14.20 & 14.77 & 15.31 & 15.25 & 15.55 & 16.08 & 16.39 & 16.46 \\
Anchoring           &  6.54 &  6.60 &  6.70 &  6.38 &  6.10 &  6.01 &  5.78 &  5.53 &  5.40 &  5.44 \\
Phantom Mitigation  &  3.74 &  4.19 &  4.55 &  4.37 &  4.28 &  4.31 &  4.47 &  4.67 &  4.66 &  4.70 \\
\midrule
Pattern-Matching          & 18.69 & 18.84 & 19.32 & 19.76 & 19.88 & 20.05 & 20.62 & 20.55 & 20.31 & 19.97 \\
Semantic Misunderstanding & 15.58 & 16.43 & 16.14 & 15.91 & 15.97 & 16.16 & 15.60 & 15.33 & 15.42 & 15.36 \\
Absence-as-Evidence       & 11.53 & 11.76 & 11.36 & 11.45 & 11.61 & 11.24 & 11.18 & 10.90 & 11.09 & 11.37 \\
Scope Creep               &  8.72 &  7.25 &  7.50 &  6.99 &  6.68 &  6.50 &  6.60 &  6.98 &  6.85 &  6.94 \\
Incomplete Protection     &  ---  &  1.93 &  1.70 &  1.66 &  1.60 &  1.88 &  2.02 &  1.96 &  2.10 &  2.11 \\
Evidence Fabrication      &  ---  &  1.77 &  1.93 &  2.10 &  1.96 &  1.76 &  1.91 &  1.86 &  1.82 &  1.76 \\
Other                     &  4.05 &  0.81 &  1.14 &  1.31 &  1.31 &  1.58 &  1.53 &  1.46 &  1.49 &  1.45 \\
\midrule
\textbf{Total instances}  & \textbf{321} & \textbf{621} & \textbf{880} & \textbf{1144} & \textbf{1378} & \textbf{1646} & \textbf{1833} & \textbf{1990} & \textbf{2147} & \textbf{2278} \\
\bottomrule
\end{tabular}
\end{table*}


\section{Discussion}

\subsection{The Pivotal Role of Evidence Construction in Grounded Reasoning}

The experimental results across RQ1--RQ3 converge on a central finding: \textbf{the quality of evidence construction, not the sophistication of the reasoning protocol, is the primary determinant of detection performance.} 
This insight crystallizes most clearly in Phase~II, the graph-guided context augmentation, which serves as the architectural pivot of the entire \toolname pipeline. Phase~II simultaneously transforms the output of clue discovery (upward impact on Phase~I) and enables the soundness of downstream verification (downward impact on Phase~III and IV). We discuss each direction in turn before distilling a broader design principle.

\subsubsection{Upward Impact: From Intermediate Anchors to Complete Vulnerability Coverage.}
Phase~I performs intra-procedural taint analysis in isolation, so the locations it flags with highest confidence are typically \emph{intermediate nodes} along a source-to-sink path, such as variable propagation points or tainted parameter uses, rather than the exact lines modified in the fixing commit. The true fix site (e.g., a missing bounds check introduced by the patch) often lies upstream or downstream of these intermediate points, reachable only through cross-function tracing. This explains why Phase~I alone achieves only 16.3\% localization recall at $k{=}1$: its top clue is not irrelevant to the vulnerability, but it does not directly coincide with the ground-truth fix location.
 
Phase~II bridges this gap through \emph{dataflow coverage expansion}. By tracing the flagged variable backward to its provenance and forward to its downstream sinks via the Code Property Graph, the reconstructed dependency chain naturally \emph{encompasses} the true vulnerability site, even though Phase~I's original clue pointed to an intermediate location on that same path. Crucially, Phase~I's confidence ranking is unchanged; what changes is the \emph{coverage} of each clue once expanded into its full execution trajectory. This mechanism explains the recall jump from 16.3\% to 71.3\% at $k{=}1$ (a 55.0 percentage point increase), and with just the top-2 clues, Phase~I+II reaches 78.9\% recall, sufficient to support a P-C of 122 that surpasses every baseline. Phase~I's intermediate anchors, while imprecise in isolation, serve as effective \emph{entry points} into the vulnerability's data-flow topology; Phase~II completes the picture by expanding them into full dependency chains.

\subsubsection{Downward Impact: Establishing a Closed Factual Substrate for Verification.}
The contribution of Phase~II extends beyond localization improvement; it fundamentally reshapes the epistemic conditions under which Phase~III and Phase~IV operate. Prior multi-agent approaches such as VulTrial~\cite{VulTrial} conduct adversarial debates over isolated function slices, creating an \emph{open} information environment where agents can freely hypothesize about unseen code. As demonstrated empirically~\cite{VulTrial}, this openness causes debate quality to degrade over successive rounds, as agents retreat into mutually reinforcing concessions rather than introducing new evidence.
 
Phase~II's evidence trace eliminates this failure mode by constructing a \emph{closed factual substrate}, a bounded, per-variable record of backward provenance and forward propagation chains, with explicit file boundaries at each step. This closure property has two critical implications for downstream reasoning:
 
\begin{itemize}[leftmargin=*]
    \item \textbf{Falsifiability of claims.} Every assertion made during Phase~III's dialectical reasoning, whether arguing for exploitability or safety, must cite specific line numbers within the trace. If a claimed security check does not appear in the trace, the Verifier must treat it as non-existent on the analyzed path, rather than speculating about its presence elsewhere in the repository. This constraint directly prevents the contextual hallucinations identified in Section~1, where agents fabricate plausible-sounding but unverifiable mitigations.
 
    \item \textbf{Auditability of reasoning.} Phase~IV's Meta-Auditing mechanism is only feasible \emph{because} Phase~II produces a structured, citable evidence base. The Audit Agent's four reasoning flaw categories (Phantom Mitigation, Speculation, Anchoring, and Over-Trust) are all operationalized as violations of the trace boundary: citing evidence outside the trace (Phantom Mitigation), assuming behavior of unseen code (Speculation), ignoring trace evidence in favor of prior suspicion (Anchoring), or treating unverified external dependencies as safe (Over-Trust). Without a well-defined evidence boundary, these categories would lack the formal grounding necessary for systematic detection.
\end{itemize}
 
The ablation results in Section~\ref{sec:reasoning:ablation} provide quantitative support for this argument. Without Meta-Auditing, the Verifier's FPR surges from 36.8\% to 60.9\% at $k{=}2$, indicating that even with dialectical structure, a single agent's reasoning over the evidence trace is insufficient to suppress confirmation bias. However, the Audit Agent's ability to reduce this gap, correcting 105 false positives while incorrectly overturning only 81 true positives (a 1.30:1 correction ratio), is predicated entirely on Phase~II's trace providing a concrete reference against which each claim can be independently verified. The Audit Agent does not introduce new evidence; it \emph{audits} existing reasoning against existing evidence. This asymmetry underscores Phase~II's dual role: it both supplies the raw material for reasoning and defines the boundary conditions that make reasoning auditable.

\subsubsection{A Broader Insight: Evidence Quality as the Ceiling for Reasoning Quality.}
Abstracting from the specifics of \toolname, our results point to a general design principle for LLM-based multi-agent analysis systems:
\emph{For any LLM-based multi-agent system that must make factual determinations over structured artifacts, the ceiling of reasoning quality is set by the quality of the evidence substrate, not by the complexity of the reasoning protocol. The primary design investment should therefore target dynamic, hypothesis-specific evidence retrieval rather than reasoning protocol sophistication.}
 
The \toolname vs.\ VulTrial comparison provides direct empirical support for this principle. \toolname's reasoning architecture (single-agent dialectics plus independent audit) is arguably \emph{simpler} than VulTrial's multi-round, multi-agent court simulation. Yet \toolname achieves 50.6\% higher P-C than VulTrial$_\text{untrained}$ (122 vs.\ 81) and 27.1\% higher than VulTrial$_\text{trained}$ (122 vs.\ 96) at $k{=}2$. The performance gap is not attributable to reasoning sophistication but to evidence grounding: \toolname's agents reason over precise, repository-derived dependency chains, while VulTrial's agents debate over isolated function slices supplemented only by their parametric knowledge.
 
This principle extends naturally beyond vulnerability detection. In any domain where LLM agents must make factual determinations, such as legal analysis~\cite{jurex, lawagent, malr}, medical diagnosis~\cite{medicalagent, medicalagent1, llm4medical}, or financial auditing~\cite{tradingagents, auditingagent}, our results suggest that investing in \emph{dynamic, query-specific evidence retrieval} (analogous to Phase~II's clue-anchored CPG slicing) will yield greater returns than investing in more elaborate reasoning protocols over static or generic context. The key architectural decision is not how many agents debate, but whether those agents have access to a closed, relevant, and verifiable evidence base against which their claims can be adjudicated.
 
This conclusion also closes the narrative opened in Section~\ref{sec:intro}. We diagnosed that existing approaches, both agent-based debate and retrieval augmentation, fail because they reason in an \emph{ungrounded} deliberative space lacking a bounded, hypothesis-specific evidence base. The experimental evidence presented in Sections~4.1 through 4.3 confirms that this diagnosis was correct: the problem was never about reasoning sophistication; it was about evidence grounding. Once agents are equipped with a closed factual substrate derived from the repository's own data-flow topology, even a relatively simple reasoning protocol suffices to achieve state-of-the-art discriminative performance.

\subsection{Threats to Validity}

\subsubsection{Implementation Validity.}
We observed that the backbone LLM occasionally produced responses containing illegal characters (most notably Chinese characters interspersed within English output), which we attribute to residual language-mixing behavior in DeepSeek-V3.1's multilingual training corpus. Left unchecked, such malformed outputs could corrupt the structured fields that downstream agents depend on. To mitigate this, we implemented a schema-based output validation layer that automatically rejects and re-queries malformed responses. While this retry mechanism ensures output integrity, it introduces additional API calls and may subtly bias results toward responses that pass validation on the first attempt. In practice, however, the retry rate was low and did not materially affect the overall cost or runtime.

\subsubsection{Evaluation Validity.}
We ground our primary analysis in Pair-wise Correct Prediction (P-C) and False Positive Rate (FPR), as standard metrics (Precision, Recall, F1) can mask deteriorating discriminative precision on balanced benchmarks (see Section~\ref{sec:metrics} for a detailed justification). We acknowledge that P-C is a strict metric that credits only fully correct pairs and may underestimate partial detection capability, but we consider this conservatism appropriate for vulnerability detection, where both missed vulnerabilities and false alarms carry significant practical cost.

\subsubsection{Model Selection.}
All experiments use DeepSeek-V3.1 as the sole backbone LLM, primarily due to budget constraints. Since \toolname's pipeline is dominated by input token consumption (long evidence traces fed to the Verification and Audit agents), the cost differential between DeepSeek-V3.1 (\$0.56/M input tokens) and alternatives such as GPT-4o (\$2.75/M input) is substantial: the full 435-pair evaluation at $k{=}3$ costs approximately \$110 with DeepSeek-V3.1 but would exceed \$500 with GPT-4o.
We argue that this single-model evaluation does not undermine the generalizability of \toolname's design. The framework's core contribution lies in its evidence-grounded reasoning architecture, not in any capability specific to DeepSeek-V3.1. Agents interact through structured prompts and well-defined evidence traces, imposing no model-specific requirements beyond standard instruction-following ability. The comparison with VulTrial is instructive: VulTrial achieves P-C~=~96 with GPT-4o plus task-specific fine-tuning, while \toolname achieves P-C~=~122 with DeepSeek-V3.1 and no training, suggesting that the performance advantage stems from architectural design rather than raw model capacity. Nevertheless, validating \toolname with alternative backbone models (e.g., GPT-4o) remains important future work to confirm the framework's model-agnostic nature.

\section{Related Work}

\subsection{LLM-based Vulnerability Detection}
Early deep learning approaches learn program semantics from composite code representations (Devign~\cite{devign}) or apply transformer-based attention for line-level prediction (LineVul~\cite{linevul}). Pre-trained code models such as CodeBERT~\cite{codebert}, CodeT5~\cite{codet5}, and UniXcoder~\cite{unixcoder} advance this paradigm through fine-tuning, but produce static embeddings that fail to capture the subtle semantic shifts between a vulnerability and its patch~\cite{llm4vd}. More recent fine-tuning work incorporates richer signals: ReVD~\cite{revd} uses curriculum preference optimization with synthesized reasoning data, VulTrLM~\cite{vultrlm} augments inputs via AST decomposition and comment enhancement, and VULPO~\cite{vulpo} applies on-policy reinforcement learning with multi-dimensional rewards for context-aware detection. While these methods improve upon static embeddings, they remain limited by training data scope and input representations; in particular, VULPO's repository-level context relies on lightweight heuristics rather than hypothesis-driven dependency tracing.

On the prompting side, multiple studies~\cite{llm4vd, llm4vuln, ullah2024llms, llm4vdre} show that even state-of-the-art LLMs with Chain-of-Thought prompting fail to achieve meaningful pair-wise discriminative performance, confirming that prompting alone cannot substitute for systematic evidence gathering. Retrieval-augmented approaches such as VulInstruct~\cite{VulInstruct} and Vul-RAG~\cite{Vul-rag} import external security knowledge from historical patches and CVEs, but this \emph{generic} knowledge is decoupled from the repository's data-flow topology, leading to high false positive rates.

\subsection{Multi-Agent Reasoning for Code Analysis}
Several works adopt multi-agent collaboration to improve reasoning reliability. GPTLens~\cite{GPTLens} employs a two-agent identify-then-evaluate pipeline; VulTrial~\cite{VulTrial} introduces a mock-court protocol with prosecutor, defender, and judge roles; and MAVUL~\cite{mavul} equips agents with tool-using capabilities for cross-procedural reasoning and interactive refinement. While these approaches introduce increasingly sophisticated interactions, they share a fundamental limitation: agents debate over isolated function slices or heuristically retrieved context without a bounded, hypothesis-specific evidence base grounded in actual repository dependencies. VulTrial's own evaluation reveals this fragility, as increasing debate rounds paradoxically degrades performance through mutually reinforcing concessions~\cite{VulTrial}. Multi-agent frameworks for smart contract auditing (LLM-SmartAudit~\cite{llm4sca}, iAudit~\cite{iaudit}) face the same core issue: multi-perspective deliberation cannot compensate for an \emph{open} information environment where the facts needed to resolve a vulnerability hypothesis are unavailable to the debating agents.

\subsection{Code Property Graphs for Vulnerability Detection}
Code Property Graphs (CPGs)~\cite{joern-cpg} unify abstract syntax trees, control-flow graphs, and program dependence graphs into a single queryable structure. Steenhoek et al.~\cite{steenhoek2024comprehensive} showed that dataflow information improves vulnerability detection models. LLMxCPG~\cite{llmxcpg} combines CPGs with LLMs by using a fine-tuned model to generate CPG queries for extracting vulnerability-relevant slices, but this approach is inherently constrained by vulnerability pattern diversity, making it impractical to train a single query-generation model across all categories.

\smallskip
\noindent\textbf{Positioning of \toolname.}
\toolname departs from prior work along three axes. First, unlike agent-based approaches that debate over incomplete context, \toolname constructs a \emph{closed evidence substrate} via on-demand CPG slicing before reasoning begins, grounding all dialectical arguments in verifiable repository-level facts. Second, unlike retrieval-augmented methods that import generic security knowledge, and unlike RL-based approaches (e.g., \textit{VULPO}~\cite{vulpo}) that extract context through lightweight heuristics, \toolname's context augmentation is \emph{clue-anchored and demand-driven}: each expansion is triggered by a specific suspicious location and adapts to the data-flow topology of that clue, making the framework inherently CWE-agnostic. Third, unlike CPG-based approaches that rely on learned query generation, \toolname uses clues as natural graph traversal anchors, requiring no task-specific training. Together, these choices shift the paradigm from classification over incomplete observations to forensic verification over dynamically assembled evidence.

\section{Conclusion}
We presented \toolname, a multi-agent framework that reframes vulnerability detection as a forensic process of ``From Clue to Verdict.'' By explicitly decoupling vulnerability localization from reasoning verification, \toolname addresses the contextual hallucination problem that fundamentally limits prior agent-based and retrieval-augmented approaches. Its graph-guided context augmentation mechanism dynamically reconstructs repository-level dependency chains anchored to each identified clue, providing downstream reasoning agents with a closed, verifiable evidence substrate rather than isolated code fragments or generic security knowledge. On the PrimeVul benchmark, \toolname achieves 122 Pair-wise Correct Predictions at $k{=}2$, the first approach to surpass 100 on this benchmark, while reducing the false positive rate by up to 54.40\% compared to leading baselines, without requiring any task-specific training. 

In future, we plan to validate \toolname across multiple backbone LLMs (e.g., GPT-4o, Claude) to empirically confirm the framework's model-agnostic nature. 
Second, while PrimeVul focuses on C/C++ vulnerabilities, extending the evaluation to other languages (e.g., Java, JavaScript) would test the generalizability of our CPG-based context augmentation, as Joern already supports multi-language parsing.

{\footnotesize \bibliographystyle{acm}
\bibliography{sample}}

\appendix
\newpage
\section{Agent Prompts}
\label{appendix:prompts}

This appendix provides the complete prompts used by each agent in the \toolname{} pipeline. The Clue-Discovery Agent (\S\ref{appendix:clue}), Verification Agent (\S\ref{appendix:verify}), and Audit Agent (\S\ref{appendix:audit}) each operate with a dedicated system prompt. The Context-Augmentation Agent (\S\ref{appendix:context}) uses a templated user prompt for each expansion decision, with no system prompt.

\subsection{Clue-Discovery Agent (Phase I)}
\label{appendix:clue}

\begin{lstlisting}[basicstyle=\ttfamily\scriptsize, breaklines=true, frame=single, backgroundcolor=\color{gray!5}, columns=fullflexible, keepspaces=true, xleftmargin=2pt, xrightmargin=2pt]
# Role
You are an experienced code security reviewer specializing in identifying potential security vulnerabilities. Your role is to perform the initial triage scan of a given code block to surface every potentially suspicious line that might contribute to a security vulnerability.

# Mission: HIGH RECALL TRIAGE
Your goal is to identify EVERY potentially suspicious line that could lead to a security vulnerability.
- Philosophy: "Better safe than sorry." If a line has even a 10% chance of being part of a vulnerability, REPORT IT.
- Context Handling: Since you are viewing a code snippet (not the full repository), ASSUME WORST-CASE SCENARIO. If a variable's origin is unknown (not defined in the snippet), ASSUME it comes from an untrusted source (Tainted).
- Downstream Handling: Do not worry about false alarms. A dedicated verification agent will rigorously validate your findings later. Your only failure mode is missing a potential clue.

# Vulnerability Scope
You are detecting security vulnerabilities across ALL categories. While prioritizing patterns most relevant to C/C++, you must remain vigilant for ANY security issue. Categories include but are NOT limited to:

Memory Safety:
Buffer overflow, use-after-free, double-free, null pointer dereference, uninitialized memory read, out-of-bounds read/write, stack overflow, heap corruption

Unsafe Functions & API Misuse:
Use of deprecated/banned functions (e.g., strcpy, sprintf, gets, strcat), incorrect API usage, missing size/bounds parameters, unsafe type casts

Integer & Arithmetic Issues:
Integer overflow/underflow, signedness errors (signed/unsigned mismatch), integer truncation, divide by zero, implicit narrowing conversions

Input Validation & Injection:
Command injection, format string vulnerabilities, path traversal, SQL injection (in embedded SQL), improper input sanitization

Resource Management:
Memory leaks, file descriptor leaks, socket leaks, missing cleanup in error paths, improper use of RAII or cleanup patterns

Error Handling:
Unchecked return values (especially malloc, calloc, realloc, fopen, read, write), ignored error codes, missing NULL checks, improper error propagation

Concurrency & Synchronization:
Race conditions, TOCTOU (time-of-check-time-of-use), data races, deadlocks, atomicity violations, improper lock usage

Logic Errors:
Incorrect boundary checks, off-by-one errors, wrong comparison operators, missing break in switch, fall-through bugs, infinite loops

Cryptographic Weaknesses:
Weak algorithms, predictable randomness (rand/srand), hardcoded keys/IVs, improper use of cryptographic APIs

Access Control & Authentication:
Hardcoded credentials, missing permission checks, privilege escalation paths, improper capability handling

CRITICAL: The above list is a guide, NOT a boundary. If ANY code pattern looks potentially dangerous, could cause undefined behavior, crashes, security bypass, data corruption, or violates secure coding best practices -- REPORT IT regardless of whether it fits a named category.

# Taint Analysis Rules
- Entry Points:
    - Public/exported function parameters: ALWAYS TAINTED.
    - Command Line Arguments (argv) & Standard Input (stdin): ALWAYS TAINTED.
    - Callback/handler parameters (e.g., void* context in signals/events): ALWAYS TAINTED.

- Internal Flow:
    - Private/static function parameters: Trace origin if visible. If origin is unknown/invisible, mark as POTENTIALLY TAINTED (Lower Priority).

- External Data:
    - Data reads from Files, Network, Environment Variables: ALWAYS TAINTED.

- Library Interaction (Crucial Distinction):
    - Content Retrieval: Functions reading raw data (e.g., recv, fread, getenv returns) -> ALWAYS TAINTED.
    - Metadata/Helpers: Return values indicating status/length (e.g., strlen, vector::size, read return code) -> Treat as SAFE unless used in arithmetic leading to Integer Overflow.

# Benign Whitelist (When to SKIP reporting)
You may skip a line if it satisfies ANY of the following conditions (Logical OR):

- Imports & Namespaces:
    - Standard #include, import, or using statements.
    - Exception: Report includes of suspicious local headers.

- Compile-Time Constants:
    - Variable declarations initialized with pure literals (e.g., const int MAX = 100;).
    - Constraint: Must NOT be a pointer to potential dynamic memory or function calls.

- Safe Logging:
    - Logging calls containing string literals only.
    - Logging of primitive types (int, bool) that are clearly not tainted sources.

- Structural & Inert Code:
    - Braces, comments, blank lines, preprocessor guards.
    - Pure control flow keywords: break;, continue;, return 0;, else.

- Pure Type Definitions (Strict Mode):
    - struct, class, enum, or typedef declarations.
    - CRITICAL EXCEPTION: Do NOT skip if the definition contains inline constructors/destructors logic or in-class member initializers calling functions.

Safety Catch: If a line involves function calls, pointer arithmetic, or array indexing and is NOT explicitly in this list -> REPORT IT (Do not skip).

# Confidence Guidelines
Assign a confidence_score (0.1 to 1.0):
- 0.8 - 1.0: Clear, well-known vulnerability pattern
- 0.5 - 0.79: Risky pattern whose safety depends on context not visible in the snippet
- 0.2 - 0.49: Potential bad practice, subtle risk, or theoretical vulnerability
- 0.1 - 0.19: Very weak signal, but worth noting for completeness

ALL confidence levels MUST be reported. Never filter out low-confidence findings.

# Multi-line & Data Flow Vulnerabilities
Some vulnerabilities span multiple lines. For such cases:
- Report the MOST CRITICAL line (typically the sink / dangerous operation) as the primary entry
- In the suspicion_reason, explicitly mention related source and intermediate lines by their line numbers
- This enables downstream agents to construct richer context via slicing from multiple points

# Output Calibration
For a typical code snippet of 100-200 lines, expect approximately 5-15 suspicious lines. Significantly fewer findings may indicate overly conservative analysis; significantly more may indicate flagging benign code. Use this as a soft reference, NOT a hard constraint.

# Output Format (STRICT)
<reasoning>
## Phase 1: Global Scan
[Identify ALL input sources, critical data flow paths, and dangerous sink points.]

## Phase 2: Line-by-Line Findings
[For each suspicious line, provide a brief rationale.]
</reasoning>
```json
[
  {
    "line_number": <int>,
    "code_line": "<the exact suspicious line of code>",
    "suspicion_reason": "<specific reason with line references>",
    "confidence_score": <float between 0.1 and 1.0>
  }
]
```
\end{lstlisting}

\subsection{Context-Augmentation Agent (Phase II)}
\label{appendix:context}

The Context-Augmentation Agent does not use a system prompt. Instead, it receives a templated user prompt for each external expansion decision. The template is shown below, where placeholders in braces are filled at runtime with the relevant file path, line number, code context, and candidate function information.

\begin{lstlisting}[basicstyle=\ttfamily\scriptsize, breaklines=true, frame=single, backgroundcolor=\color{gray!5}, columns=fullflexible, keepspaces=true, xleftmargin=2pt, xrightmargin=2pt]
You are performing a security analysis.

[Section 1: Target Issue]
File: {file_path}
Line: {line_number}
Code: `{code_line}`
Reason: {suspicion_reason}

[Section 2: Current Analysis Context]
Here is the code context and data flow traces retrieved from all relevant files so far:

{current_context}

[Section 3: Decision Point]
We encountered a call to external function `{target_func_name}` defined in `{target_file_path}`.

[Question]
Based on the data flow shown above, is it necessary to inspect `{target_func_name}` to confirm the vulnerability?

Answer strictly with "YES" or "NO".
\end{lstlisting}

\subsection{Verification Agent (Phase III)}
\label{appendix:verify}

\begin{lstlisting}[basicstyle=\ttfamily\scriptsize, breaklines=true, frame=single, backgroundcolor=\color{gray!5}, columns=fullflexible, keepspaces=true, xleftmargin=2pt, xrightmargin=2pt]
# Role
You are a Senior Security Verification Analyst operating as the deep analysis stage in a vulnerability detection pipeline. An upstream triage scanner has flagged suspicious code lines with preliminary reasons. Your task is to rigorously evaluate each flagged line against its repo-level code context to determine whether it represents a true exploitable vulnerability or a false positive, through structured adversarial reasoning.

# Understanding Your Input
1. Suspicious Code Line: The specific line flagged by the upstream triage scanner.
2. Suspicion Reason: The scanner's preliminary rationale. Treat this as a hypothesis to verify, not a conclusion.
3. Project-level Code Context: Surrounding code from the execution path, constructed via CPG. Includes annotated markers: [FUNCTION ENTRY], [TARGET], and relevant execution path lines.
4. Data Flow Trace: Variable propagation chain with markers: [SOURCE], [PROPAGATION], [SINK], [ALIAS], [COND], [CALL], [TARGET].

Context Completeness Assumption: The provided trace represents the relevant execution path. If a security check or mitigation is not visible in the trace, assume it does not exist on this path.

# Analysis Framework: Adversarial Verification

## Phase 1: Comprehension
Establish the factual foundation before any judgment:
- What does this code do?
- What is the data origin? Is the source trusted or untrusted?
- What would make this dangerous?
- What mitigation would make this safe?

## Phase 2: Devil's Advocate -- Adversarial Debate

### 2A: The Case for VULNERABLE (Red Team)
Assume this IS a vulnerability. Build the strongest possible attack argument:
- What specific input or condition would trigger the vulnerability?
- What existing checks can be bypassed, and how?
- What is the concrete security impact?
- Cite specific lines from the trace as evidence.

### 2B: The Case for NOT_VULNERABLE (Blue Team)
Assume this is NOT a vulnerability. Build the strongest possible defense argument:
- What mitigations are present in the trace? Cite specific lines.
- Are there implicit safety guarantees from the API, type system, or language semantics?
- Is the data source actually trusted despite appearing tainted?
- Would the "attack" require conditions that are practically impossible?

### 2C: Verdict Adjudication
Compare the two arguments head-to-head:
- Which side has concrete evidence from the trace vs. speculation?
- Are the Red Team's attack conditions realistic?
- Are the Blue Team's defenses actually on the execution path?
- What is the weakest link in each argument?

## Phase 3: Final Verdict
- VULNERABLE: Red Team argument is supported by concrete evidence; Blue Team cannot demonstrate sufficient mitigation.
- NOT_VULNERABLE: Blue Team demonstrates clear mitigation; or Red Team's attack scenario requires unrealistic conditions.

Confidence Calibration:
- 0.85 - 1.0: One side's argument is overwhelmingly stronger.
- 0.65 - 0.84: One side is stronger but the other has partial merit.
- 0.45 - 0.64: Gray zone -- both sides have comparable arguments.
- 0.25 - 0.44: Weak signal -- limited trace information.
- Below 0.25: Essentially guessing.

# Key Principles
1. Evidence Over Intuition: Every claim must reference specific lines or trace entries.
2. Context Over Pattern: Never flag on pattern alone -- always verify context.
3. Trace Is Ground Truth: If a check is not in the trace, it does not exist on this path.
4. Library Semantics Matter: Use knowledge of documented library behavior. Note uncertainty explicitly.
5. Anti-Anchoring: The upstream suspicion reason is a starting hint, not a diagnosis.

# Output Format
<thinking>
## Phase 1: Comprehension
[Factual summary]

## Phase 2A: Case for VULNERABLE (Red Team)
[Strongest attack argument with line references]

## Phase 2B: Case for NOT_VULNERABLE (Blue Team)
[Strongest defense argument with line references]

## Phase 2C: Verdict Adjudication
[Head-to-head comparison]

## Phase 3: Final Verdict
[Verdict and confidence reasoning]
</thinking>
```json
{
  "verdict": "VULNERABLE or NOT_VULNERABLE",
  "confidence": <float>,
  "cwe_id": "CWE-XXX or null",
  "vulnerability_type": "Brief description or null",
  "key_evidence": "Most decisive evidence with line citation"
}
```
\end{lstlisting}

\subsection{Audit Agent (Phase IV)}
\label{appendix:audit}

\begin{lstlisting}[basicstyle=\ttfamily\scriptsize, breaklines=true, frame=single, backgroundcolor=\color{gray!5}, columns=fullflexible, keepspaces=true, xleftmargin=2pt, xrightmargin=2pt]
# Role
You are a Senior Security Audit Analyst. An upstream verification agent has analyzed a suspicious code line against its repo-level context and produced a verdict (VULNERABLE or NOT_VULNERABLE) with detailed adversarial reasoning. You are tasked with conducting an independent, impartial second review.

# Mission: IMPARTIAL REASONING AUDIT
- Philosophy: You are a neutral judge with no preference for either outcome. Your only commitment is to evidence-based correctness.
- Scope: You audit the verifier's reasoning process AND independently evaluate the code and trace.
- Standard: A verdict is correct if and only if it is supported by concrete evidence in the provided code context, data flow trace, and well-established knowledge of programming languages, standard libraries, and widely-used frameworks.

# Understanding Your Input

## Part 1: Original Analysis Input
1. Suspicious Code Line: The flagged line with file location and line number.
2. Suspicion Reason: The upstream scanner's preliminary rationale.
3. Project-level Code Context: Surrounding code from the execution path, constructed via CPG.
4. Data Flow Trace: Variable propagation chain with step-by-step markers.

## Part 2: Verifier's Reasoning
The verifier's complete analysis including Phase 1 (Comprehension), Phase 2A (Red Team), Phase 2B (Blue Team), Phase 2C (Adjudication), Phase 3 (Final Verdict), and JSON output.

# Audit Framework

## Step 1: Independent Comprehension
Before examining the verifier's reasoning, form your own understanding:
- What does the target code do?
- What are the data sources and trust levels?
- What security property could be violated?
- What specific mitigation would be required for safety?

## Step 2: Evidence Cross-Check
Verify that the verifier's factual claims match the actual code and trace:
- Does every line reference actually exist in the provided context?
- Do the cited lines actually contain what the verifier claims?
- Are there relevant lines the verifier overlooked?

## Step 3: Reasoning Quality Audit
Evaluate both arguments for common reasoning flaws:

Flaws that INCORRECTLY support VULNERABLE:
- Speculation Flaw: Asserting vulnerability based on what might happen outside the trace.
- Pattern-Matching Flaw: Flagging a dangerous pattern without verifying exploitability.
- Anchoring Flaw: Shaped by upstream suspicion rather than independent analysis.
- Absence-as-Evidence Flaw: Concluding VULNERABLE primarily because no mitigation is visible.

Flaws that INCORRECTLY support NOT_VULNERABLE:
- Phantom Mitigation Flaw: Citing a security check that doesn't exist in the trace.
- Over-Trust Flaw: Assuming callers validate input without evidence.
- Incomplete Protection Flaw: Citing mitigation that only partially addresses the risk.

Flaws that can support either direction:
- Semantic Misunderstanding Flaw: Misinterpreting API behavior or language semantics.
- Scope Creep Flaw: Arguing about code outside the provided execution path.
- Evidence Fabrication Flaw: Claiming something exists in the trace when it does not.

## Step 4: Independent Verdict
- AGREE: Verifier's reasoning is sound and well-supported. Final verdict matches original.
- DISAGREE: A specific, material reasoning flaw changes the conclusion. You MUST list at least one flaw that directly supports overturning the verdict. Final verdict is the opposite.
- DEFER: Concerns exist but no concrete flaw materially changes the outcome. Original verdict preserved.

Decision Standard:
- VULNERABLE requires: a concrete, traceable attack path with no sufficient mitigation visible.
- NOT_VULNERABLE requires: concrete mitigation visible, or attack conditions are impossible.

# Evidence Rules

Permitted Evidence:
- Code in trace: Anything explicitly present in the provided context and data flow trace.
- Well-established language/library/framework knowledge.

Prohibited Evidence:
- Assumed caller behavior, assumed runtime configuration, assumed unseen code, unverifiable library claims.

Handling Incomplete Traces:
An incomplete trace is neutral. If evidence is genuinely ambiguous and you cannot identify a specific reasoning flaw, use DEFER.

# Output Format
<audit_reasoning>
## Independent Comprehension
[Your own understanding, formed before reviewing verifier's reasoning]

## Evidence Cross-Check
[Verify verifier's factual claims against actual code and trace]

## Reasoning Quality Audit
[Identify specific reasoning flaws in BOTH arguments]

## Independent Verdict
[Your conclusion and whether it agrees with the verifier]
</audit_reasoning>
```json
{
  "audit_verdict": "AGREE or DISAGREE or DEFER",
  "original_verdict": "VULNERABLE or NOT_VULNERABLE",
  "final_verdict": "VULNERABLE or NOT_VULNERABLE",
  "confidence": <float>,
  "audit_rationale": "One-sentence summary",
  "reasoning_flaws_found": ["list of flaw types, or empty"]
}
```
\end{lstlisting}

\newpage
\section{Case Study: End-to-End AEGIS Pipeline Execution}
\label{app:case_study}

To better illustrate the dynamic reasoning process of AEGIS, this appendix provides a complete, end-to-end execution example of our pipeline. This case study demonstrates how a local code anomaly is identified, how its cross-function context is dynamically augmented, and how the dialectical verification and meta-auditing agents reach a final verdict.

\subsection{Target Input}
\textbf{Commit URL:} \url{https://git.kernel.org/cgit/linux/kernel/git/davem/net.git/commit/?id=7892032cfe67f4bde6fc2ee967e45a8fbaf33756} \\
\textbf{Target File:} \texttt{net/ipv6/ip6\_gre.c} \\
\textbf{Target Function:} \texttt{ip6gre\_err}

\begin{lstlisting}[language=C, basicstyle=\ttfamily\scriptsize, breaklines=true, frame=single, backgroundcolor=\color{gray!5}, columns=fullflexible, keepspaces=true, xleftmargin=2pt, xrightmargin=2pt]
static void ip6gre_err(struct sk_buff *skb, struct inet6_skb_parm *opt,
                u8 type, u8 code, int offset, __be32 info)
{
        const struct ipv6hdr *ipv6h = (const struct ipv6hdr *)skb->data;
        __be16 *p = (__be16 *)(skb->data + offset);
        int grehlen = offset + 4;
        struct ip6_tnl *t;
        __be16 flags;

        flags = p[0];
        if (flags&(GRE_CSUM|GRE_KEY|GRE_SEQ|GRE_ROUTING|GRE_VERSION)) {
                if (flags&(GRE_VERSION|GRE_ROUTING))
                        return;
                if (flags&GRE_KEY) {
                        grehlen += 4;
                        if (flags&GRE_CSUM)
                                grehlen += 4;
                }
        }

        /* If only 8 bytes returned, keyed message will be dropped here */
        if (!pskb_may_pull(skb, grehlen))
                return;
        ipv6h = (const struct ipv6hdr *)skb->data;
        p = (__be16 *)(skb->data + offset);

        t = ip6gre_tunnel_lookup(skb->dev, &ipv6h->daddr, &ipv6h->saddr,
                                flags & GRE_KEY ?
                                *(((__be32 *)p) + (grehlen / 4) - 1) : 0,
                                p[1]);
        if (!t)
                return;

        switch (type) {
                __u32 teli;
                struct ipv6_tlv_tnl_enc_lim *tel;
                __u32 mtu;
        case ICMPV6_DEST_UNREACH:
                net_dbg_ratelimited("%s: Path to destination invalid or inactive!\n",
                                    t->parms.name);
                break;
        case ICMPV6_TIME_EXCEED:
                if (code == ICMPV6_EXC_HOPLIMIT) {
                        net_dbg_ratelimited("%s: Too small hop limit or routing loop in tunnel!\n",
                                            t->parms.name);
                }
                break;
        case ICMPV6_PARAMPROB:
                teli = 0;
                if (code == ICMPV6_HDR_FIELD)
                        teli = ip6_tnl_parse_tlv_enc_lim(skb, skb->data);

                if (teli && teli == be32_to_cpu(info) - 2) {
                        tel = (struct ipv6_tlv_tnl_enc_lim *) &skb->data[teli];
                        if (tel->encap_limit == 0) {
                                net_dbg_ratelimited("%s: Too small encapsulation limit or routing loop in tunnel!\n",
                                                    t->parms.name);
                        }
                } else {
                        net_dbg_ratelimited("%s: Recipient unable to parse tunneled packet!\n",
                                            t->parms.name);
                }
                break;
        case ICMPV6_PKT_TOOBIG:
                mtu = be32_to_cpu(info) - offset;
                if (mtu < IPV6_MIN_MTU)
                        mtu = IPV6_MIN_MTU;
                t->dev->mtu = mtu;
                break;
        }

        if (time_before(jiffies, t->err_time + IP6TUNNEL_ERR_TIMEO))
                t->err_count++;
        else
                t->err_count = 1;
        t->err_time = jiffies;
}
\end{lstlisting}

\subsection{Phase I: Clue Discovery}
Operating under the Worst-Case Taint Assumption, the Clue-Discovery Agent scans the target function in isolation and identifies the following suspicious anomalies:

\begin{agentbox}[Clue-Discovery Agent Output]{blue}
\textbf{Clue C1:}
\begin{itemize}
    \item \textbf{Line Number:} 397
    \item \textbf{Code Line:} \texttt{*(((\_\_be32 *)p) + (grehlen / 4) - 1) : 0,}
    \item \textbf{Reason:} Complex pointer arithmetic with division; 'grehlen' derived from network data; potential miscalculation leading to out-of-bounds memory access.
    \item \textbf{Confidence Score:} 0.8
\end{itemize}

\textbf{Clue C2:}
\begin{itemize}
    \item \textbf{Line Number:} 373
    \item \textbf{Code Line:} \texttt{\_\_be16 *p = (\_\_be16 *)(skb->data + offset);}
    \item \textbf{Reason:} Pointer arithmetic with 'offset' parameter (potentially tainted) and 'skb->data' (network data); no bounds checking; potential out-of-bounds access.
    \item \textbf{Confidence Score:} 0.7
\end{itemize}
\end{agentbox}

\subsection{Phase II: Graph-Guided Context Augmentation}
Triggered by the boundary variables and external calls in \textbf{Clue C2}, the Context-Augmentation Agent dynamically stitches the CPG to expand the cross-file dependencies. The reconstructed repository-level context is presented in two parts: the expanded code blocks retrieved via CPG slicing, and the resulting structured data flow trace.

\vspace{0.5em}
\noindent\textbf{Part 1: Expanded Code Context}

\begin{lstlisting}[language=C, basicstyle=\ttfamily\scriptsize, breaklines=true, frame=single, backgroundcolor=\color{gray!5}, columns=fullflexible, keepspaces=true, xleftmargin=2pt, xrightmargin=2pt]
static void ip6gre_err(struct sk_buff *skb, struct inet6_skb_parm *opt,  // [FUNCTION ENTRY]
                u8 type, u8 code, int offset, __be32 info)
        const struct ipv6hdr *ipv6h = (const struct ipv6hdr *)skb->data;
        __be16 *p = (__be16 *)(skb->data + offset);  // [TARGET]
        int grehlen = offset + 4;
        flags = p[0];
        if (flags&(GRE_CSUM|GRE_KEY|GRE_SEQ|GRE_ROUTING|GRE_VERSION)) {
                if (flags&(GRE_VERSION|GRE_ROUTING))
                if (flags&GRE_KEY) {
                        grehlen += 4;
                        if (flags&GRE_CSUM)
                                grehlen += 4;
        if (!pskb_may_pull(skb, grehlen))  // [CROSS-FILE CALL]
        ipv6h = (const struct ipv6hdr *)skb->data;
        p = (__be16 *)(skb->data + offset);
        t = ip6gre_tunnel_lookup(skb->dev, &ipv6h->daddr, &ipv6h->saddr,
                                flags & GRE_KEY ?
                                *(((__be32 *)p) + (grehlen / 4) - 1) : 0,
                                p[1]);
        if (!t)
        switch (type) {
                                    t->parms.name);
                if (code == ICMPV6_EXC_HOPLIMIT) {
                                            t->parms.name);
                if (code == ICMPV6_HDR_FIELD)
                        teli = ip6_tnl_parse_tlv_enc_lim(skb, skb->data);  // [CROSS-FILE CALL]
                if (teli && teli == be32_to_cpu(info) - 2) {
                        tel = (struct ipv6_tlv_tnl_enc_lim *) &skb->data[teli];
                        if (tel->encap_limit == 0) {
                                                    t->parms.name);
                                            t->parms.name);
                mtu = be32_to_cpu(info) - offset;
                if (mtu < IPV6_MIN_MTU)
                        mtu = IPV6_MIN_MTU;
                t->dev->mtu = mtu;
        if (time_before(jiffies, t->err_time + IP6TUNNEL_ERR_TIMEO))
                t->err_count++;
                t->err_count = 1;
        t->err_time = jiffies;
\end{lstlisting}

\begin{lstlisting}[language=C, basicstyle=\ttfamily\scriptsize, breaklines=true, frame=single, backgroundcolor=\color{gray!5}, columns=fullflexible, keepspaces=true, xleftmargin=2pt, xrightmargin=2pt]
__u16 ip6_tnl_parse_tlv_enc_lim(struct sk_buff *skb, __u8 *raw)  // [FUNCTION ENTRY]
 {
        const struct ipv6hdr *ipv6h = (const struct ipv6hdr *)raw;
        unsigned int nhoff = raw - skb->data;
        unsigned int off = nhoff + sizeof(*ipv6h);
        u8 next, nexthdr = ipv6h->nexthdr;
 
        while (ipv6_ext_hdr(nexthdr) && nexthdr != NEXTHDR_NONE) {
                struct ipv6_opt_hdr *hdr;
                u16 optlen;
 
                if (!pskb_may_pull(skb, off + sizeof(*hdr)))
                        break;
 
                hdr = (struct ipv6_opt_hdr *)(skb->data + off);
                if (nexthdr == NEXTHDR_FRAGMENT) {
                        struct frag_hdr *frag_hdr = (struct frag_hdr *) hdr;
                        if (frag_hdr->frag_off)
                                break;
                        optlen = 8;
                } else if (nexthdr == NEXTHDR_AUTH) {
                        optlen = (hdr->hdrlen + 2) << 2;
                } else {
                        optlen = ipv6_optlen(hdr);
                }
                /* cache hdr->nexthdr, since pskb_may_pull() might
                 * invalidate hdr
                 */
                next = hdr->nexthdr;
                if (nexthdr == NEXTHDR_DEST) {
                        u16 i = 2;
 
                        /* Remember : hdr is no longer valid at this point. */
                        if (!pskb_may_pull(skb, off + optlen))
                                break;
 
                        while (1) {
                                struct ipv6_tlv_tnl_enc_lim *tel;
 
                                /* No more room for encapsulation limit */
                                if (i + sizeof(*tel) > optlen)
                                        break;
                                tel = (struct ipv6_tlv_tnl_enc_lim *)(skb->data + off + i);
                                if (tel->type == IPV6_TLV_TNL_ENCAP_LIMIT &&
                                    tel->length == 1)
                                        return i + off - nhoff;  // [CROSS-FILE CALL]
                                if (tel->type)
                                        i += tel->length + 2;
                                        i++;
                nexthdr = next;
                off += optlen;
        return 0;  // [CROSS-FILE CALL]
\end{lstlisting}

\begin{lstlisting}[language=C, basicstyle=\ttfamily\scriptsize, breaklines=true, frame=single, backgroundcolor=\color{gray!5}, columns=fullflexible, keepspaces=true, xleftmargin=2pt, xrightmargin=2pt]
static inline unsigned int skb_headlen(const struct sk_buff *skb)  // [FUNCTION ENTRY]
 {
        return skb->len - skb->data_len;
 }
 static inline int pskb_may_pull(struct sk_buff *skb, unsigned int len)  // [FUNCTION ENTRY]
 {
        if (likely(len <= skb_headlen(skb)))
                return 1;
        if (unlikely(len > skb->len))
                return 0;
        return __pskb_pull_tail(skb, len - skb_headlen(skb)) != NULL;
 }
\end{lstlisting}

\vspace{0.5em}
\noindent\textbf{Part 2: Reconstructed Data Flow Trace}

\begin{agentbox}[By-Variable Data Flow Trace]{green}

\ttfamily\scriptsize

\textbf{Variable: \texttt{p}} \\
\textcolor[RGB]{0,0,255}{[SOURCE]} ipv6/ip6\_gre.c:369 (\texttt{`static void ip6gre\_err(struct sk\_buff *skb, struct inet6\_skb\_parm *opt,`}) \\
\textcolor[RGB]{105,105,105}{[PROP]} ipv6/ip6\_gre.c:370 (\texttt{`u8 type, u8 code, int offset, \_\_be32 info)`}) \\
\textcolor[RGB]{255,0,0}{[TARGET]} ipv6/ip6\_gre.c:373 (\texttt{`\_\_be16 *p = (\_\_be16 *)(skb->data + offset);`}) \\
\textcolor[RGB]{105,105,105}{[PROP]} ipv6/ip6\_gre.c:395 (\texttt{`t = ip6gre\_tunnel\_lookup(skb->dev, \&ipv6h->daddr, \&ipv6h->saddr,`}) \\
\textcolor[RGB]{105,105,105}{[PROP]} ipv6/ip6\_gre.c:396 (\texttt{`flags \& GRE\_KEY ?`}) \\
\textcolor[RGB]{105,105,105}{[PROP]} ipv6/ip6\_gre.c:397 (\texttt{`*(((\_\_be32 *)p) + (grehlen / 4) - 1) : 0,`}) \\
\textcolor[RGB]{105,105,105}{[PROP]} ipv6/ip6\_gre.c:398 (\texttt{`p[1]);`}) \\
\textcolor[RGB]{184,134,11}{[COND]} ipv6/ip6\_gre.c:399 (\texttt{`if (!t)`}) \\
\textcolor[RGB]{0,128,0}{[CALL]} ipv6/ip6\_gre.c:408 (\texttt{`t->parms.name);`}) \\
\textcolor[RGB]{184,134,11}{[COND]} ipv6/ip6\_tunnel.c:412 (\texttt{`if (!pskb\_may\_pull(skb, off + sizeof(*hdr)))`}) \\
\textcolor[RGB]{0,128,0}{[CALL]} ipv6/ip6\_gre.c:413 (\texttt{`t->parms.name);`}) \\
\textcolor[RGB]{0,128,0}{[CALL]} ipv6/ip6\_tunnel.c:415 (\texttt{`hdr = (struct ipv6\_opt\_hdr *)(skb->data + off);`}) \\
\textcolor[RGB]{0,128,0}{[CALL]} ipv6/ip6\_tunnel.c:417 (\texttt{`struct frag\_hdr *frag\_hdr = (struct frag\_hdr *) hdr;`}) \\
\textcolor[RGB]{184,134,11}{[COND]} ipv6/ip6\_tunnel.c:418 (\texttt{`if (frag\_hdr->frag\_off)`}) \\
\textcolor[RGB]{0,128,0}{[CALL]} ipv6/ip6\_gre.c:419 (\texttt{`teli = ip6\_tnl\_parse\_tlv\_enc\_lim(skb, skb->data);`}) \\
\textcolor[RGB]{0,128,0}{[CALL]} ipv6/ip6\_tunnel.c:420 (\texttt{`optlen = 8;`}) \\
\textcolor[RGB]{0,128,0}{[CALL]} ipv6/ip6\_tunnel.c:421 (\texttt{`\} else if (nexthdr == NEXTHDR\_AUTH) \{`}) \\
\textcolor[RGB]{184,134,11}{[COND]} ipv6/ip6\_gre.c:421 (\texttt{`if (teli \&\& teli == be32\_to\_cpu(info) - 2) \{`}) \\
\textcolor[RGB]{0,128,0}{[CALL]} ipv6/ip6\_tunnel.c:422 (\texttt{`optlen = (hdr->hdrlen + 2) << 2;`}) \\
\textcolor[RGB]{0,128,0}{[CALL]} ipv6/ip6\_gre.c:422 (\texttt{`tel = (struct ipv6\_tlv\_tnl\_enc\_lim *) \&skb->data[teli];`}) \\
\textcolor[RGB]{184,134,11}{[COND]} ipv6/ip6\_gre.c:423 (\texttt{`if (tel->encap\_limit == 0) \{`}) \\
\textcolor[RGB]{0,128,0}{[CALL]} ipv6/ip6\_tunnel.c:424 (\texttt{`optlen = ipv6\_optlen(hdr);`}) \\
\textcolor[RGB]{0,128,0}{[CALL]} ipv6/ip6\_gre.c:425 (\texttt{`t->parms.name);`}) \\
\textcolor[RGB]{0,128,0}{[CALL]} ipv6/ip6\_tunnel.c:429 (\texttt{`next = hdr->nexthdr;`}) \\
\textcolor[RGB]{0,128,0}{[CALL]} ipv6/ip6\_gre.c:429 (\texttt{`t->parms.name);`}) \\
\textcolor[RGB]{184,134,11}{[COND]} ipv6/ip6\_tunnel.c:430 (\texttt{`if (nexthdr == NEXTHDR\_DEST) \{`}) \\
\textcolor[RGB]{184,134,11}{[COND]} ipv6/ip6\_tunnel.c:434 (\texttt{`if (!pskb\_may\_pull(skb, off + optlen))`}) \\
\textcolor[RGB]{0,128,0}{[CALL]} ipv6/ip6\_gre.c:436 (\texttt{`t->dev->mtu = mtu;`}) \\
\textcolor[RGB]{184,134,11}{[COND]} ipv6/ip6\_gre.c:440 (\texttt{`if (time\_before(jiffies, t->err\_time + IP6TUNNEL\_ERR\_TIMEO))`}) \\
\textcolor[RGB]{0,128,0}{[CALL]} ipv6/ip6\_gre.c:441 (\texttt{`t->err\_count++;`}) \\
\textcolor[RGB]{105,105,105}{[PROP]} ipv6/ip6\_gre.c:443 (\texttt{`t->err\_count = 1;`}) \\
\textcolor[RGB]{0,128,0}{[CALL]} ipv6/ip6\_tunnel.c:444 (\texttt{`tel = (struct ipv6\_tlv\_tnl\_enc\_lim *)(skb->data + off + i);`}) \\
\textcolor[RGB]{105,105,105}{[PROP]} ipv6/ip6\_gre.c:444 (\texttt{`t->err\_time = jiffies;`}) \\
\textcolor[RGB]{184,134,11}{[COND]} ipv6/ip6\_tunnel.c:446 (\texttt{`if (tel->type == IPV6\_TLV\_TNL\_ENCAP\_LIMIT \&\&`}) \\
\textcolor[RGB]{0,128,0}{[CALL]} ipv6/ip6\_tunnel.c:447 (\texttt{`tel->length == 1)`}) \\
\textcolor[RGB]{0,128,0}{[CALL]} ipv6/ip6\_tunnel.c:448 (\texttt{`return i + off - nhoff;`}) \\
\textcolor[RGB]{184,134,11}{[COND]} ipv6/ip6\_tunnel.c:450 (\texttt{`if (tel->type)`}) \\
\textcolor[RGB]{0,128,0}{[CALL]} ipv6/ip6\_tunnel.c:451 (\texttt{`i += tel->length + 2;`}) \\
\textcolor[RGB]{0,128,0}{[CALL]} ipv6/ip6\_tunnel.c:453 (\texttt{`i++;`}) \\
\textbf{Variable: \texttt{skb}} \\
\textcolor[RGB]{0,0,255}{[SOURCE]} ipv6/ip6\_gre.c:369 (\texttt{`static void ip6gre\_err(struct sk\_buff *skb, struct inet6\_skb\_parm *opt,`}) \\
\textcolor[RGB]{255,140,0}{[RET]} ipv6/ip6\_tunnel.c:401 (\texttt{`\_\_u16 ip6\_tnl\_parse\_tlv\_enc\_lim(struct sk\_buff *skb, \_\_u8 *raw)`}) \\
\textcolor[RGB]{0,128,0}{[CALL]} ipv6/ip6\_tunnel.c:404 (\texttt{`unsigned int nhoff = raw - skb->data;`}) \\
\textcolor[RGB]{0,128,0}{[CALL]} ipv6/ip6\_tunnel.c:405 (\texttt{`unsigned int off = nhoff + sizeof(*ipv6h);`}) \\
\textcolor[RGB]{184,134,11}{[COND]} ipv6/ip6\_tunnel.c:408 (\texttt{`while (ipv6\_ext\_hdr(nexthdr) \&\& nexthdr != NEXTHDR\_NONE) \{`}) \\
\textcolor[RGB]{184,134,11}{[COND]} ipv6/ip6\_tunnel.c:412 (\texttt{`if (!pskb\_may\_pull(skb, off + sizeof(*hdr)))`}) \\
\textcolor[RGB]{0,128,0}{[CALL]} ipv6/ip6\_tunnel.c:415 (\texttt{`hdr = (struct ipv6\_opt\_hdr *)(skb->data + off);`}) \\
\textcolor[RGB]{184,134,11}{[COND]} ipv6/ip6\_tunnel.c:416 (\texttt{`if (nexthdr == NEXTHDR\_FRAGMENT) \{`}) \\
\textcolor[RGB]{0,128,0}{[CALL]} ipv6/ip6\_tunnel.c:417 (\texttt{`struct frag\_hdr *frag\_hdr = (struct frag\_hdr *) hdr;`}) \\
\textcolor[RGB]{184,134,11}{[COND]} ipv6/ip6\_tunnel.c:418 (\texttt{`if (frag\_hdr->frag\_off)`}) \\
\textcolor[RGB]{0,128,0}{[CALL]} ipv6/ip6\_tunnel.c:422 (\texttt{`optlen = (hdr->hdrlen + 2) << 2;`}) \\
\textcolor[RGB]{0,128,0}{[CALL]} ipv6/ip6\_tunnel.c:424 (\texttt{`optlen = ipv6\_optlen(hdr);`}) \\
\textcolor[RGB]{0,128,0}{[CALL]} ipv6/ip6\_tunnel.c:429 (\texttt{`next = hdr->nexthdr;`}) \\
\textcolor[RGB]{184,134,11}{[COND]} ipv6/ip6\_tunnel.c:434 (\texttt{`if (!pskb\_may\_pull(skb, off + optlen))`}) \\
\textcolor[RGB]{184,134,11}{[COND]} ipv6/ip6\_tunnel.c:441 (\texttt{`if (i + sizeof(*tel) > optlen)`}) \\
\textcolor[RGB]{0,128,0}{[CALL]} ipv6/ip6\_tunnel.c:444 (\texttt{`tel = (struct ipv6\_tlv\_tnl\_enc\_lim *)(skb->data + off + i);`}) \\
\textcolor[RGB]{184,134,11}{[COND]} ipv6/ip6\_tunnel.c:446 (\texttt{`if (tel->type == IPV6\_TLV\_TNL\_ENCAP\_LIMIT \&\&`}) \\
\textcolor[RGB]{0,128,0}{[CALL]} ipv6/ip6\_tunnel.c:447 (\texttt{`tel->length == 1)`}) \\
\textcolor[RGB]{0,128,0}{[CALL]} ipv6/ip6\_tunnel.c:448 (\texttt{`return i + off - nhoff;`}) \\
\textcolor[RGB]{184,134,11}{[COND]} ipv6/ip6\_tunnel.c:450 (\texttt{`if (tel->type)`}) \\
\textcolor[RGB]{0,128,0}{[CALL]} ipv6/ip6\_tunnel.c:451 (\texttt{`i += tel->length + 2;`}) \\
\textcolor[RGB]{0,128,0}{[CALL]} ipv6/ip6\_tunnel.c:456 (\texttt{`nexthdr = next;`}) \\
\textcolor[RGB]{0,128,0}{[CALL]} ipv6/ip6\_tunnel.c:457 (\texttt{`off += optlen;`}) \\
\textcolor[RGB]{255,140,0}{[RET]} linux/skbuff.h:1804 (\texttt{`static inline unsigned int skb\_headlen(const struct sk\_buff *skb)`}) \\
\textcolor[RGB]{0,128,0}{[CALL]} linux/skbuff.h:1806 (\texttt{`return skb->len - skb->data\_len;`}) \\
\textcolor[RGB]{255,140,0}{[RET]} linux/skbuff.h:1967 (\texttt{`static inline int pskb\_may\_pull(struct sk\_buff *skb, unsigned int len)`}) \\
\textcolor[RGB]{184,134,11}{[COND]} linux/skbuff.h:1969 (\texttt{`if (likely(len <= skb\_headlen(skb)))`}) \\
\textcolor[RGB]{184,134,11}{[COND]} linux/skbuff.h:1971 (\texttt{`if (unlikely(len > skb->len))`}) \\
\textcolor[RGB]{0,128,0}{[CALL]} linux/skbuff.h:1973 (\texttt{`return \_\_pskb\_pull\_tail(skb, len - skb\_headlen(skb)) != NULL;`}) \\
\textbf{Variable: \texttt{data}} \\
\textcolor[RGB]{255,0,0}{[TARGET]} ipv6/ip6\_gre.c:373 (\texttt{`\_\_be16 *p = (\_\_be16 *)(skb->data + offset);`}) \\
\textcolor[RGB]{105,105,105}{[PROP]} ipv6/ip6\_tunnel.c:415 (\texttt{`hdr = (struct ipv6\_opt\_hdr *)(skb->data + off);`}) \\
\textcolor[RGB]{105,105,105}{[PROP]} ipv6/ip6\_tunnel.c:444 (\texttt{`tel = (struct ipv6\_tlv\_tnl\_enc\_lim *)(skb->data + off + i);`}) \\
\textbf{Variable: \texttt{offset}} \\
\textcolor[RGB]{0,0,255}{[SOURCE]} ipv6/ip6\_gre.c:369 (\texttt{`static void ip6gre\_err(struct sk\_buff *skb, struct inet6\_skb\_parm *opt,`}) \\
\textcolor[RGB]{105,105,105}{[PROP]} ipv6/ip6\_gre.c:370 (\texttt{`u8 type, u8 code, int offset, \_\_be32 info)`}) \\
\textcolor[RGB]{0,128,0}{[CALL]} ipv6/ip6\_gre.c:393 (\texttt{`p = (\_\_be16 *)(skb->data + offset);`}) \\
\textcolor[RGB]{105,105,105}{[PROP]} ipv6/ip6\_gre.c:395 (\texttt{`t = ip6gre\_tunnel\_lookup(skb->dev, \&ipv6h->daddr, \&ipv6h->saddr,`}) \\
\textcolor[RGB]{105,105,105}{[PROP]} ipv6/ip6\_gre.c:396 (\texttt{`flags \& GRE\_KEY ?`}) \\
\textcolor[RGB]{105,105,105}{[PROP]} ipv6/ip6\_gre.c:397 (\texttt{`*(((\_\_be32 *)p) + (grehlen / 4) - 1) : 0,`}) \\
\textcolor[RGB]{105,105,105}{[PROP]} ipv6/ip6\_gre.c:398 (\texttt{`p[1]);`}) \\
\textcolor[RGB]{184,134,11}{[COND]} ipv6/ip6\_gre.c:399 (\texttt{`if (!t)`}) \\
\textcolor[RGB]{0,128,0}{[CALL]} ipv6/ip6\_gre.c:408 (\texttt{`t->parms.name);`}) \\
\textcolor[RGB]{0,128,0}{[CALL]} ipv6/ip6\_gre.c:413 (\texttt{`t->parms.name);`}) \\
\textcolor[RGB]{0,128,0}{[CALL]} ipv6/ip6\_tunnel.c:415 (\texttt{`hdr = (struct ipv6\_opt\_hdr *)(skb->data + off);`}) \\
\textcolor[RGB]{184,134,11}{[COND]} ipv6/ip6\_tunnel.c:416 (\texttt{`if (nexthdr == NEXTHDR\_FRAGMENT) \{`}) \\
\textcolor[RGB]{0,128,0}{[CALL]} ipv6/ip6\_tunnel.c:417 (\texttt{`struct frag\_hdr *frag\_hdr = (struct frag\_hdr *) hdr;`}) \\
\textcolor[RGB]{184,134,11}{[COND]} ipv6/ip6\_tunnel.c:418 (\texttt{`if (frag\_hdr->frag\_off)`}) \\
\textcolor[RGB]{0,128,0}{[CALL]} ipv6/ip6\_gre.c:419 (\texttt{`teli = ip6\_tnl\_parse\_tlv\_enc\_lim(skb, skb->data);`}) \\
\textcolor[RGB]{0,128,0}{[CALL]} ipv6/ip6\_tunnel.c:420 (\texttt{`optlen = 8;`}) \\
\textcolor[RGB]{0,128,0}{[CALL]} ipv6/ip6\_tunnel.c:421 (\texttt{`\} else if (nexthdr == NEXTHDR\_AUTH) \{`}) \\
\textcolor[RGB]{184,134,11}{[COND]} ipv6/ip6\_gre.c:421 (\texttt{`if (teli \&\& teli == be32\_to\_cpu(info) - 2) \{`}) \\
\textcolor[RGB]{0,128,0}{[CALL]} ipv6/ip6\_gre.c:422 (\texttt{`tel = (struct ipv6\_tlv\_tnl\_enc\_lim *) \&skb->data[teli];`}) \\
\textcolor[RGB]{0,128,0}{[CALL]} ipv6/ip6\_tunnel.c:422 (\texttt{`optlen = (hdr->hdrlen + 2) << 2;`}) \\
\textcolor[RGB]{0,128,0}{[CALL]} ipv6/ip6\_gre.c:422 (\texttt{`tel = (struct ipv6\_tlv\_tnl\_enc\_lim *) \&skb->data[teli];`}) \\
\textcolor[RGB]{184,134,11}{[COND]} ipv6/ip6\_gre.c:423 (\texttt{`if (tel->encap\_limit == 0) \{`}) \\
\textcolor[RGB]{0,128,0}{[CALL]} ipv6/ip6\_tunnel.c:424 (\texttt{`optlen = ipv6\_optlen(hdr);`}) \\
\textcolor[RGB]{0,128,0}{[CALL]} ipv6/ip6\_gre.c:425 (\texttt{`t->parms.name);`}) \\
\textcolor[RGB]{0,128,0}{[CALL]} ipv6/ip6\_tunnel.c:429 (\texttt{`next = hdr->nexthdr;`}) \\
\textcolor[RGB]{0,128,0}{[CALL]} ipv6/ip6\_gre.c:429 (\texttt{`t->parms.name);`}) \\
\textcolor[RGB]{184,134,11}{[COND]} ipv6/ip6\_tunnel.c:430 (\texttt{`if (nexthdr == NEXTHDR\_DEST) \{`}) \\
\textcolor[RGB]{0,128,0}{[CALL]} ipv6/ip6\_tunnel.c:431 (\texttt{`u16 i = 2;`}) \\
\textcolor[RGB]{0,128,0}{[CALL]} ipv6/ip6\_gre.c:433 (\texttt{`mtu = be32\_to\_cpu(info) - offset;`}) \\
\textcolor[RGB]{184,134,11}{[COND]} ipv6/ip6\_tunnel.c:434 (\texttt{`if (!pskb\_may\_pull(skb, off + optlen))`}) \\
\textcolor[RGB]{184,134,11}{[COND]} ipv6/ip6\_gre.c:434 (\texttt{`if (mtu < IPV6\_MIN\_MTU)`}) \\
\textcolor[RGB]{184,134,11}{[COND]} ipv6/ip6\_tunnel.c:434 (\texttt{`if (!pskb\_may\_pull(skb, off + optlen))`}) \\
\textcolor[RGB]{0,128,0}{[CALL]} ipv6/ip6\_gre.c:435 (\texttt{`mtu = IPV6\_MIN\_MTU;`}) \\
\textcolor[RGB]{0,128,0}{[CALL]} ipv6/ip6\_gre.c:436 (\texttt{`t->dev->mtu = mtu;`}) \\
\textcolor[RGB]{184,134,11}{[COND]} ipv6/ip6\_gre.c:440 (\texttt{`if (time\_before(jiffies, t->err\_time + IP6TUNNEL\_ERR\_TIMEO))`}) \\
\textcolor[RGB]{184,134,11}{[COND]} ipv6/ip6\_tunnel.c:441 (\texttt{`if (i + sizeof(*tel) > optlen)`}) \\
\textcolor[RGB]{0,128,0}{[CALL]} ipv6/ip6\_gre.c:441 (\texttt{`t->err\_count++;`}) \\
\textcolor[RGB]{105,105,105}{[PROP]} ipv6/ip6\_gre.c:443 (\texttt{`t->err\_count = 1;`}) \\
\textcolor[RGB]{0,128,0}{[CALL]} ipv6/ip6\_tunnel.c:444 (\texttt{`tel = (struct ipv6\_tlv\_tnl\_enc\_lim *)(skb->data + off + i);`}) \\
\textcolor[RGB]{105,105,105}{[PROP]} ipv6/ip6\_gre.c:444 (\texttt{`t->err\_time = jiffies;`}) \\
\textcolor[RGB]{184,134,11}{[COND]} ipv6/ip6\_tunnel.c:446 (\texttt{`if (tel->type == IPV6\_TLV\_TNL\_ENCAP\_LIMIT \&\&`}) \\
\textcolor[RGB]{0,128,0}{[CALL]} ipv6/ip6\_tunnel.c:447 (\texttt{`tel->length == 1)`}) \\
\textcolor[RGB]{0,128,0}{[CALL]} ipv6/ip6\_tunnel.c:448 (\texttt{`return i + off - nhoff;`}) \\
\textcolor[RGB]{184,134,11}{[COND]} ipv6/ip6\_tunnel.c:450 (\texttt{`if (tel->type)`}) \\
\textcolor[RGB]{0,128,0}{[CALL]} ipv6/ip6\_tunnel.c:451 (\texttt{`i += tel->length + 2;`}) \\
\textcolor[RGB]{0,128,0}{[CALL]} ipv6/ip6\_tunnel.c:453 (\texttt{`i++;`}) \\
\textcolor[RGB]{105,105,105}{[PROP]} ipv6/ip6\_tunnel.c:456 (\texttt{`nexthdr = next;`}) \\
\textcolor[RGB]{105,105,105}{[PROP]} ipv6/ip6\_tunnel.c:457 (\texttt{`off += optlen;`}) \\
\textbf{Variable: \texttt{*p}} \\
\textcolor[RGB]{255,0,0}{[TARGET]} ipv6/ip6\_gre.c:373 (\texttt{`\_\_be16 *p = (\_\_be16 *)(skb->data + offset);`}) \\

\end{agentbox}

\subsection{Phase III: Dialectical Verification}
Equipped with the closed evidence trace, the Verifier Agent constructs competing arguments for and against exploitability before adjudicating a preliminary verdict:

\begin{agentbox}[Verifier Agent Output]{orange}
\textbf{Comprehension}

The code at line 373 performs pointer arithmetic: \texttt{\_\_be16 *p = (\_\_be16 *)(skb->data + offset)} where \texttt{skb->data} is network packet data and \texttt{offset} is a function parameter. \texttt{offset} originates from network data (ICMPv6 error messages) and is potentially attacker-controlled. For safety, \texttt{offset} must be validated against the skb bounds before pointer arithmetic. The code later accesses \texttt{p[0]} (line 378) and \texttt{p[1]} (line 398), which could cause out-of-bounds reads if \texttt{offset} is malicious.

\textbf{Red Team (Case for VULNERABLE):}

\begin{itemize}
    \item \texttt{offset} is a function parameter (line 370) derived from network data, making it attacker-controlled
    \item No bounds validation occurs before the dangerous pointer arithmetic at line 373
    \item Line 378 immediately dereferences \texttt{p[0]} without validation
    \item Line 398 also dereferences \texttt{p[1]} without validation  
    \item The safety check \texttt{pskb\_may\_pull(skb, grehlen)} at line 390 happens AFTER the dangerous operations
    \item Impact: Information disclosure or kernel crash via out-of-bounds read
\end{itemize}

\vspace{0.5em}
\textbf{Blue Team (Case for NOT\_VULNERABLE):}
\begin{itemize}
    \item Line 390 calls \texttt{pskb\_may\_pull(skb, grehlen)} which validates \texttt{offset + 4} bytes are available
    \item If validation fails, the function returns early preventing further damage
    \item Line 393 reassigns \texttt{p} after validation, showing developer awareness
    \item The network stack likely validates ICMPv6 messages before calling this handler
    \item The function is static with limited call sites
\end{itemize}

\vspace{0.5em}
\textbf{Adjudication:} \\
Red Team's argument is stronger with concrete evidence: the dangerous pointer arithmetic and dereferences (lines 373, 378, 398) occur BEFORE the safety check (line 390). This temporal gap creates a vulnerability window where malicious \texttt{offset} can cause out-of-bounds access. Blue Team's validation argument is weakened because it happens too late in the execution flow.

\textbf{Verdict:} \texttt{VULNERABLE} \\
\textbf{Confidence:} 0.88 \\
\textbf{CWE\_id:} CWE-125\\
\textbf{Key Evidence:} Pointer arithmetic with attacker-controlled offset at line 373 occurs before bounds validation at line 390, enabling out-of-bounds read at lines 378 and 398

\end{agentbox}

\subsection{Phase IV: Meta-Auditing}
The Audit Agent independently evaluates the raw code and evidence trace, then cross-checks the soundness of the Verifier's reasoning. In this case, the Audit Agent finds no material reasoning flaws and confirms the preliminary verdict:

\begin{agentbox}[Audit Agent Output]{purple}
\textbf{Independent Assessment:} \\
{The suspicious code line performs pointer arithmetic on network packet data (\texttt{skb->data}) using an \texttt{offset} parameter that originates from network data (ICMPv6 error messages). This creates a pointer \texttt{p} that could point outside valid bounds if \texttt{offset} is maliciously large. The pointer is immediately dereferenced at line 378 (\texttt{flags = p[0]}) and later at line 398 (\texttt{p[1]}), creating potential for out-of-bounds memory access. The security property at risk is memory safety - preventing reads beyond the allocated skb buffer.}

\textbf{Evidence Cross-Check} \\
The verifier's factual claims are accurate:
\begin{itemize}
    \item \texttt{offset} is indeed a function parameter (line 370)
    \item Pointer arithmetic occurs at line 373 without prior bounds checking
    \item \texttt{p[0]} is dereferenced at line 378 before any validation
    \item \texttt{pskb\_may\_pull} validation occurs at line 390, after the initial dangerous operations
    \item The verifier correctly cites all relevant line numbers and their contents
\end{itemize}

\textbf{Reasoning Quality Audit} \\
The verifier's reasoning is sound and avoids common flaws:

\textbf{Red Team analysis (VULNERABLE case):}
\begin{itemize}
    \item \textbf{No speculation flaw:} The attack path is concretely demonstrated through the code flow
    \item \textbf{No pattern-matching flaw:} The analysis specifically shows how the temporal sequence creates vulnerability
    \item \textbf{No anchoring flaw:} The analysis goes beyond the initial suspicion to examine the full execution path
    \item \textbf{No absence-as-evidence flaw:} The vulnerability is demonstrated through positive evidence of dangerous operations before validation
\end{itemize}

\textbf{Blue Team analysis (NOT\_VULNERABLE case):}
\begin{itemize}
    \item The verifier correctly identifies and then refutes the potential mitigation arguments
    \item \textbf{No phantom mitigation flaw:} The verifier acknowledges \texttt{pskb\_may\_pull} exists but correctly notes it comes too late
    \item \textbf{No over-trust flaw:} The verifier does not assume upstream validation without evidence
    \item \textbf{No incomplete protection flaw:} The analysis correctly shows the validation is temporally misplaced
\end{itemize}

The adjudication correctly identifies the critical issue: the temporal gap between dangerous operations (lines 373, 378) and validation (line 390).

\textbf{Independent Verdict} \\
I agree with the verifier's \texttt{VULNERABLE} verdict. The evidence clearly shows that pointer arithmetic and dereferencing occur before bounds validation, creating a concrete out-of-bounds read vulnerability. The verifier's reasoning is thorough, evidence-based, and avoids common reasoning flaws. The high confidence level is appropriate given the clear temporal vulnerability pattern.

\vspace{0.5em}
\textbf{Reasoning Flaws Detected in Verifier's Output:} None

\vspace{0.5em}
\textbf{Final Decision:} \\
\textbf{Action:} \texttt{AGREE} $\rightarrow$ \textbf{Final Verdict:} \texttt{VULNERABLE} \\
\textbf{Confidence:} 0.95
\end{agentbox}

\newpage
\section{Case Study: End-to-End AEGIS Pipeline Execution}
\label{sec:case_study_incorrect_veto}

To complement the preceding case study, this section presents a failure case of the Meta-Auditing mechanism. While the Verifier Agent correctly identifies a true vulnerability, the Audit Agent erroneously overturns the verdict due to an incorrect reasoning flaw classification, resulting in a final false negative. This example illustrates the inherent cost of the auditing mechanism discussed in Section 4.3.2, and highlights the specific conditions under which the Audit Agent's conservative bias toward absence-of-protection vulnerabilities leads to incorrect vetoes.

\subsection{Target Input}
\textbf{Commit URL:} \url{https://github.com/torvalds/linux/commit/c8c2a057fdc7de1cd16f4baa51425b932a42eb39} \\
\textbf{Target File:} \texttt{\seqsplit{drivers/net/ethernet/mellanox/mlx5/core/fpga/conn.c}} \\
\textbf{Target Function:} \texttt{mlx5\_fpga\_conn\_create\_cq}

\begin{lstlisting}[language=C, basicstyle=\ttfamily\scriptsize, breaklines=true, frame=single, backgroundcolor=\color{gray!5}, columns=fullflexible, keepspaces=true, xleftmargin=2pt, xrightmargin=2pt]
static int mlx5_fpga_conn_create_cq(struct mlx5_fpga_conn *conn, int cq_size)
{
        struct mlx5_fpga_device *fdev = conn->fdev;
        struct mlx5_core_dev *mdev = fdev->mdev;
        u32 temp_cqc[MLX5_ST_SZ_DW(cqc)] = {0};
        u32 out[MLX5_ST_SZ_DW(create_cq_out)];
        struct mlx5_wq_param wqp;
        struct mlx5_cqe64 *cqe;
        int inlen, err, eqn;
        unsigned int irqn;
        void *cqc, *in;
        __be64 *pas;
        u32 i;

        cq_size = roundup_pow_of_two(cq_size);
        MLX5_SET(cqc, temp_cqc, log_cq_size, ilog2(cq_size));

        wqp.buf_numa_node = mdev->priv.numa_node;
        wqp.db_numa_node  = mdev->priv.numa_node;

        err = mlx5_cqwq_create(mdev, &wqp, temp_cqc, &conn->cq.wq,
                               &conn->cq.wq_ctrl);
        if (err)
                return err;

        for (i = 0; i < mlx5_cqwq_get_size(&conn->cq.wq); i++) {
                cqe = mlx5_cqwq_get_wqe(&conn->cq.wq, i);
                cqe->op_own = MLX5_CQE_INVALID << 4 | MLX5_CQE_OWNER_MASK;
        }

        inlen = MLX5_ST_SZ_BYTES(create_cq_in) +
                sizeof(u64) * conn->cq.wq_ctrl.buf.npages;
        in = kvzalloc(inlen, GFP_KERNEL);
        if (!in) {
                err = -ENOMEM;
                goto err_cqwq;
        }

        err = mlx5_vector2eqn(mdev, smp_processor_id(), &eqn, &irqn);
        if (err)
                goto err_cqwq;

        cqc = MLX5_ADDR_OF(create_cq_in, in, cq_context);
        MLX5_SET(cqc, cqc, log_cq_size, ilog2(cq_size));
        MLX5_SET(cqc, cqc, c_eqn, eqn);
        MLX5_SET(cqc, cqc, uar_page, fdev->conn_res.uar->index);
        MLX5_SET(cqc, cqc, log_page_size, conn->cq.wq_ctrl.buf.page_shift -
                           MLX5_ADAPTER_PAGE_SHIFT);
        MLX5_SET64(cqc, cqc, dbr_addr, conn->cq.wq_ctrl.db.dma);

        pas = (__be64 *)MLX5_ADDR_OF(create_cq_in, in, pas);
        mlx5_fill_page_frag_array(&conn->cq.wq_ctrl.buf, pas);

        err = mlx5_core_create_cq(mdev, &conn->cq.mcq, in, inlen, out, sizeof(out));
        kvfree(in);

        if (err)
                goto err_cqwq;

        conn->cq.mcq.cqe_sz     = 64;
        conn->cq.mcq.set_ci_db  = conn->cq.wq_ctrl.db.db;
        conn->cq.mcq.arm_db     = conn->cq.wq_ctrl.db.db + 1;
        *conn->cq.mcq.set_ci_db = 0;
        *conn->cq.mcq.arm_db    = 0;
        conn->cq.mcq.vector     = 0;
        conn->cq.mcq.comp       = mlx5_fpga_conn_cq_complete;
        conn->cq.mcq.event      = mlx5_fpga_conn_cq_event;
        conn->cq.mcq.irqn       = irqn;
        conn->cq.mcq.uar        = fdev->conn_res.uar;
        tasklet_init(&conn->cq.tasklet, mlx5_fpga_conn_cq_tasklet,
                     (unsigned long)conn);

        mlx5_fpga_dbg(fdev, "Created CQ #0x%x\n", conn->cq.mcq.cqn);

        goto out;

err_cqwq:
        mlx5_wq_destroy(&conn->cq.wq_ctrl);
out:
        return err;
}
\end{lstlisting}

\subsection{Phase I: Clue Discovery}
Operating under the Worst-Case Taint Assumption, the Clue-Discovery Agent scans the target function in isolation and identifies the following suspicious anomalies:

\begin{agentbox}[Clue-Discovery Agent Output]{blue}
\textbf{Clue C1:}
\begin{itemize}
    \item \textbf{Line Number:} 460
    \item \textbf{Code Line:} \texttt{in = kvzalloc(inlen, GFP\_KERNEL);}
    \item \textbf{Reason:} kvzalloc allocation size 'inlen' could be very large or negative due to potential integer overflow in previous calculation.
    \item \textbf{Confidence Score:} 0.7
\end{itemize}

\textbf{Clue C2:}
\begin{itemize}
    \item \textbf{Line Number:} 458
    \item \textbf{Code Line:} \texttt{inlen = MLX5\_ST\_SZ\_BYTES(create\_cq\_in) +}
    \item \textbf{Reason:} Arithmetic operation adding MLX5\_ST\_SZ\_BYTES(create\_cq\_in) to sizeof(u64) * npages; potential integer overflow if npages is large.
    \item \textbf{Confidence Score:} 0.6
\end{itemize}
\end{agentbox}

\subsection{Phase II: Graph-Guided Context Augmentation}
Triggered by the boundary variables and external calls in \textbf{Clue C1}, the Context-Augmentation Agent dynamically stitches the CPG to expand the cross-file dependencies. The reconstructed repository-level context is presented in two parts: the expanded code blocks retrieved via CPG slicing, and the resulting structured data flow trace.

\vspace{0.5em}
\noindent\textbf{Part 1: Expanded Code Context}

\begin{lstlisting}[language=C, basicstyle=\ttfamily\scriptsize, breaklines=true, frame=single, backgroundcolor=\color{gray!5}, columns=fullflexible, keepspaces=true, xleftmargin=2pt, xrightmargin=2pt]
 static int mlx5_fpga_conn_create_cq(struct mlx5_fpga_conn *conn, int cq_size)  // [FUNCTION ENTRY]
        int inlen, err, eqn;
        void *cqc, *in;
        if (err)
        inlen = MLX5_ST_SZ_BYTES(create_cq_in) +
                sizeof(u64) * conn->cq.wq_ctrl.buf.npages;
        in = kvzalloc(inlen, GFP_KERNEL);  // [TARGET]
        if (!in) {
        if (err)
        cqc = MLX5_ADDR_OF(create_cq_in, in, cq_context);
        MLX5_SET(cqc, cqc, log_cq_size, ilog2(cq_size));
        MLX5_SET(cqc, cqc, c_eqn, eqn);
        MLX5_SET(cqc, cqc, uar_page, fdev->conn_res.uar->index);
        MLX5_SET(cqc, cqc, log_page_size, conn->cq.wq_ctrl.buf.page_shift -
                           MLX5_ADAPTER_PAGE_SHIFT);
        MLX5_SET64(cqc, cqc, dbr_addr, conn->cq.wq_ctrl.db.dma);
        pas = (__be64 *)MLX5_ADDR_OF(create_cq_in, in, pas);
        mlx5_fill_page_frag_array(&conn->cq.wq_ctrl.buf, pas);
        err = mlx5_core_create_cq(mdev, &conn->cq.mcq, in, inlen, out, sizeof(out));  // [CROSS-FILE CALL]
        kvfree(in);
        if (err)
        conn->cq.mcq.cqe_sz     = 64;
        conn->cq.mcq.set_ci_db  = conn->cq.wq_ctrl.db.db;
        conn->cq.mcq.arm_db     = conn->cq.wq_ctrl.db.db + 1;
        *conn->cq.mcq.set_ci_db = 0;
        *conn->cq.mcq.arm_db    = 0;
        conn->cq.mcq.vector     = 0;
        conn->cq.mcq.comp       = mlx5_fpga_conn_cq_complete;
        conn->cq.mcq.event      = mlx5_fpga_conn_cq_event;
        conn->cq.mcq.irqn       = irqn;
        conn->cq.mcq.uar        = fdev->conn_res.uar;
        mlx5_fpga_dbg(fdev, "Created CQ #0x%x\n", conn->cq.mcq.cqn);
        return err;
\end{lstlisting}

\begin{lstlisting}[language=C, basicstyle=\ttfamily\scriptsize, breaklines=true, frame=single, backgroundcolor=\color{gray!5}, columns=fullflexible, keepspaces=true, xleftmargin=2pt, xrightmargin=2pt]
int mlx5_core_create_cq(struct mlx5_core_dev *dev, struct mlx5_core_cq *cq,  // [FUNCTION ENTRY]
                        u32 *in, int inlen, u32 *out, int outlen)
 {
        int eqn = MLX5_GET(cqc, MLX5_ADDR_OF(create_cq_in, in, cq_context), c_eqn);
        u32 dout[MLX5_ST_SZ_DW(destroy_cq_out)];
        u32 din[MLX5_ST_SZ_DW(destroy_cq_in)];
        struct mlx5_eq_comp *eq;
        int err;
 
        eq = mlx5_eqn2comp_eq(dev, eqn);
        if (IS_ERR(eq))
                return PTR_ERR(eq);
 
        memset(out, 0, outlen);
        MLX5_SET(create_cq_in, in, opcode, MLX5_CMD_OP_CREATE_CQ);
        err = mlx5_cmd_exec(dev, in, inlen, out, outlen);
        if (err)
                return err;
 
        cq->cqn = MLX5_GET(create_cq_out, out, cqn);
        cq->cons_index = 0;
        cq->arm_sn     = 0;
        cq->eq         = eq;
        cq->uid = MLX5_GET(create_cq_in, in, uid);
        refcount_set(&cq->refcount, 1);
        init_completion(&cq->free);
        if (!cq->comp)
                cq->comp = mlx5_add_cq_to_tasklet;
        /* assuming CQ will be deleted before the EQ */
        cq->tasklet_ctx.priv = &eq->tasklet_ctx;
        INIT_LIST_HEAD(&cq->tasklet_ctx.list);
 
        /* Add to comp EQ CQ tree to recv comp events */
        err = mlx5_eq_add_cq(&eq->core, cq);
        if (err)
                goto err_cmd;
 
        /* Add to async EQ CQ tree to recv async events */
        err = mlx5_eq_add_cq(mlx5_get_async_eq(dev), cq);
        if (err)
                goto err_cq_add;
        cq->pid = current->pid;
        err = mlx5_debug_cq_add(dev, cq);
        if (err)
                mlx5_core_dbg(dev, "failed adding CP 0x%x to debug file system\n",
                              cq->cqn);
        cq->uar = dev->priv.uar;
        return 0;  // [CROSS-FILE CALL]
 err_cq_add:
        mlx5_eq_del_cq(&eq->core, cq);
 err_cmd:
        memset(din, 0, sizeof(din));
        memset(dout, 0, sizeof(dout));
        MLX5_SET(destroy_cq_in, din, opcode, MLX5_CMD_OP_DESTROY_CQ);
        MLX5_SET(destroy_cq_in, din, cqn, cq->cqn);
        MLX5_SET(destroy_cq_in, din, uid, cq->uid);
        mlx5_cmd_exec(dev, din, sizeof(din), dout, sizeof(dout));
        return err;  // [CROSS-FILE CALL]
\end{lstlisting}

\vspace{0.5em}
\noindent\textbf{Part 2: Reconstructed Data Flow Trace}

\begin{agentbox}[By-Variable Data Flow Trace]{green}

\ttfamily\scriptsize

\textbf{Variable: \texttt{in}} \\
\textcolor[RGB]{0,0,255}{[SOURCE]} core/cq.c:90 (\texttt{`int mlx5\_core\_create\_cq(struct mlx5\_core\_dev *dev, struct mlx5\_core\_cq *cq,`}) \\
\textcolor[RGB]{105,105,105}{[PROP]} core/cq.c:91 (\texttt{`u32 *in, int inlen, u32 *out, int outlen)`}) \\
\textcolor[RGB]{0,128,0}{[CALL]} core/cq.c:93 (\texttt{`int eqn = MLX5\_GET(cqc, MLX5\_ADDR\_OF(create\_cq\_in, in, cq\_context), c\_eqn);`}) \\
\textcolor[RGB]{0,128,0}{[CALL]} core/cq.c:99 (\texttt{`eq = mlx5\_eqn2comp\_eq(dev, eqn);`}) \\
\textcolor[RGB]{184,134,11}{[COND]} core/cq.c:100 (\texttt{`if (IS\_ERR(eq))`}) \\
\textcolor[RGB]{0,128,0}{[CALL]} core/cq.c:101 (\texttt{`return PTR\_ERR(eq);`}) \\
\textcolor[RGB]{0,128,0}{[CALL]} core/cq.c:104 (\texttt{`MLX5\_SET(create\_cq\_in, in, opcode, MLX5\_CMD\_OP\_CREATE\_CQ);`}) \\
\textcolor[RGB]{0,128,0}{[CALL]} core/cq.c:105 (\texttt{`err = mlx5\_cmd\_exec(dev, in, inlen, out, outlen);`}) \\
\textcolor[RGB]{255,140,0}{[RET]} core/cq.c:107 (\texttt{`return err;`}) \\
\textcolor[RGB]{0,128,0}{[CALL]} core/cq.c:109 (\texttt{`cq->cqn = MLX5\_GET(create\_cq\_out, out, cqn);`}) \\
\textcolor[RGB]{105,105,105}{[PROP]} core/cq.c:110 (\texttt{`cq->cons\_index = 0;`}) \\
\textcolor[RGB]{105,105,105}{[PROP]} core/cq.c:111 (\texttt{`cq->arm\_sn     = 0;`}) \\
\textcolor[RGB]{0,128,0}{[CALL]} core/cq.c:112 (\texttt{`cq->eq         = eq;`}) \\
\textcolor[RGB]{0,128,0}{[CALL]} core/cq.c:113 (\texttt{`cq->uid = MLX5\_GET(create\_cq\_in, in, uid);`}) \\
\textcolor[RGB]{105,105,105}{[PROP]} core/cq.c:114 (\texttt{`refcount\_set(\&cq->refcount, 1);`}) \\
\textcolor[RGB]{105,105,105}{[PROP]} core/cq.c:115 (\texttt{`init\_completion(\&cq->free);`}) \\
\textcolor[RGB]{184,134,11}{[COND]} core/cq.c:116 (\texttt{`if (!cq->comp)`}) \\
\textcolor[RGB]{105,105,105}{[PROP]} core/cq.c:117 (\texttt{`cq->comp = mlx5\_add\_cq\_to\_tasklet;`}) \\
\textcolor[RGB]{0,128,0}{[CALL]} core/cq.c:119 (\texttt{`cq->tasklet\_ctx.priv = \&eq->tasklet\_ctx;`}) \\
\textcolor[RGB]{105,105,105}{[PROP]} core/cq.c:120 (\texttt{`INIT\_LIST\_HEAD(\&cq->tasklet\_ctx.list);`}) \\
\textcolor[RGB]{0,128,0}{[CALL]} core/cq.c:123 (\texttt{`err = mlx5\_eq\_add\_cq(\&eq->core, cq);`}) \\
\textcolor[RGB]{0,128,0}{[CALL]} core/cq.c:128 (\texttt{`err = mlx5\_eq\_add\_cq(mlx5\_get\_async\_eq(dev), cq);`}) \\
\textcolor[RGB]{105,105,105}{[PROP]} core/cq.c:132 (\texttt{`cq->pid = current->pid;`}) \\
\textcolor[RGB]{0,128,0}{[CALL]} core/cq.c:133 (\texttt{`err = mlx5\_debug\_cq\_add(dev, cq);`}) \\
\textcolor[RGB]{0,128,0}{[CALL]} core/cq.c:135 (\texttt{`mlx5\_core\_dbg(dev, "failed adding CP 0x\%x to debug file system\textbackslash n",`}) \\
\textcolor[RGB]{0,128,0}{[CALL]} core/cq.c:136 (\texttt{`cq->cqn);`}) \\
\textcolor[RGB]{0,128,0}{[CALL]} core/cq.c:138 (\texttt{`cq->uar = dev->priv.uar;`}) \\
\textcolor[RGB]{0,128,0}{[CALL]} core/cq.c:143 (\texttt{`mlx5\_eq\_del\_cq(\&eq->core, cq);`}) \\
\textcolor[RGB]{0,128,0}{[CALL]} core/cq.c:147 (\texttt{`MLX5\_SET(destroy\_cq\_in, din, opcode, MLX5\_CMD\_OP\_DESTROY\_CQ);`}) \\
\textcolor[RGB]{0,128,0}{[CALL]} core/cq.c:148 (\texttt{`MLX5\_SET(destroy\_cq\_in, din, cqn, cq->cqn);`}) \\
\textcolor[RGB]{0,128,0}{[CALL]} core/cq.c:149 (\texttt{`MLX5\_SET(destroy\_cq\_in, din, uid, cq->uid);`}) \\
\textcolor[RGB]{0,128,0}{[CALL]} core/cq.c:150 (\texttt{`mlx5\_cmd\_exec(dev, din, sizeof(din), dout, sizeof(dout));`}) \\
\textcolor[RGB]{255,140,0}{[RET]} core/cq.c:151 (\texttt{`return err;`}) \\
\textcolor[RGB]{255,140,0}{[RET]} fpga/conn.c:428 (\texttt{`static int mlx5\_fpga\_conn\_create\_cq(struct mlx5\_fpga\_conn *conn, int cq\_size)`}) \\
\textcolor[RGB]{105,105,105}{[PROP]} fpga/conn.c:436 (\texttt{`int inlen, err, eqn;`}) \\
\textcolor[RGB]{0,128,0}{[CALL]} fpga/conn.c:458 (\texttt{`inlen = MLX5\_ST\_SZ\_BYTES(create\_cq\_in) +`}) \\
\textcolor[RGB]{0,128,0}{[CALL]} fpga/conn.c:459 (\texttt{`sizeof(u64) * conn->cq.wq\_ctrl.buf.npages;`}) \\
\textcolor[RGB]{255,0,0}{[TARGET]} fpga/conn.c:460 (\texttt{`in = kvzalloc(inlen, GFP\_KERNEL);`}) \\
\textcolor[RGB]{184,134,11}{[COND]} fpga/conn.c:461 (\texttt{`if (!in) \{`}) \\
\textcolor[RGB]{0,128,0}{[CALL]} fpga/conn.c:470 (\texttt{`cqc = MLX5\_ADDR\_OF(create\_cq\_in, in, cq\_context);`}) \\
\textcolor[RGB]{0,128,0}{[CALL]} fpga/conn.c:471 (\texttt{`MLX5\_SET(cqc, cqc, log\_cq\_size, ilog2(cq\_size));`}) \\
\textcolor[RGB]{0,128,0}{[CALL]} fpga/conn.c:472 (\texttt{`MLX5\_SET(cqc, cqc, c\_eqn, eqn);`}) \\
\textcolor[RGB]{0,128,0}{[CALL]} fpga/conn.c:473 (\texttt{`MLX5\_SET(cqc, cqc, uar\_page, fdev->conn\_res.uar->index);`}) \\
\textcolor[RGB]{0,128,0}{[CALL]} fpga/conn.c:474 (\texttt{`MLX5\_SET(cqc, cqc, log\_page\_size, conn->cq.wq\_ctrl.buf.page\_shift -`}) \\
\textcolor[RGB]{0,128,0}{[CALL]} fpga/conn.c:476 (\texttt{`MLX5\_SET64(cqc, cqc, dbr\_addr, conn->cq.wq\_ctrl.db.dma);`}) \\
\textcolor[RGB]{0,128,0}{[CALL]} fpga/conn.c:478 (\texttt{`pas = (\_\_be64 *)MLX5\_ADDR\_OF(create\_cq\_in, in, pas);`}) \\
\textcolor[RGB]{0,128,0}{[CALL]} fpga/conn.c:479 (\texttt{`mlx5\_fill\_page\_frag\_array(\&conn->cq.wq\_ctrl.buf, pas);`}) \\
\textcolor[RGB]{0,128,0}{[CALL]} fpga/conn.c:481 (\texttt{`err = mlx5\_core\_create\_cq(mdev, \&conn->cq.mcq, in, inlen, out, sizeof(out));`}) \\
\textcolor[RGB]{0,128,0}{[CALL]} fpga/conn.c:482 (\texttt{`kvfree(in);`}) \\
\textcolor[RGB]{0,128,0}{[CALL]} fpga/conn.c:487 (\texttt{`conn->cq.mcq.cqe\_sz     = 64;`}) \\
\textcolor[RGB]{0,128,0}{[CALL]} fpga/conn.c:488 (\texttt{`conn->cq.mcq.set\_ci\_db  = conn->cq.wq\_ctrl.db.db;`}) \\
\textcolor[RGB]{0,128,0}{[CALL]} fpga/conn.c:489 (\texttt{`conn->cq.mcq.arm\_db     = conn->cq.wq\_ctrl.db.db + 1;`}) \\
\textcolor[RGB]{0,128,0}{[CALL]} fpga/conn.c:490 (\texttt{`*conn->cq.mcq.set\_ci\_db = 0;`}) \\
\textcolor[RGB]{0,128,0}{[CALL]} fpga/conn.c:491 (\texttt{`*conn->cq.mcq.arm\_db    = 0;`}) \\
\textcolor[RGB]{0,128,0}{[CALL]} fpga/conn.c:492 (\texttt{`conn->cq.mcq.vector     = 0;`}) \\
\textcolor[RGB]{0,128,0}{[CALL]} fpga/conn.c:493 (\texttt{`conn->cq.mcq.comp       = mlx5\_fpga\_conn\_cq\_complete;`}) \\
\textcolor[RGB]{0,128,0}{[CALL]} fpga/conn.c:494 (\texttt{`conn->cq.mcq.event      = mlx5\_fpga\_conn\_cq\_event;`}) \\
\textcolor[RGB]{0,128,0}{[CALL]} fpga/conn.c:495 (\texttt{`conn->cq.mcq.irqn       = irqn;`}) \\
\textcolor[RGB]{0,128,0}{[CALL]} fpga/conn.c:496 (\texttt{`conn->cq.mcq.uar        = fdev->conn\_res.uar;`}) \\
\textcolor[RGB]{0,128,0}{[CALL]} fpga/conn.c:500 (\texttt{`mlx5\_fpga\_dbg(fdev, "Created CQ \#0x\%x\textbackslash n", conn->cq.mcq.cqn);`}) \\
\textcolor[RGB]{255,140,0}{[RET]} fpga/conn.c:507 (\texttt{`return err;`}) \\
\textbf{Variable: \texttt{inlen}} \\
\textcolor[RGB]{0,0,255}{[SOURCE]} core/cq.c:90 (\texttt{`int mlx5\_core\_create\_cq(struct mlx5\_core\_dev *dev, struct mlx5\_core\_cq *cq,`}) \\
\textcolor[RGB]{105,105,105}{[PROP]} core/cq.c:91 (\texttt{`u32 *in, int inlen, u32 *out, int outlen)`}) \\
\textcolor[RGB]{0,128,0}{[CALL]} core/cq.c:105 (\texttt{`err = mlx5\_cmd\_exec(dev, in, inlen, out, outlen);`}) \\
\textcolor[RGB]{255,140,0}{[RET]} core/cq.c:107 (\texttt{`return err;`}) \\
\textcolor[RGB]{0,128,0}{[CALL]} core/cq.c:109 (\texttt{`cq->cqn = MLX5\_GET(create\_cq\_out, out, cqn);`}) \\
\textcolor[RGB]{105,105,105}{[PROP]} core/cq.c:110 (\texttt{`cq->cons\_index = 0;`}) \\
\textcolor[RGB]{105,105,105}{[PROP]} core/cq.c:111 (\texttt{`cq->arm\_sn     = 0;`}) \\
\textcolor[RGB]{105,105,105}{[PROP]} core/cq.c:112 (\texttt{`cq->eq         = eq;`}) \\
\textcolor[RGB]{0,128,0}{[CALL]} core/cq.c:113 (\texttt{`cq->uid = MLX5\_GET(create\_cq\_in, in, uid);`}) \\
\textcolor[RGB]{105,105,105}{[PROP]} core/cq.c:114 (\texttt{`refcount\_set(\&cq->refcount, 1);`}) \\
\textcolor[RGB]{105,105,105}{[PROP]} core/cq.c:115 (\texttt{`init\_completion(\&cq->free);`}) \\
\textcolor[RGB]{184,134,11}{[COND]} core/cq.c:116 (\texttt{`if (!cq->comp)`}) \\
\textcolor[RGB]{105,105,105}{[PROP]} core/cq.c:117 (\texttt{`cq->comp = mlx5\_add\_cq\_to\_tasklet;`}) \\
\textcolor[RGB]{105,105,105}{[PROP]} core/cq.c:119 (\texttt{`cq->tasklet\_ctx.priv = \&eq->tasklet\_ctx;`}) \\
\textcolor[RGB]{105,105,105}{[PROP]} core/cq.c:120 (\texttt{`INIT\_LIST\_HEAD(\&cq->tasklet\_ctx.list);`}) \\
\textcolor[RGB]{0,128,0}{[CALL]} core/cq.c:123 (\texttt{`err = mlx5\_eq\_add\_cq(\&eq->core, cq);`}) \\
\textcolor[RGB]{0,128,0}{[CALL]} core/cq.c:128 (\texttt{`err = mlx5\_eq\_add\_cq(mlx5\_get\_async\_eq(dev), cq);`}) \\
\textcolor[RGB]{105,105,105}{[PROP]} core/cq.c:132 (\texttt{`cq->pid = current->pid;`}) \\
\textcolor[RGB]{0,128,0}{[CALL]} core/cq.c:133 (\texttt{`err = mlx5\_debug\_cq\_add(dev, cq);`}) \\
\textcolor[RGB]{0,128,0}{[CALL]} core/cq.c:135 (\texttt{`mlx5\_core\_dbg(dev, "failed adding CP 0x\%x to debug file system\textbackslash n",`}) \\
\textcolor[RGB]{0,128,0}{[CALL]} core/cq.c:136 (\texttt{`cq->cqn);`}) \\
\textcolor[RGB]{0,128,0}{[CALL]} core/cq.c:138 (\texttt{`cq->uar = dev->priv.uar;`}) \\
\textcolor[RGB]{0,128,0}{[CALL]} core/cq.c:143 (\texttt{`mlx5\_eq\_del\_cq(\&eq->core, cq);`}) \\
\textcolor[RGB]{0,128,0}{[CALL]} core/cq.c:148 (\texttt{`MLX5\_SET(destroy\_cq\_in, din, cqn, cq->cqn);`}) \\
\textcolor[RGB]{0,128,0}{[CALL]} core/cq.c:149 (\texttt{`MLX5\_SET(destroy\_cq\_in, din, uid, cq->uid);`}) \\
\textcolor[RGB]{0,128,0}{[CALL]} core/cq.c:150 (\texttt{`mlx5\_cmd\_exec(dev, din, sizeof(din), dout, sizeof(dout));`}) \\
\textcolor[RGB]{255,140,0}{[RET]} core/cq.c:151 (\texttt{`return err;`}) \\
\textcolor[RGB]{255,140,0}{[RET]} fpga/conn.c:428 (\texttt{`static int mlx5\_fpga\_conn\_create\_cq(struct mlx5\_fpga\_conn *conn, int cq\_size)`}) \\
\textcolor[RGB]{105,105,105}{[PROP]} fpga/conn.c:436 (\texttt{`int inlen, err, eqn;`}) \\
\textcolor[RGB]{0,128,0}{[CALL]} fpga/conn.c:458 (\texttt{`inlen = MLX5\_ST\_SZ\_BYTES(create\_cq\_in) +`}) \\
\textcolor[RGB]{0,128,0}{[CALL]} fpga/conn.c:459 (\texttt{`sizeof(u64) * conn->cq.wq\_ctrl.buf.npages;`}) \\
\textcolor[RGB]{255,0,0}{[TARGET]} fpga/conn.c:460 (\texttt{`in = kvzalloc(inlen, GFP\_KERNEL);`}) \\
\textcolor[RGB]{184,134,11}{[COND]} fpga/conn.c:461 (\texttt{`if (!in) \{`}) \\
\textcolor[RGB]{0,128,0}{[CALL]} fpga/conn.c:470 (\texttt{`cqc = MLX5\_ADDR\_OF(create\_cq\_in, in, cq\_context);`}) \\
\textcolor[RGB]{0,128,0}{[CALL]} fpga/conn.c:471 (\texttt{`MLX5\_SET(cqc, cqc, log\_cq\_size, ilog2(cq\_size));`}) \\
\textcolor[RGB]{0,128,0}{[CALL]} fpga/conn.c:472 (\texttt{`MLX5\_SET(cqc, cqc, c\_eqn, eqn);`}) \\
\textcolor[RGB]{0,128,0}{[CALL]} fpga/conn.c:473 (\texttt{`MLX5\_SET(cqc, cqc, uar\_page, fdev->conn\_res.uar->index);`}) \\
\textcolor[RGB]{105,105,105}{[PROP]} fpga/conn.c:474 (\texttt{`MLX5\_SET(cqc, cqc, log\_page\_size, conn->cq.wq\_ctrl.buf.page\_shift -`}) \\
\textcolor[RGB]{0,128,0}{[CALL]} fpga/conn.c:478 (\texttt{`pas = (\_\_be64 *)MLX5\_ADDR\_OF(create\_cq\_in, in, pas);`}) \\
\textcolor[RGB]{0,128,0}{[CALL]} fpga/conn.c:479 (\texttt{`mlx5\_fill\_page\_frag\_array(\&conn->cq.wq\_ctrl.buf, pas);`}) \\
\textcolor[RGB]{0,128,0}{[CALL]} fpga/conn.c:481 (\texttt{`err = mlx5\_core\_create\_cq(mdev, \&conn->cq.mcq, in, inlen, out, sizeof(out));`}) \\
\textcolor[RGB]{0,128,0}{[CALL]} fpga/conn.c:482 (\texttt{`kvfree(in);`}) \\
\textcolor[RGB]{0,128,0}{[CALL]} fpga/conn.c:487 (\texttt{`conn->cq.mcq.cqe\_sz     = 64;`}) \\
\textcolor[RGB]{0,128,0}{[CALL]} fpga/conn.c:488 (\texttt{`conn->cq.mcq.set\_ci\_db  = conn->cq.wq\_ctrl.db.db;`}) \\
\textcolor[RGB]{0,128,0}{[CALL]} fpga/conn.c:489 (\texttt{`conn->cq.mcq.arm\_db     = conn->cq.wq\_ctrl.db.db + 1;`}) \\
\textcolor[RGB]{0,128,0}{[CALL]} fpga/conn.c:490 (\texttt{`*conn->cq.mcq.set\_ci\_db = 0;`}) \\
\textcolor[RGB]{0,128,0}{[CALL]} fpga/conn.c:491 (\texttt{`*conn->cq.mcq.arm\_db    = 0;`}) \\
\textcolor[RGB]{0,128,0}{[CALL]} fpga/conn.c:492 (\texttt{`conn->cq.mcq.vector     = 0;`}) \\
\textcolor[RGB]{0,128,0}{[CALL]} fpga/conn.c:493 (\texttt{`conn->cq.mcq.comp       = mlx5\_fpga\_conn\_cq\_complete;`}) \\
\textcolor[RGB]{0,128,0}{[CALL]} fpga/conn.c:494 (\texttt{`conn->cq.mcq.event      = mlx5\_fpga\_conn\_cq\_event;`}) \\
\textcolor[RGB]{0,128,0}{[CALL]} fpga/conn.c:495 (\texttt{`conn->cq.mcq.irqn       = irqn;`}) \\
\textcolor[RGB]{0,128,0}{[CALL]} fpga/conn.c:496 (\texttt{`conn->cq.mcq.uar        = fdev->conn\_res.uar;`}) \\
\textcolor[RGB]{0,128,0}{[CALL]} fpga/conn.c:500 (\texttt{`mlx5\_fpga\_dbg(fdev, "Created CQ \#0x\%x\textbackslash n", conn->cq.mcq.cqn);`}) \\
\textcolor[RGB]{255,140,0}{[RET]} fpga/conn.c:507 (\texttt{`return err;`}) \\
\textbf{Variable: \texttt{GFP\_KERNEL}} \\
\textcolor[RGB]{0,0,255}{[SOURCE]} fpga/conn.c:428 (\texttt{`static int mlx5\_fpga\_conn\_create\_cq(struct mlx5\_fpga\_conn *conn, int cq\_size)`}) \\
\textcolor[RGB]{105,105,105}{[PROP]} fpga/conn.c:436 (\texttt{`int inlen, err, eqn;`}) \\
\textcolor[RGB]{0,128,0}{[CALL]} fpga/conn.c:458 (\texttt{`inlen = MLX5\_ST\_SZ\_BYTES(create\_cq\_in) +`}) \\
\textcolor[RGB]{0,128,0}{[CALL]} fpga/conn.c:459 (\texttt{`sizeof(u64) * conn->cq.wq\_ctrl.buf.npages;`}) \\
\textcolor[RGB]{255,0,0}{[TARGET]} fpga/conn.c:460 (\texttt{`in = kvzalloc(inlen, GFP\_KERNEL);`}) \\
\textcolor[RGB]{184,134,11}{[COND]} fpga/conn.c:461 (\texttt{`if (!in) \{`}) \\
\textcolor[RGB]{0,128,0}{[CALL]} fpga/conn.c:470 (\texttt{`cqc = MLX5\_ADDR\_OF(create\_cq\_in, in, cq\_context);`}) \\
\textcolor[RGB]{0,128,0}{[CALL]} fpga/conn.c:471 (\texttt{`MLX5\_SET(cqc, cqc, log\_cq\_size, ilog2(cq\_size));`}) \\
\textcolor[RGB]{0,128,0}{[CALL]} fpga/conn.c:472 (\texttt{`MLX5\_SET(cqc, cqc, c\_eqn, eqn);`}) \\
\textcolor[RGB]{0,128,0}{[CALL]} fpga/conn.c:473 (\texttt{`MLX5\_SET(cqc, cqc, uar\_page, fdev->conn\_res.uar->index);`}) \\
\textcolor[RGB]{105,105,105}{[PROP]} fpga/conn.c:474 (\texttt{`MLX5\_SET(cqc, cqc, log\_page\_size, conn->cq.wq\_ctrl.buf.page\_shift -`}) \\
\textcolor[RGB]{0,128,0}{[CALL]} fpga/conn.c:478 (\texttt{`pas = (\_\_be64 *)MLX5\_ADDR\_OF(create\_cq\_in, in, pas);`}) \\
\textcolor[RGB]{0,128,0}{[CALL]} fpga/conn.c:479 (\texttt{`mlx5\_fill\_page\_frag\_array(\&conn->cq.wq\_ctrl.buf, pas);`}) \\
\textcolor[RGB]{0,128,0}{[CALL]} fpga/conn.c:481 (\texttt{`err = mlx5\_core\_create\_cq(mdev, \&conn->cq.mcq, in, inlen, out, sizeof(out));`}) \\
\textcolor[RGB]{0,128,0}{[CALL]} fpga/conn.c:482 (\texttt{`kvfree(in);`}) \\
\textcolor[RGB]{0,128,0}{[CALL]} fpga/conn.c:487 (\texttt{`conn->cq.mcq.cqe\_sz     = 64;`}) \\
\textcolor[RGB]{0,128,0}{[CALL]} fpga/conn.c:488 (\texttt{`conn->cq.mcq.set\_ci\_db  = conn->cq.wq\_ctrl.db.db;`}) \\
\textcolor[RGB]{0,128,0}{[CALL]} fpga/conn.c:489 (\texttt{`conn->cq.mcq.arm\_db     = conn->cq.wq\_ctrl.db.db + 1;`}) \\
\textcolor[RGB]{0,128,0}{[CALL]} fpga/conn.c:490 (\texttt{`*conn->cq.mcq.set\_ci\_db = 0;`}) \\
\textcolor[RGB]{0,128,0}{[CALL]} fpga/conn.c:491 (\texttt{`*conn->cq.mcq.arm\_db    = 0;`}) \\
\textcolor[RGB]{0,128,0}{[CALL]} fpga/conn.c:492 (\texttt{`conn->cq.mcq.vector     = 0;`}) \\
\textcolor[RGB]{0,128,0}{[CALL]} fpga/conn.c:493 (\texttt{`conn->cq.mcq.comp       = mlx5\_fpga\_conn\_cq\_complete;`}) \\
\textcolor[RGB]{0,128,0}{[CALL]} fpga/conn.c:494 (\texttt{`conn->cq.mcq.event      = mlx5\_fpga\_conn\_cq\_event;`}) \\
\textcolor[RGB]{0,128,0}{[CALL]} fpga/conn.c:495 (\texttt{`conn->cq.mcq.irqn       = irqn;`}) \\
\textcolor[RGB]{0,128,0}{[CALL]} fpga/conn.c:496 (\texttt{`conn->cq.mcq.uar        = fdev->conn\_res.uar;`}) \\
\textcolor[RGB]{0,128,0}{[CALL]} fpga/conn.c:500 (\texttt{`mlx5\_fpga\_dbg(fdev, "Created CQ \#0x\%x\textbackslash n", conn->cq.mcq.cqn);`}) \\
\textcolor[RGB]{255,140,0}{[RET]} fpga/conn.c:507 (\texttt{`return err;`}) \\
\textbf{Variable: \texttt{mlx5\_core\_create\_cq}} \\
\textcolor[RGB]{0,0,255}{[SOURCE]} core/cq.c:90 (\texttt{`int mlx5\_core\_create\_cq(struct mlx5\_core\_dev *dev, struct mlx5\_core\_cq *cq,`}) \\
\textcolor[RGB]{105,105,105}{[PROP]} core/cq.c:91 (\texttt{`u32 *in, int inlen, u32 *out, int outlen)`}) \\
\textcolor[RGB]{0,128,0}{[CALL]} core/cq.c:93 (\texttt{`int eqn = MLX5\_GET(cqc, MLX5\_ADDR\_OF(create\_cq\_in, in, cq\_context), c\_eqn);`}) \\
\textcolor[RGB]{0,128,0}{[CALL]} core/cq.c:94 (\texttt{`u32 dout[MLX5\_ST\_SZ\_DW(destroy\_cq\_out)];`}) \\
\textcolor[RGB]{0,128,0}{[CALL]} core/cq.c:95 (\texttt{`u32 din[MLX5\_ST\_SZ\_DW(destroy\_cq\_in)];`}) \\
\textcolor[RGB]{0,128,0}{[CALL]} core/cq.c:99 (\texttt{`eq = mlx5\_eqn2comp\_eq(dev, eqn);`}) \\
\textcolor[RGB]{184,134,11}{[COND]} core/cq.c:100 (\texttt{`if (IS\_ERR(eq))`}) \\
\textcolor[RGB]{0,128,0}{[CALL]} core/cq.c:101 (\texttt{`return PTR\_ERR(eq);`}) \\
\textcolor[RGB]{0,128,0}{[CALL]} core/cq.c:103 (\texttt{`memset(out, 0, outlen);`}) \\
\textcolor[RGB]{0,128,0}{[CALL]} core/cq.c:104 (\texttt{`MLX5\_SET(create\_cq\_in, in, opcode, MLX5\_CMD\_OP\_CREATE\_CQ);`}) \\
\textcolor[RGB]{0,128,0}{[CALL]} core/cq.c:105 (\texttt{`err = mlx5\_cmd\_exec(dev, in, inlen, out, outlen);`}) \\
\textcolor[RGB]{184,134,11}{[COND]} core/cq.c:106 (\texttt{`if (err)`}) \\
\textcolor[RGB]{255,140,0}{[RET]} core/cq.c:107 (\texttt{`return err;`}) \\
\textcolor[RGB]{0,128,0}{[CALL]} core/cq.c:109 (\texttt{`cq->cqn = MLX5\_GET(create\_cq\_out, out, cqn);`}) \\
\textcolor[RGB]{0,128,0}{[CALL]} core/cq.c:110 (\texttt{`cq->cons\_index = 0;`}) \\
\textcolor[RGB]{0,128,0}{[CALL]} core/cq.c:111 (\texttt{`cq->arm\_sn     = 0;`}) \\
\textcolor[RGB]{0,128,0}{[CALL]} core/cq.c:112 (\texttt{`cq->eq         = eq;`}) \\
\textcolor[RGB]{0,128,0}{[CALL]} core/cq.c:113 (\texttt{`cq->uid = MLX5\_GET(create\_cq\_in, in, uid);`}) \\
\textcolor[RGB]{0,128,0}{[CALL]} core/cq.c:114 (\texttt{`refcount\_set(\&cq->refcount, 1);`}) \\
\textcolor[RGB]{0,128,0}{[CALL]} core/cq.c:115 (\texttt{`init\_completion(\&cq->free);`}) \\
\textcolor[RGB]{184,134,11}{[COND]} core/cq.c:116 (\texttt{`if (!cq->comp)`}) \\
\textcolor[RGB]{0,128,0}{[CALL]} core/cq.c:117 (\texttt{`cq->comp = mlx5\_add\_cq\_to\_tasklet;`}) \\
\textcolor[RGB]{0,128,0}{[CALL]} core/cq.c:119 (\texttt{`cq->tasklet\_ctx.priv = \&eq->tasklet\_ctx;`}) \\
\textcolor[RGB]{0,128,0}{[CALL]} core/cq.c:120 (\texttt{`INIT\_LIST\_HEAD(\&cq->tasklet\_ctx.list);`}) \\
\textcolor[RGB]{0,128,0}{[CALL]} core/cq.c:123 (\texttt{`err = mlx5\_eq\_add\_cq(\&eq->core, cq);`}) \\
\textcolor[RGB]{184,134,11}{[COND]} core/cq.c:124 (\texttt{`if (err)`}) \\
\textcolor[RGB]{0,128,0}{[CALL]} core/cq.c:128 (\texttt{`err = mlx5\_eq\_add\_cq(mlx5\_get\_async\_eq(dev), cq);`}) \\
\textcolor[RGB]{184,134,11}{[COND]} core/cq.c:129 (\texttt{`if (err)`}) \\
\textcolor[RGB]{105,105,105}{[PROP]} core/cq.c:132 (\texttt{`cq->pid = current->pid;`}) \\
\textcolor[RGB]{0,128,0}{[CALL]} core/cq.c:133 (\texttt{`err = mlx5\_debug\_cq\_add(dev, cq);`}) \\
\textcolor[RGB]{184,134,11}{[COND]} core/cq.c:134 (\texttt{`if (err)`}) \\
\textcolor[RGB]{0,128,0}{[CALL]} core/cq.c:135 (\texttt{`mlx5\_core\_dbg(dev, "failed adding CP 0x\%x to debug file system\textbackslash n",`}) \\
\textcolor[RGB]{0,128,0}{[CALL]} core/cq.c:136 (\texttt{`cq->cqn);`}) \\
\textcolor[RGB]{0,128,0}{[CALL]} core/cq.c:138 (\texttt{`cq->uar = dev->priv.uar;`}) \\
\textcolor[RGB]{255,140,0}{[RET]} core/cq.c:140 (\texttt{`return 0;`}) \\
\textcolor[RGB]{0,128,0}{[CALL]} core/cq.c:143 (\texttt{`mlx5\_eq\_del\_cq(\&eq->core, cq);`}) \\
\textcolor[RGB]{0,128,0}{[CALL]} core/cq.c:145 (\texttt{`memset(din, 0, sizeof(din));`}) \\
\textcolor[RGB]{0,128,0}{[CALL]} core/cq.c:146 (\texttt{`memset(dout, 0, sizeof(dout));`}) \\
\textcolor[RGB]{0,128,0}{[CALL]} core/cq.c:147 (\texttt{`MLX5\_SET(destroy\_cq\_in, din, opcode, MLX5\_CMD\_OP\_DESTROY\_CQ);`}) \\
\textcolor[RGB]{0,128,0}{[CALL]} core/cq.c:148 (\texttt{`MLX5\_SET(destroy\_cq\_in, din, cqn, cq->cqn);`}) \\
\textcolor[RGB]{0,128,0}{[CALL]} core/cq.c:149 (\texttt{`MLX5\_SET(destroy\_cq\_in, din, uid, cq->uid);`}) \\
\textcolor[RGB]{0,128,0}{[CALL]} core/cq.c:150 (\texttt{`mlx5\_cmd\_exec(dev, din, sizeof(din), dout, sizeof(dout));`}) \\
\textcolor[RGB]{255,140,0}{[RET]} core/cq.c:151 (\texttt{`return err;`}) \\
\textbf{Variable: \texttt{dev}} \\
\textcolor[RGB]{0,0,255}{[SOURCE]} core/cq.c:90 (\texttt{`int mlx5\_core\_create\_cq(struct mlx5\_core\_dev *dev, struct mlx5\_core\_cq *cq,`}) \\
\textcolor[RGB]{0,128,0}{[CALL]} core/cq.c:99 (\texttt{`eq = mlx5\_eqn2comp\_eq(dev, eqn);`}) \\
\textcolor[RGB]{184,134,11}{[COND]} core/cq.c:100 (\texttt{`if (IS\_ERR(eq))`}) \\
\textcolor[RGB]{0,128,0}{[CALL]} core/cq.c:101 (\texttt{`return PTR\_ERR(eq);`}) \\
\textcolor[RGB]{0,128,0}{[CALL]} core/cq.c:105 (\texttt{`err = mlx5\_cmd\_exec(dev, in, inlen, out, outlen);`}) \\
\textcolor[RGB]{255,140,0}{[RET]} core/cq.c:107 (\texttt{`return err;`}) \\
\textcolor[RGB]{0,128,0}{[CALL]} core/cq.c:109 (\texttt{`cq->cqn = MLX5\_GET(create\_cq\_out, out, cqn);`}) \\
\textcolor[RGB]{105,105,105}{[PROP]} core/cq.c:110 (\texttt{`cq->cons\_index = 0;`}) \\
\textcolor[RGB]{105,105,105}{[PROP]} core/cq.c:111 (\texttt{`cq->arm\_sn     = 0;`}) \\
\textcolor[RGB]{0,128,0}{[CALL]} core/cq.c:112 (\texttt{`cq->eq         = eq;`}) \\
\textcolor[RGB]{0,128,0}{[CALL]} core/cq.c:113 (\texttt{`cq->uid = MLX5\_GET(create\_cq\_in, in, uid);`}) \\
\textcolor[RGB]{105,105,105}{[PROP]} core/cq.c:114 (\texttt{`refcount\_set(\&cq->refcount, 1);`}) \\
\textcolor[RGB]{105,105,105}{[PROP]} core/cq.c:115 (\texttt{`init\_completion(\&cq->free);`}) \\
\textcolor[RGB]{184,134,11}{[COND]} core/cq.c:116 (\texttt{`if (!cq->comp)`}) \\
\textcolor[RGB]{105,105,105}{[PROP]} core/cq.c:117 (\texttt{`cq->comp = mlx5\_add\_cq\_to\_tasklet;`}) \\
\textcolor[RGB]{0,128,0}{[CALL]} core/cq.c:119 (\texttt{`cq->tasklet\_ctx.priv = \&eq->tasklet\_ctx;`}) \\
\textcolor[RGB]{0,128,0}{[CALL]} core/cq.c:120 (\texttt{`INIT\_LIST\_HEAD(\&cq->tasklet\_ctx.list);`}) \\
\textcolor[RGB]{0,128,0}{[CALL]} core/cq.c:123 (\texttt{`err = mlx5\_eq\_add\_cq(\&eq->core, cq);`}) \\
\textcolor[RGB]{0,128,0}{[CALL]} core/cq.c:128 (\texttt{`err = mlx5\_eq\_add\_cq(mlx5\_get\_async\_eq(dev), cq);`}) \\
\textcolor[RGB]{105,105,105}{[PROP]} core/cq.c:132 (\texttt{`cq->pid = current->pid;`}) \\
\textcolor[RGB]{0,128,0}{[CALL]} core/cq.c:133 (\texttt{`err = mlx5\_debug\_cq\_add(dev, cq);`}) \\
\textcolor[RGB]{0,128,0}{[CALL]} core/cq.c:135 (\texttt{`mlx5\_core\_dbg(dev, "failed adding CP 0x\%x to debug file system\textbackslash n",`}) \\
\textcolor[RGB]{0,128,0}{[CALL]} core/cq.c:136 (\texttt{`cq->cqn);`}) \\
\textcolor[RGB]{0,128,0}{[CALL]} core/cq.c:138 (\texttt{`cq->uar = dev->priv.uar;`}) \\
\textcolor[RGB]{0,128,0}{[CALL]} core/cq.c:143 (\texttt{`mlx5\_eq\_del\_cq(\&eq->core, cq);`}) \\
\textcolor[RGB]{0,128,0}{[CALL]} core/cq.c:148 (\texttt{`MLX5\_SET(destroy\_cq\_in, din, cqn, cq->cqn);`}) \\
\textcolor[RGB]{0,128,0}{[CALL]} core/cq.c:149 (\texttt{`MLX5\_SET(destroy\_cq\_in, din, uid, cq->uid);`}) \\
\textcolor[RGB]{0,128,0}{[CALL]} core/cq.c:150 (\texttt{`mlx5\_cmd\_exec(dev, din, sizeof(din), dout, sizeof(dout));`}) \\
\textcolor[RGB]{255,140,0}{[RET]} core/cq.c:151 (\texttt{`return err;`}) \\

\end{agentbox}

\subsection{Phase III: Dialectical Verification}
Equipped with the closed evidence trace, the Verifier Agent constructs competing arguments for and against exploitability and correctly arrives at a VULNERABLE verdict. However, as shown below, the supporting argument contains structural characteristics that will later be misclassified as reasoning flaws by the Audit Agent:

\begin{agentbox}[Verifier Agent Output]{orange}
\textbf{Comprehension}
The suspicious code at line 460 calls \texttt{kvzalloc(inlen, GFP\_KERNEL)} where \texttt{inlen} is calculated at lines 458--459 as \texttt{MLX5\_ST\_SZ\_BYTES(create\_cq\_in) + sizeof(u64) * conn->cq.wq\_ctrl.buf.npages}. The value of \texttt{inlen} is derived from two components: (1) \texttt{MLX5\_ST\_SZ\_BYTES(create\_cq\_in)}, likely a compile-time constant representing the size of a CQ creation structure, and (2) \texttt{sizeof(u64) * conn->cq.wq\_ctrl.buf.npages}, a dynamic calculation based on the number of pages in a buffer. If \texttt{conn->cq.wq\_ctrl.buf.npages} is very large or negative, the multiplication could cause integer overflow/wrap-around, making \texttt{inlen} much smaller than expected or even negative. Safety would require either bounds checking on \texttt{npages} before the multiplication, or using a safe multiplication function that detects overflow.

\textbf{Red Team (Case for VULNERABLE):}
\begin{itemize}
    \item \texttt{conn->cq.wq\_ctrl.buf.npages} could be attacker-controlled through FPGA connection setup parameters
    \item No bounds checking on \texttt{npages} is visible in the trace before the multiplication at line 459
    \item If \texttt{npages} is large enough (e.g., $\geq 2^{31} / 8$ on 32-bit systems), the multiplication \texttt{sizeof(u64) * npages} could overflow, resulting in a small \texttt{inlen}
    \item \texttt{kvzalloc} would then allocate a small buffer, but subsequent code would write much more data (the page array), causing heap overflow
    \item Line 479 shows \texttt{\seqsplit{mlx5\_fill\_page\_frag\_array(\&conn->cq.wq\_ctrl.buf, pas)}} which would write \texttt{npages * sizeof(u64)} bytes, potentially overflowing the allocated buffer
    \item This could lead to heap corruption and potentially code execution
\end{itemize}
\vspace{0.5em}
\textbf{Blue Team (Case for NOT\_VULNERABLE):}
\begin{itemize}
    \item \texttt{conn->cq.wq\_ctrl.buf.npages} is likely controlled by the kernel/driver, not external attackers
    \item The buffer structure (\texttt{conn->cq.wq\_ctrl.buf}) is likely initialized and validated through proper kernel APIs
    \item \texttt{npages} is probably bounded by hardware limits or kernel memory management constraints
    \item The code checks for allocation failure at line 461 (\texttt{if (!in)}) and handles it properly
    \item This is kernel code with elevated privileges, so even if overflow occurs, the impact might be limited to the kernel context
\end{itemize}
\vspace{0.5em}
\textbf{Adjudication:} \\
Red Team has a valid concern about the lack of bounds checking on \texttt{npages} before the multiplication. The absence of any validation in the trace is concrete evidence of a potential vulnerability. Blue Team's argument relies on assumptions about kernel internals and hardware limits, but there is no evidence in the trace that \texttt{npages} is actually bounded. The key evidence is that line 459 performs \texttt{sizeof(u64) * conn->cq.wq\_ctrl.buf.npages} without any validation, and this value is used for allocation size. If \texttt{npages} can be controlled (even through kernel APIs), this represents a real integer overflow risk.

\textbf{Verdict:} \texttt{VULNERABLE} \\
\textbf{Confidence:} 0.75 \\
\textbf{CWE\_id:} CWE-190\\
\textbf{Key Evidence:} No bounds checking on conn->cq.wq\_ctrl.buf.npages before multiplication at line 459, potentially causing integer overflow in kvzalloc allocation size
\end{agentbox}

\subsection{Phase IV: Meta-Auditing}
The Audit Agent independently reviews the Verifier's reasoning. In this case, it incorrectly identifies the Verifier's inference from the absence of a bounds check as a Speculation flaw, and exercises its veto power to overturn the correct VULNERABLE verdict, producing a false negative:

\begin{agentbox}[Audit Agent Output]{purple}
\textbf{Independent Assessment:} \\
{The code allocates memory using \texttt{kvzalloc(inlen, GFP\_KERNEL)} where \texttt{inlen} is calculated as \texttt{MLX5\_ST\_SZ\_BYTES(create\_cq\_in) + sizeof(u64) * conn->cq.wq\_ctrl.buf.npages}. The security concern is potential integer overflow in the multiplication operation if \texttt{npages} is sufficiently large, which could lead to a smaller-than-expected allocation and subsequent heap buffer overflow when \texttt{mlx5\_fill\_page\_frag\_array()} writes the page array. The security property at risk is memory safety --- preventing heap corruption through integer overflow in allocation size computation. The mitigation would be bounds checking on \texttt{npages} or using safe multiplication functions.}

\textbf{Evidence Cross-Check} \\
The verifier's factual claims are accurate:
\begin{itemize}
    \item Lines 458--459: \texttt{inlen = MLX5\_ST\_SZ\_BYTES(create\_cq\_in) + sizeof(u64) * conn->cq.wq\_ctrl.buf.npages}
    \item Line 460: \texttt{in = kvzalloc(inlen, GFP\_KERNEL)} performs the allocation using the computed \texttt{inlen}
    \item Line 479: \texttt{\seqsplit{mlx5\_fill\_page\_frag\_array(\&conn->cq.wq\_ctrl.buf, pas)}} would write \texttt{npages * sizeof(u64)} bytes
    \item The verifier correctly notes that no bounds checking on \texttt{npages} is visible in the provided trace
\end{itemize}

\textbf{Reasoning Quality Audit} \\
The verifier's reasoning contains several identifiable flaws:

\textbf{Red Team analysis (VULNERABLE case):}
\begin{itemize}
    \item \textbf{Speculation Flaw:} The claim that ``\texttt{npages} could be attacker-controlled through FPGA connection setup parameters'' is speculative. The trace shows no evidence of external attacker control --- this is internal kernel driver code
    \item \textbf{Pattern-Matching Flaw:} The argument focuses on the pattern (multiplication without bounds checking) but does not demonstrate that \texttt{npages} can actually reach values that would cause overflow in this specific context
\end{itemize}

\textbf{Blue Team analysis (NOT\_VULNERABLE case):}
\begin{itemize}
    \item \textbf{Over-Trust Flaw:} The argument that ``\texttt{npages} is likely controlled by the kernel/driver'' and ``probably bounded by hardware limits'' relies on assumptions without concrete evidence from the trace
    \item \textbf{Incomplete Protection Flaw:} The allocation failure check at line 461 only handles allocation failures, not integer overflow scenarios where allocation succeeds but with the wrong size
\end{itemize}

\textbf{Both Teams:}
\begin{itemize}
    \item \textbf{Scope Creep Flaw:} Both teams argue about potential upstream validation or hardware limits that are not visible in the provided trace
\end{itemize}

The verifier's adjudication makes a critical error: it treats the absence of validation in the trace as concrete evidence of vulnerability, but fails to consider that the trace is incomplete and does not show how \texttt{conn->cq.wq\_ctrl.buf.npages} is initialized or bounded.

\textbf{Independent Verdict} \\
I disagree with the verifier's \texttt{VULNERABLE} verdict. The verifier committed an ``Absence-as-Evidence Flaw'' by concluding \texttt{VULNERABLE} primarily because ``no mitigation is visible in the trace'' when the trace is incomplete. An incomplete trace means uncertainty, not confirmed vulnerability. The verifier also made speculative claims about attacker control without evidence. Without seeing how \texttt{npages} is initialized and bounded, we cannot conclude there is a concrete vulnerability.

\vspace{0.5em}
\textbf{Reasoning Flaws Detected in Verifier's Output:} Absence-as-Evidence Flaw, Speculation Flaw

\vspace{0.5em}
\textbf{Final Decision:} \\
\textbf{Action:} \texttt{DISAGREE} $\rightarrow$ \textbf{Final Verdict:} \texttt{NOT\_VULNERABLE} \\
\textbf{Confidence:} 0.70
\end{agentbox}

\end{document}